\documentclass[3p]{elsarticle}
\usepackage{graphicx}
\usepackage{amsmath,amsfonts,amssymb,amsthm}
\usepackage{bm}
\usepackage{tabularx}
\usepackage{comment}
\usepackage[lined]{algorithm2e}
\usepackage{ulem}
\usepackage[hidelinks]{hyperref}
\usepackage{float,caption,subcaption}
\usepackage{color, xcolor}
\DeclareMathOperator*{\arglocmin}{arg\,loc\,min}
\DeclareMathOperator*{\argmin}{arg\,min}

\definecolor{ETH_gruen}{HTML}{627313}
\definecolor{ETH_rot}{HTML}{B7352D}

\usepackage{booktabs}
\usepackage{pifont}
\usepackage{array}
\usepackage{multirow}
\newcommand{\cmark}{\textcolor{ETH_gruen}{\checkmark}}
\newcommand{\xmark}{\textcolor{ETH_rot}{\ding{55}}}

\newtheorem{remark}{Remark}

\newtheorem{properties}{Properties}
\newtheorem{proposition}{Proposition}

\begin{document}

\begin{frontmatter}
\title{Iterative convergence in phase-field brittle fracture computations:\\exact line search is all you need}
\journal{arXiv}

\author{J. Heinzmann}
\author{F. Vicentini}
\author{P. Carrara}
\author{L. De Lorenzis}
\address{Computational Mechanics Group, ETH Zürich, Tannenstrasse 3, 8092 Zürich, Switzerland}

\begin{abstract}
    Variational phase-field models of brittle fracture pose a local constrained minimization problem of a non-convex energy functional.
    In the discrete setting, the problem is most often solved by alternate minimization, exploiting the separate convexity of the energy with respect to the two unknowns.
    This approach is theoretically guaranteed to converge, provided each of the individual subproblems is solved successfully.
    However, strong non-linearities of the energy functional may lead to failure of iterative convergence within one or both subproblems.
    We analyze and visualize the energy along Newton directions to illustrate why Newton's method without line search fails.
    Motivated by this, we propose to employ an exact line search algorithm based on bisection, which (under certain conditions) can guarantee global convergence of Newton's method for each subproblem and consequently the successful determination of critical points of the energy through the alternate minimization scheme.
    Through several benchmark tests computed with various strain energy decompositions and two strategies for the enforcement of the irreversibility constraint in two and three dimensions, we demonstrate the robustness of the approach and assess its efficiency in comparison with other commonly used line search algorithms.
    With the outlined approach, we are able to compute the especially demanding Brazilian test featuring contact in 3D with the star-convex model.
\end{abstract}


\end{frontmatter}

\section{Introduction}
The phase-field approach to fracture has emerged as a powerful framework to model the nucleation and propagation of cracks in brittle materials.
It was originally proposed in \cite{bourdin_numerical_2000} as the regularization of the variational reformulation of Griffith's theory \cite{griffith_phenomena_1921,francfort_revisiting_1998} based on the Ambrosio–Tortorelli functional \cite{ambrosio_approximation_1992}.
The approach is based on the introduction of a phase-field variable to describe the smooth transition between the intact and the fully broken material states.
As shown in \cite{pham_gradient_2011}, the resulting model can be interpreted as a special family of gradient damage models, whereby the phase-field variable can be interpreted as a damage variable.
This interpretation motivates an irreversibility constraint, i.e. the phase-field variable is only allowed to increase over time.
Following variational principles, the evolution of the state of the system -- as represented by the displacement and the phase fields -- is obtained through the incremental local constrained minimization of a total energy functional \cite{bourdin_numerical_2000}.

The necessary conditions of the minimization problem (i.e. the stationarity conditions for the energy) deliver a system of coupled non-linear partial differential equations (PDEs) which are solved numerically, typically with the finite element (FE) method. The numerical solution is known to be challenging for two main reasons \cite{gerasimov_penalization_2019}.
The first issue is the \textit{computational cost}, as the regularized crack representation necessitates a spatially fine discretization (at least locally).
The second issue is \textit{robustness}, as the total energy functional is non-convex \cite{Svolos_convexity_2023} and generally strongly non-linear with respect to the coupled displacement and phase-field variables.
As a result, iterative solution algorithms may need a significant number of iterations, or even fail to reach convergence.
To tackle the first issue, various strategies have been proposed, e.g. adaptive mesh refinement and/or time stepping \cite{Freddi_mesh_2022,Gupta_adaptive_2022,Kristensen_phase_2020,Wu_BFGS_2020,Rohracker_efficient_2025}, pre-conditioning strategies \cite{Badri_preconditioning_2021,kopanicakova_nonlinear_2022,Jodlbauer_parallel_2020}, and massive parallelization \cite{chafia_massively_2025,ZiaeiRad_massive_2016}.
In this paper, we focus on the second issue.

For the solution of the governing PDEs, the two main options are the \textit{monolithic} and the \textit{staggered} solution strategies.
With the \textit{monolithic} approach, the coupled system of equations is simultaneously solved for both fields, typically with Newton's method.
Unfortunately, the tangent stiffness matrix may become indefinite, and iterative convergence issues typically arise due to the non-convexity of the total energy \cite{gerasimov_linesearch_2016}.
As a remedy, the employment of globalization strategies has been proposed, e.g. line search \cite{gerasimov_linesearch_2016,kopanicakova_nonlinear_2022,Lampron_efficient_2021} or trust-region methods \cite{Kopanicakova_recursive_2020}.
To deal with the potential indefiniteness of the stiffness matrix, several modifications to Newton's method have been attempted, e.g. inertia corrections to the stiffness matrix \cite{Lampron_efficient_2021}, freezing of the phase-field variable to an extrapolated value in the displacement subproblem \cite{Heister_primal_2015}, or a dynamic modification of the coupling term \cite{Wick_modified_2017}.
A comparison of various improvement strategies of Newton's method within the monolithic solution approach can be found in \cite{Wick_modified_2017}, where it is concluded that designing a robust monolithic solution scheme remains challenging.

The \textit{staggered} solution scheme, also known as \textit{alternate minimization}, exploits the separate convexity of the energy functional with respect to each of the arguments by solving alternately for one of them while keeping the other one fixed, until convergence is reached.
While this strategy is typically more robust than the monolithic approach, a large number of staggered iterations may be needed to reach convergence.
Hence, various acceleration methods have been proposed \cite{KirkesaetherBrun_iterative_2020,Luo_fast_2023,farell_linear_2017,Storvik_accelerated_2021}.
Importantly, the staggered scheme is theoretically guaranteed to converge to a stationary point of the energy, of course provided that the individual subproblems, which in turn require an iterative scheme (typically with Newton's method), are solved successfully \cite{bourdin_numerical_2007,Burke_adaptive_2010}.
Unfortunately, due to their strong non-linearity, the individual subproblems may not reach iterative convergence \cite{Vicentini_energy_2024}.
Convergence issues are particularly likely in the displacement subproblem when adopting strain energy decompositions to describe fracture under multiaxial stress states \cite{amor_regularized_2009,miehe2010thermodynamically,freddi_regularized_2010,de_lorenzis_nucleation_2022,Vicentini_energy_2024}, or in the phase-field subproblem when adopting a penalization method to enforce irreversibility \cite{gerasimov_penalization_2019}.

In this study, our goal is to overcome the iterative convergence issues of Newton's method for the individual subproblems within the staggered solution scheme.
With these issues solved, this scheme becomes a robust approach which, under certain conditions, is theoretically guaranteed (and practically observed) to converge.
To achieve our goal, we propose to enhance Newton's method with an \textit{exact line search} to find the minimum of the energy along the Newton direction for each iteration.
Following classical textbooks, e.g. \cite{Nocedal_numerical_2006}, we denote as \textit{exact line search} an algorithm which minimizes some objective function along the search direction (as opposed to an \textit{inexact line search} which accepts a step length that is 'good enough' according to some preset conditions).
In our proposed strategy, we employ a simple and robust bisection algorithm to seek the root of the directional derivative of the energy along the Newton direction.
Building upon the theoretical guarantee of convergence for the bisection algorithm, it is possible to establish a global convergence property of Newton's method with exact line search.
As a result, the overall staggered solution scheme is guaranteed to converge for convex subproblems.
We also study one practical case of non-convexity in the displacement subproblem, namely, the star-convex energy decomposition \cite{Vicentini_energy_2024} for some values of its tuning parameter.
For this case, we establish the robustness of the solution approach empirically through various numerical tests.
Further, we provide insights on the behavior of the phase-field energy functional sampled along the Newton direction.
Finally, we evaluate the computational efficiency of our Newton's method enhanced with exact line search in comparison to other commonly used line search algorithms.
As such, the main contribution of this paper lies in the thorough evaluation, practical validation, and detailed analysis of Newton's method with an exact line search algorithm in the specific and demanding context of variational phase-field fracture models.
We recently contributed the proposed line search algorithm to \texttt{PETSc} \cite{balay_2024a,balay_2024b,balay_1997}, and all our implementations are publicly available at \url{https://github.com/jonas-heinzmann/phase_field_exact_linesearch}.

The remainder of this paper is structured as follows.
We briefly review variational phase-field models of brittle fracture in Section~\ref{sec:formulation_convergence_issues}.
After introducing the FE discretization and the standard staggered solution approach, we demonstrate the iterative convergence issues of the individual subproblems with a set of benchmark tests.
In Section~\ref{sec:bisection_linesearch} we introduce our proposed exact line search algorithm and establish its convergence properties theoretically.
In Section~\ref{sec:assessment}, we demonstrate them numerically on the benchmark tests of Section~\ref{sec:formulation_convergence_issues} and assess the performance of the algorithm more in detail, including a comparison to other commonly used line search algorithms in terms of computational efficiency.
Finally, we showcase the robustness of our method by computing the challenging 3D Brazilian test in Section~\ref{sec:brazilian3D}.
We summarize the main conclusions in Section~\ref{sec:conclusions}.

\section{Phase-field modeling of brittle fracture: formulation, discretization and convergence issues}\label{sec:formulation_convergence_issues}
In this section, we briefly review the phase-field model for brittle fracture used in this study and its FE discretization.
We then describe the standard staggered solution approach and demonstrate the iterative convergence issues in the solution of the displacement and phase-field subproblems by a set of numerical benchmarks.

\subsection{Total energy functional in phase-field modeling of brittle fracture}
We consider a homogeneous isotropic linearly elastic material occupying the domain $\Omega \subset \mathbb{R}^d$, where $d$ is the spatial dimension.
The state at point $\boldsymbol{x} \in \Omega$ is described by the displacement $\boldsymbol{u}$ and the scalar phase-field variable $\alpha \in [0,1]$, with $\alpha=0$ denoting the pristine material and $\alpha=1$ the fully damaged material.
We neglect inertia effects and assume linearized kinematics, i.e. our strain measure is the infinitesimal strain tensor $\boldsymbol{\varepsilon} = \nabla_{\text{sym}} \boldsymbol{u}$.
We formulate the evolution problem in the time-discrete setting \cite{baldelli2021numerical}, where subscripts $n$ and $n-1$ respectively refer to the current time step $t_n$ and to the previous time step $t_{n-1}<t_n$.
For notational brevity, in the following we mostly drop the index $n$.
The \textit{total energy} functional is given by
\begin{equation}\label{eq:energy_functional}
    \mathcal{E}_n(\boldsymbol{u}, \alpha) =
    \int_\Omega \psi(\boldsymbol{\varepsilon}(\boldsymbol{u}), \alpha)\, \mathrm{d}\boldsymbol{x}
    + \int_\Omega \frac{G_\text{c}}{c_w} \left( \frac{w(\alpha)}{\ell} + \ell |\nabla \alpha|^2 \right) \,\mathrm{d}\boldsymbol{x} \quad\text{,}
\end{equation}
where $G_\text{c}$ denotes the fracture toughness, $0<\ell \ll \text{diam}(\Omega)$ the regularization length, $c_w$ is a normalization constant, and the local dissipation function $w(\alpha)$ is a monotonically increasing function with $w(0)=0$, $w(1)=1$.
For simplicity, we do not consider non-homogeneous Neumann boundary conditions in this work.

The spaces of admissible displacement and damage fields are defined as
\begin{equation}
    \mathcal{U}_n = \{ \boldsymbol{u} \in H^1 (\Omega; \mathbb{R}^d) : \boldsymbol{u} = \bar{\boldsymbol{u}}_n\;\text{on}\; \partial_{u} \Omega \}
    \quad\text{and}\quad
    \mathcal{D}_n = \{ \alpha \in H^1 (\Omega) : \alpha= \bar{\alpha}_n\;\text{on}\;\partial_\alpha \Omega,\,\alpha \geq \alpha_{n-1} \} \quad\text{,}
\end{equation}
where $\bar{\boldsymbol{u}}_n$ denotes the prescribed displacement on the displacement Dirichlet boundary $\partial_u \Omega$, $\bar{\alpha}_n$ is the prescribed damage on the damage Dirichlet boundary $\partial\Omega_{\alpha}$, and in $\mathcal{D}_n$ we enforce the \textit{irreversibility} of damage $\alpha\geq\alpha_{n-1}$.
Note that, with an initial damage value $\alpha_0 \geq 0$, the irreversibility condition guarantees that the lower bound for the damage variable, $\alpha\geq0$, is satisfied.

To account for multi-axial stress states, we adopt the \textit{energy decomposition} approach \cite{Vicentini_energy_2024}, in which the elastic energy density $\psi(\boldsymbol{\varepsilon}, \alpha)$ is written as
\begin{equation}\label{eq:split}
    \psi (\boldsymbol{\varepsilon}, \alpha)
    = a(\alpha)\,\psi_{\text{D}} (\boldsymbol{\varepsilon}) + \psi_{\text{R}} (\boldsymbol{\varepsilon}) \quad\text{,} 
    \quad \text{with} \quad \psi_{\text{D}} (\boldsymbol{\varepsilon}) + \psi_{\text{R}} (\boldsymbol{\varepsilon}) = \psi_0(\boldsymbol{\varepsilon}) \quad\text{,}
\end{equation}
where the \textit{degradation function} $a(\alpha)$ is a monotonically decreasing function of $\alpha$ such that $a(0)=1+a_0$ and $a(1)=a_0>0$, with $a_0=o(\ell)$ which we set to $10^{-6}$.
For simplicity, in this paper we adopt the classical second-order polynomial degradation function
\begin{equation}\label{eq:degradation_fun}
    a(\alpha) = (1-\alpha)^2 + a_0 \quad\text{,}
\end{equation}
as introduced in~\cite{bourdin_numerical_2000}. The line-search algorithm that will be proposed, however, is equally applicable to other (convex) degradation functions, including those used in~\cite{lorentz2011convergence, WU201820, zolesi2024stability}. 
The associated local dissipation function $w(\alpha)$ is commonly chosen as either linear or quadratic~\cite{bourdin_numerical_2000, ambrosio_approximation_1992, pham_gradient_2011}, known as \texttt{AT1} and \texttt{AT2}, respectively, i.e.,
\begin{equation}\label{eq:dissipation_fun}
    \texttt{AT1: } w(\alpha) = \alpha, \quad \texttt{AT2: } w(\alpha) = \alpha^2 \quad\text{.}
\end{equation}
For \texttt{AT1}, $c_w=8/3$, while for \texttt{AT2} $c_w=2$~\cite{gerasimov_penalization_2019}. In the theoretical developments that follow, we assume $w(\alpha)$ to be either \texttt{AT1} or \texttt{AT2}.

In \eqref{eq:split}, $\psi_\text{D}$ and $\psi_\text{R}$ denote, respectively, the \textit{degradable} and the \textit{residual} components of the elastic energy density, whose sum is equal to the elastic energy density of the pristine material, $\psi_0(\boldsymbol{\varepsilon}) = \tfrac{\kappa_0}{2}\,\text{tr}^2(\boldsymbol{\varepsilon}) + \mu_0\,\|\boldsymbol{\varepsilon}_\text{dev}\|^2$, with $\kappa_0$ and $\mu_0$ as the bulk and shear moduli, respectively.
The underlying idea of this approach is that only a part of the elastic energy density, namely $\psi_\text{D}$, drives the evolution of damage.
At the same time, it remains possible to store part of the elastic energy density, i.e., $\psi_\text{R}$, even when damage is fully developed ($\alpha=1$).

The variational energy decompositions proposed in the literature include the volumetric--deviatoric \cite{amor_regularized_2009}, the spectral \cite{miehe2010thermodynamically}, the no-tension \cite{freddi_regularized_2010} and the Drucker-Prager (DP)-like \cite{de_lorenzis_nucleation_2022} splits, whose formulations are reported in Appendix~\ref{app:splits}. For the following mathematical developments, 
we assume to be adopting one of these 'classical' models.
In the numerical computations of this work we also employ the \textit{star-convex} formulation proposed in \cite{Vicentini_energy_2024}, see Appendix~\ref{app:splits}, for which the applicability of the following mathematical developments needs some additional specifications, as discussed later in Section~\ref{star-convex}.

With the 'classical' energy decompositions and the previous definitions, the following properties hold:
\begin{properties}\label{prop:psi}
    The elastic energy density $\psi(\boldsymbol{\varepsilon},\alpha)$ is:
    \begin{itemize}
        \item positively homogeneous of degree $2$ and strictly convex with respect to $\boldsymbol{\varepsilon}$;
        \item quadratic and strictly convex with respect to $\alpha$.
    \end{itemize}
\end{properties}
The first follows from the expressions of the strain energy decompositions used in this work, see Appendix~\ref{app:splits}.
The second follows directly from the choice \eqref{eq:degradation_fun}.
For a more detailed analysis and proofs of the mathematical properties, we refer the reader e.g. to \cite{Graeser_truncated_2021,Svolos_convexity_2023,bourdin_variational_2008}.

\subsection{FE discretization}\label{subsec:discretization}
Let the domain $\Omega$ be partitioned into $N_e$ finite elements, each occupying a subdomain $\Omega_e$, $e=1,\dots,N_e$. The displacement and damage fields at time $t_n$ are approximated by their FE counterparts
\begin{equation}
\boldsymbol{u}(\boldsymbol{x}) \simeq \boldsymbol{u}_n^{h}(\boldsymbol{x}) = \boldsymbol{N}_u(\boldsymbol{x}) \, \mathbf{u}^e_n, \quad 
\alpha(\boldsymbol{x}) \simeq \alpha_n^{h}(\boldsymbol{x}) = \boldsymbol{N}_\alpha(\boldsymbol{x}) \, \boldsymbol{\alpha}^e_n 
\quad \text{in } \Omega_e,
\end{equation}
where $\boldsymbol{N}_u$ and $\boldsymbol{N}_\alpha$ are matrices containing the shape functions, and $\mathbf{u}^e_n$, $\bm{\alpha}^e_n$ are the vectors of the element degrees of freedom (DOFs).
The global DOFs are obtained via \textit{assembly}, denoted by $\bigcup_{e=1}^{N_e}$
\begin{equation}
    \begin{bmatrix} \mathbf{u} \\ \bar{\mathbf{u}}_n \end{bmatrix} = \bigcup_{e=1}^{N_e} \mathbf{u}^e_n
    \quad\text{, and}\quad
    \begin{bmatrix} \bm{\alpha} \\ \bar{\bm{\alpha}}_n \end{bmatrix} = \bigcup_{e=1}^{N_e} \bm{\alpha}^e_n,
\end{equation}
with $\mathbf{u}$ and $\bm{\alpha}$ denoting the \textit{unknown} DOFs, and $\bar{\mathbf{u}}_n$ and $\bar{\bm{\alpha}}_n$ the \textit{known} DOFs associated with the Dirichlet boundary conditions on $\partial_u\Omega$ and $\partial_\alpha\Omega$, respectively.
In particular, $\mathbf{u} = [u_1, \dots, u_{N_u}]^\intercal$ collects the $N_u$ unknown displacement DOFs, while $\bm{\alpha} = [\alpha_1, \dots, \alpha_{N_\alpha}]^\intercal$ collects the $N_\alpha$ unknown damage DOFs.
The global vector of unknowns is denoted as $\mathbf{z} = [\mathbf{u}, \bm{\alpha}]^\intercal \in \mathbb{R}^N$, where $N = N_u + N_\alpha$ is the total number of DOFs.

The total energy of the discretized system is expressed in terms of the unknown DOFs as
\begin{equation}\label{eq:energy_FE}
    E_n(\mathbf{z}) = \mathcal{E}_n(\boldsymbol{u}_n^h, \bm{\alpha}_n^h) \quad\text{.}
\end{equation}
Assuming that $E_n$ is a sufficiently smooth function of $\mathbf{z}$, we can write its second-order Taylor expansion around $\mathbf{z}^*=[\mathbf{u}^*, \bm{\alpha}^*]^\intercal$, for any $\mathbf{y} \in \mathbb{R}^N$ and $\epsilon\in\mathbb{R}$, as
\begin{equation}
    E_n(\mathbf{z}^*+ \epsilon\, \mathbf{y}) - E_n(\mathbf{z}^*) = \epsilon\, \mathbf{y}^\intercal\,\mathbf{R}(\mathbf{z}^*) + \frac{\epsilon^2}{2}\, \mathbf{y}^\intercal\,\mathbf{K}(\mathbf{z}^*)\, \mathbf{y} + o\left(\epsilon^2\right) \quad\text{.}
\end{equation}
Here, $\mathbf{R}(\mathbf{z}^*) \in \mathbb{R}^N$ is the \textit{residual vector} composed of the sub-vectors $\mathbf{R}_{\mathbf{u}}(\mathbf{z}^*)  \in \mathbb{R}^{N_u}$ and $\mathbf{R}_{\bm{\alpha}}(\mathbf{z}^*) \in \mathbb{R}^{N_\alpha}$, which correspond to the residuals for the displacement and phase-field degrees of freedom, respectively, i.e.,
\begin{equation}
    \mathbf{R}(\mathbf{z}^*) =
    \begin{bmatrix}
        \mathbf{R}_{\mathbf{u}}(\mathbf{z}^*) \\
        \mathbf{R}_{\bm{\alpha}}(\mathbf{z}^*)
    \end{bmatrix}
    \quad \text{with} \quad
    \mathbf{R}_{\mathbf{u}}(\mathbf{z}^{*}) = \left( \dfrac{\partial E_n(\mathbf{z}^*)}{\partial \mathbf{u}} \right)^\intercal\quad\text{,}
    \quad
    \mathbf{R}_{\bm{\alpha}}(\mathbf{z}^*) = \left( \dfrac{\partial E_n(\mathbf{z}^*)}{\partial \bm{\alpha}} \right)^\intercal\quad\text{,}
\end{equation}
while $\mathbf{K}(\mathbf{z}^*) \in \mathbb{R}^{N \times N}$ is the symmetric \textit{stiffness matrix} composed as
\begin{equation}
    \mathbf{K}(\mathbf{z}^*) =
    \begin{bmatrix}
        \mathbf{K}_{\mathbf{uu}}(\mathbf{z}^*) & \mathbf{K}_{\mathbf{u}\bm{\alpha}}(\mathbf{z}^*) \\[8pt]
        \mathbf{K}_{\bm{\alpha}\mathbf{u}}(\mathbf{z}^*) & \mathbf{K}_{\bm{\alpha\alpha}}(\mathbf{u}^*)
    \end{bmatrix}\quad\text{,}
\end{equation}
where $\mathbf{K}_{\mathbf{uu}}(\mathbf{z}^*)$ and $\mathbf{K}_{\bm{\alpha\alpha}}(\mathbf{u}^*)$ are the symmetric stiffness matrices of the displacement and damage problems, respectively, i.e.,
\begin{equation}
    \mathbf{K}_{\mathbf{uu}}(\mathbf{z}^*) = \dfrac{\partial^2 E_n(\mathbf{z}^*)}{\partial \mathbf{u}^2} \in \mathbb{R}^{N_u \times N_u}, \quad \mathbf{K}_{\bm{\alpha\alpha}}(\mathbf{u}^*)= \dfrac{\partial^2 E_n(\mathbf{z}^*)}{\partial \bm{\alpha}^2} \in \mathbb{R}^{N_\alpha \times N_\alpha}
\end{equation}
and $\mathbf{K}_{\mathbf{u}\bm{\alpha}}(\mathbf{z}^*)$ and $\mathbf{K}_{\mathbf{u}\bm{\alpha}}(\mathbf{z}^*)$ are the mixed-term matrices given by
\begin{equation}
    \mathbf{K}_{\mathbf{u}\bm{\alpha}}(\mathbf{z}^*) = \dfrac{\partial^2 E_n(\mathbf{z}^*)}{\partial \mathbf{u} \partial \bm{\alpha}} \in \mathbb{R}^{N_u \times N_\alpha},\quad \mathbf{K}_{\bm{\alpha}\mathbf{u}}(\mathbf{z}^*)=\dfrac{\partial^2 E_n(\mathbf{z}^*)}{ \partial \bm{\alpha}\partial \mathbf{u}}= \left( \mathbf{K}_{\mathbf{u}\bm{\alpha}}(\mathbf{z}^*) \right)^\intercal\in \mathbb{R}^{N_\alpha \times N_u} \quad\text{.}
\end{equation}

For these vectors and matrices, the following properties hold:
\begin{properties}\label{prop:R_K}
    From the choices {\eqref{eq:split}-\eqref{eq:dissipation_fun}} and Properties~\ref{prop:psi}, we obtain:
    \begin{itemize}
        \item $E_n(\mathbf{u},\bm{\alpha})$ is strictly convex with respect to $\mathbf{u}$, hence $\mathbf{K}_{\mathbf{uu}}(\mathbf{u},\bm{\alpha})$ is symmetric positive definite;
        \item $\mathbf{R}_{\mathbf{u}}(\mathbf{u},\bm{\alpha})$ is generally not affine in $\mathbf{u}$ because of the energy decomposition.
            It reduces to an affine form in the absence of an energy decomposition, i.e. for $\psi_{\text{D}}=\psi_0$ and $\psi_{\text{R}}=0$;
        \item $E_n(\mathbf{u},\bm{\alpha})$ is strictly convex with respect to $\bm{\alpha}$, hence $\mathbf{K}_{\bm{\alpha}\bm{\alpha}}(\mathbf{u})$ is symmetric positive definite. Additionally, $\mathbf{K}_{\bm{\alpha}\bm{\alpha}}(\mathbf{u})$ is independent of $\bm{\alpha}$;
        \item  $\mathbf{R}_{\bm{\alpha}}(\mathbf{u},\bm{\alpha})$ is affine with respect to $\bm{\alpha}$, i.e., $\mathbf{R}_{\bm{\alpha}}(\mathbf{u},\bm{\alpha}) = \mathbf{K}_{\bm{\alpha}\bm{\alpha}}(\mathbf{u})\,\bm{\alpha} + \mathbf{R}_{\bm{\alpha}}(\mathbf{u},\mathbf{0})$.
    \end{itemize}
\end{properties}
For a more detailed analysis, we refer e.g. to \cite{Graeser_truncated_2021,Svolos_convexity_2023,bourdin_variational_2008}.

\subsection{Discrete evolution problem}\label{subsec:evolution_problem}
We employ first-order Lagrange elements for the phase-field discretization.
In this way, the pointwise unilateral constraint \( \alpha \geq \alpha_{n-1} \) naturally translates to the componentwise condition $\bm{\alpha} \geq \bm{\alpha}_{n-1}$.
Accordingly, we define the set of admissible states $\mathcal{Z}_n$ as
\begin{equation}
    \mathcal{Z}_n = \left\{ \mathbf{z} = [\mathbf{u}, \bm{\alpha}]^\intercal \in \mathbb{R}^N \equiv \mathbb{R}^{N_u + N_\alpha} \; : \; \bm{\alpha} \geq \bm{\alpha}_{n-1} \right\} \quad\text{.}
\end{equation}
The evolution of the system follows the variational principle of local minimization of the total energy $E_n(\mathbf{z})$ defined in \eqref{eq:energy_FE} over all $\mathbf{z}\in \mathcal{Z}_n$, namely
\begin{equation}\label{eq:full_prob}
    \mathbf{z}_n = [\mathbf{u}_n, \bm{\alpha}_n]^\intercal = \arglocmin_{[\mathbf{u}, \bm{\alpha}]^\intercal \in \mathcal{Z}_n} E_n(\mathbf{u},\bm{\alpha})\quad\text{.}
\end{equation}
This is equivalent to finding $\mathbf{z}_n \in \mathcal{Z}_n$ such that
\begin{equation}
    \forall \mathbf{y}\in \mathcal{Z}_n,\;\exists\,\bar{\epsilon}>0:\quad 
    E_n(\mathbf{z}_n+\epsilon\,(\mathbf{y}-\mathbf{z}_n)) - E_n(\mathbf{z}_n) \geq 0 
    \quad \forall\, \epsilon \in [0,\bar{\epsilon}] \quad\text{.}
\end{equation}
A necessary condition for local stability is given by the first-order stability conditions
\begin{subequations}\label{eq:1_order}
    \begin{align}
        &\mathbf{R}_{\mathbf{u}}(\mathbf{u}_n, \bm{\alpha}_n) = \boldsymbol{0} \quad\text{,} \label{eq:1_order_a}\\
        &\mathbf{R}_{\bm{\alpha}}(\mathbf{u}_n,\bm{\alpha}_n) \geq \boldsymbol{0}\quad\text{,}\qquad 
        \bm{\alpha}_n \geq \bm{\alpha}_{n-1},\qquad 
        (\bm{\alpha}_n-\bm{\alpha}_{n-1})^\intercal \mathbf{R}_{\bm{\alpha}}(\mathbf{u}_n,\bm{\alpha}_n) = 0 \quad\text{,} \label{eq:1_order_b}
    \end{align}
\end{subequations}
which correspond to the discrete version of the \textit{equilibrium} equation and of the \textit{damage Karush–Kuhn–Tucker} (KKT) conditions, respectively.

\begin{remark}
    As it is common in the numerical practice, in the remainder of this work we restrict our attention to the first-order problem~\eqref{eq:1_order}.
    This allows us to identify stationary (critical) points of the energy $E_n(\mathbf{z})$, which are not necessarily solutions of problem~\eqref{eq:full_prob}.
    Addressing the first-order problem already entails significant challenges, which are the focus of this work.
    A second-order analysis, as proposed in \cite{baldelli2021numerical}, can subsequently be employed to assess or achieve local minimality; while not applied here, it remains complementary to the present approach.
\end{remark}

\subsection{Alternate minimization with Newton solvers}\label{sec:standard}
The standard procedure to solve the problem in~\eqref{eq:1_order_a}, \eqref{eq:1_order_b} consists in using a staggered (or alternate minimization) scheme, as proposed in~\cite{bourdin_numerical_2000}.
As summarized in Algorithm~\ref{alg:altmin}, this means that at the $i$-th staggered iteration
\begin{itemize}
    \item we first determine the displacement vector $\mathbf{u}^i$ that satisfies the equilibrium equation~\eqref{eq:1_order_a} for a fixed damage field $\boldsymbol{\alpha}^{i-1}$;
        we refer to this step as the \textit{mechanical (sub)problem}.
    \item Then, with the updated displacement solution $\mathbf{u}^i$ held constant, we find the damage variable $\boldsymbol{\alpha}^i$ that satisfies the KKT conditions~\eqref{eq:1_order_b}; this is referred to as the \textit{damage (sub)problem}.
\end{itemize}
These two steps are repeated in an iterative manner until convergence, i.e., until we obtain the vector $\mathbf{z}_n = [\mathbf{u}_n, \boldsymbol{\alpha}_n]^\intercal$ that simultaneously satisfies both~\eqref{eq:1_order_a} and~\eqref{eq:1_order_b}.
In the numerical context, we determine convergence of the alternate minimization scheme based on the norm of the re-evaluated residual of the displacement field with the updated damage, that is $\|\mathbf{R}_{\mathbf{u}} (\mathbf{u}^{i}, \bm{\alpha}^{i}) \|_2 \leq \mathtt{TOL}_{\mathtt{AM}}$ for which we set $\mathtt{TOL}_{\mathtt{AM}} = 10^{-6}$.
Further, we set a maximum number of staggered iterations, $i_{\max} = 5\,000$.

\begin{algorithm}[t]
    \DontPrintSemicolon
    \KwIn{$\mathbf{u}_{n-1}$, $\bm{\alpha}_{n-1}$}
    \KwOut{$\mathbf{z}_n = [\mathbf{u}_n, \bm{\alpha}_n]^\intercal$}
    initialize $i = 0$, $\mathbf{u}^0 \gets \mathbf{u}_{n-1}$, $\bm{\alpha}^0 \gets \bm{\alpha}_{n-1}$\;
    \While{$i<i_{\max}$}{
        \vspace{1mm}
        $i \gets i + 1$\;
        $\mathbf{u}^i \gets \mathbf{u}: \mathbf{R}_{\mathbf{u}}(\mathbf{u}, \boldsymbol{\alpha}^{i-1}) = \boldsymbol{0}$ \quad (mechanical problem)\;
        \vspace{1mm}
        $\boldsymbol{\alpha}^i \gets \boldsymbol{\alpha}: \mathbf{R}_{\bm{\alpha}}(\mathbf{u}^i, \bm{\alpha}) \geq \boldsymbol{0},\;
        \bm{\alpha} \geq \bm{\alpha}_{n-1},\;
        (\bm{\alpha} - \bm{\alpha}_{n-1})^\intercal \mathbf{R}_{\bm{\alpha}}(\mathbf{u}^i, \bm{\alpha}) = 0$
        \quad (damage problem)\;
        \vspace{1mm}
        exit if $\|\mathbf{R}_{\mathbf{u}} (\mathbf{u}^{i}, \bm{\alpha}^{i}) \|_2 \leq \mathtt{TOL}_{\mathtt{AM}}$
        }
    $(\mathbf{u}_n, \boldsymbol{\alpha}_n) \gets (\mathbf{u}^i, \boldsymbol{\alpha}^i)$
    \caption{Alternate minimization loop at time step $n$}
    \label{alg:altmin}
\end{algorithm}

\begin{properties}
\label{prop:vectorE}
Owing to Properties~\ref{prop:R_K}, the mechanical and damage problems have the following properties:
\begin{itemize}
    \item the mechanical problem is equivalent to the convex minimization problem
    \begin{equation}
        \text{find}\quad\mathbf{u}^i = \argmin_{\mathbf{v}\in\mathbb{R}^{N_u}} E_n(\mathbf{v},\bm{\alpha}^{i-1})
    \end{equation} and, given the strict convexity and coercivity of $E_n(\mathbf{u},\bm{\alpha}^i)$ in $\mathbf{u}$, this problem has a unique solution;
    \item the mechanical problem is generally non-linear in $\mathbf{u}$ due to the energy decomposition.
        It reduces to a linear problem in the absence of energy decomposition;
    \item the damage problem is equivalent to the convex constrained minimization problem
      \begin{equation}
        \text{find}\quad\bm{\alpha}^i = \argmin_{\bm{\beta}\in\mathbb{R}^{N_{\alpha}},\bm{\beta} \geq \bm{\alpha}_{n-1}} E_n(\mathbf{u}^{i},\bm{\beta})
    \end{equation}
        and, given the strict convexity of $E_n(\mathbf{u}^i, \bm{\alpha})$ in $\bm{\alpha}$, this problem has a unique solution;
    \item the damage problem corresponds to a linear complementarity problem \cite{marengo2021rigorous}.
\end{itemize}
\end{properties}
Since both subproblems correspond to convex minimization problems, the staggered scheme can be proven to converge to a critical point of the total energy, see \cite[Theorem~1]{bourdin_numerical_2007}, and \cite{Burke_adaptive_2010}.
Clearly, the success of the overall algorithm relies on the successful solution of the two subproblems.

\subsubsection{Newton's method}\label{subsec:Newtons_method}
Both mechanical and damage problems are solved iteratively with Newton's method.
For the mechanical problem, the solution at staggered iteration $i$ and Newton iteration $k$ reads
\begin{equation}\label{eq:Newton_update}
    \mathbf{u}^{i,k+1} = \mathbf{u}^{i,k} + \Delta \mathbf{u}^{i,k} \quad\text{.}
\end{equation}
The solution update $\Delta \mathbf{u}^{i,k}$ is obtained by solving the linear system
\begin{equation}\label{eq:Newton_step}
    \mathbf{K}^{i,k}_{\mathbf{uu}} \Delta \mathbf{u}^{i,k} = -\mathbf{R}_{\mathbf{u}}^{i,k} \qquad\text{.}
\end{equation}
We determine convergence of Newton's method based on the norm of the residual vector, i.e., $\|\mathbf{R}_{\mathbf{u}}^{i,k}\|_2 \leq \mathtt{TOL}_{\mathtt{NM}}$ for which we set $\mathtt{TOL}_{\mathtt{NM}} = 10^{-8}$, while we allow for at most $5\,000$ Newton iterations\footnote{
    A maximum number of iterations equal to $5\,000$ is excessively large for any practical application. However,
    in this work we want to ensure that failure of Newton's method is not due to reaching the threshold on the maximum number of iterations.
    In practice, we do not observe such high iteration counts, as reported in Sections~\ref{sec:assessment} and \ref{sec:brazilian3D} for our numerical examples.
}.
Note that for Newton's method only \textit{local} convergence (i.e., in proximity of the solution) can be proven in general, see e.g. \cite[Theorem~5.2.1]{dennis_numerical_1996}.

\subsubsection{Treatment of contraints}\label{subsec:treatment_of_constraints}
Newton's method cannot directly be applied to solve the KKT system of the damage problem in \eqref{eq:1_order_b}.
Various approaches to solve it include history variable \cite{miehe_phase_2010}, augmented Lagrange \cite{Wheeler_augmented_2014,Wick_modified_2017,Wick_error_2017}, and interior-point methods \cite{Wambacq_interior_2021}.
In this work, we adopt two further alternatives: the \textit{reduced-space active set} method and the penalty method.

\paragraph{Reduced-space active set strategy}
Proposed by \cite{benson_flexible_2006}, this strategy modifies Newton's method based on active and inactive sets (in the sense of the constraint being active or inactive).
At staggered iteration $i$ and Newton iteration $j$ for the damage problem, the active set $\mathcal{A}$ and the inactive set $\mathcal{I}$ are defined as
\begin{subequations}\label{eq:rs_sets}
    \begin{align}
        &\mathcal{A}^{i,j} = \mathcal{A} (\bm{\alpha}^{i,j}, \bm{\alpha}_{n-1}) = \left\{ q \in \{1,\dots,N_\alpha\} : [\bm{\alpha}^{i,j}]_q = [\bm{\alpha}_{n-1}]_q \,\text{and}\, [\mathbf{R}_{\bm{\alpha}}^{i,j}]_q > 0 \right\} \label{eq:rs_sets_A} \qquad\text{and}\\
        &\mathcal{I}^{i,j} = \mathcal{I} (\bm{\alpha}^{i,j}, \bm{\alpha}_{n-1}) = \left\{ q \in \{1,\dots,N_\alpha\} : [\bm{\alpha}^{i,j}]_q > [\bm{\alpha}_{n-1}]_q \,\text{or}\, [\mathbf{R}_{\bm{\alpha}}^{i,j}]_q \leq 0 \right\} \qquad\text{,}\label{eq:rs_sets_B}
    \end{align}
\end{subequations}
where we denote with $[\mathbf{v}]_q$ the $q$-th component of vector $\mathbf{v}$.
With the sets \eqref{eq:rs_sets} at hand, the modified Newton solution update is computed as
\begin{equation}\label{eq:RS_Newtonstep}
    [\Delta \bm{\alpha}^{i,j}]_{\mathcal{I}^{i,j}} = -\left([\mathbf{K}_{\bm{\alpha\alpha}}^{i,j}]_{\mathcal{I}^{i,j}\mathcal{I}^{i,j}}\right)^{-1} [\mathbf{R}_{\bm{\alpha}}^{i,j}]_{\mathcal{I}^{i,j}}
    \qquad\text{and}\qquad
    [\Delta \bm{\alpha}^{i,j}]_{\mathcal{A}^{i,j}} = \mathbf{0} \quad\text{,}
\end{equation}
while the solution update is obtained as
\begin{equation}
    \bm{\alpha}^{i,j+1} = \pi \left( \bm{\alpha}^{i,j} + \Delta \bm{\alpha}^{i,j} \right)
    \label{eq:RS_Newtonupdate}
\end{equation}
with the operator $\pi$ projecting the solution onto the admissible space, i.e. $\pi \left( [\bm{\alpha}^{i,j} + \Delta \bm{\alpha}^{i,j}]_q \right) = [\bm{\alpha}^{i,j} + \Delta \bm{\alpha}^{i,j}]_q$ if $[\bm{\alpha}^{i,j} + \Delta \bm{\alpha}^{i,j}]_q \geq [\bm{\alpha}_{n-1}]_q$ and $\pi \left( [\bm{\alpha}^{i,j} + \Delta \bm{\alpha}^{i,j}]_q \right) = [\bm{\alpha}_{n-1}]_q$ if $[\bm{\alpha}^{i,j} + \Delta \bm{\alpha}^{i,j}]_q < [\bm{\alpha}_{n-1}]_q$.
Finally, for the convergence check of Newton's method, we must consider the norm of a modified residual vector, namely
\begin{equation}\label{eq:RS_residualconvergence}
    [\check{\mathbf{R}}_{\bm{\alpha}}^{i,j}]_q =
    \begin{cases}
        [\mathbf{R}_{\bm{\alpha}}^{i,j}]_q &[\bm{\alpha}^{i,j}]_q > [\bm{\alpha}_{n-1}]_q\\
        \min \{ [\mathbf{R}_{\bm{\alpha}}^{i,j}]_q, 0 \} &[\bm{\alpha}^{i,j}]_q = [\bm{\alpha}_{n-1}]_q\\
    \end{cases}\qquad\text{,}
\end{equation}
i.e. at convergence it must be $\|\check{\mathbf{R}}_{\bm{\alpha}}^{i,j} \|_2 \leq \mathtt{TOL}_{\mathtt{NM}}$.
\begin{properties}
    As outlined in \cite{benson_flexible_2006}, this approach has the following properties:
    \begin{itemize}
        \item by construction, all solution iterates $\bm{\alpha}^{i,j}$ stay within the admissible bounds, and
        \item the properties of symmetry and positive definiteness of the 'full' stiffness matrix $\mathbf{K}_{\bm{\alpha\alpha}}^{i,j}$ are retained also by the reduced stiffness matrix $[\mathbf{K}_{\bm{\alpha\alpha}}^{i,j}]_{\mathcal{I}^{i,j}\mathcal{I}^{i,j}}$.
    \end{itemize}
\end{properties}
To the best of our knowledge, no general proof of convergence is available for this method (as also noted in \cite[p.161]{benson_flexible_2006}).
Nevertheless, we adopt it due to its algorithmic simplicity and since it requires only a few additional operations.
Further, the algorithm is highly efficient for the \texttt{AT1} model, since the linear system has to be solved only on the inactive set $\mathcal{I}^{i,j}$, which is significantly smaller than $N_{\alpha}$ due to the finite support of the phase field.
As we demonstrate later, and as also reported e.g. in \cite{farell_linear_2017}, this algorithm is very robust for the phase-field fracture problem and does not fail to reach convergence in any of our tested cases.
Alternatives with a theoretical guarantee of convergence include the one presented in \cite{Graeser_truncated_2021} and the penalization discussed in the next subsection.
Another similar strategy is the semi-smooth algorithm in \cite{benson_flexible_2006}, which however entails a modification of the stiffness matrix.
With the chosen approach, the convexity properties of the underlying problem remain unchanged, which is beneficial for the later developments in this work.
Due to the fact that the stiffness matrix needs no modification, and since the phase-field stiffness matrix is independent of $\bm{\alpha}$ (Properties~\ref{prop:R_K}), the stiffness matrix can be computed only once at the first iteration and then reused, thereby reducing the computational cost.
Only the active and inactive sets change during the Newton iterations.
\begin{remark}
    The extension of the method to also include an upper bound is straightforward, as implemented in \texttt{PETSc} \cite{balay_2024a,balay_2024b}.
    For this, the definitions of the active and inactive sets in \eqref{eq:rs_sets}, the projection operator in \eqref{eq:RS_Newtonupdate} and the residual vector for convergence checks in \eqref{eq:RS_residualconvergence} must be adjusted to also account for the upper bound.
    We enforce component-wise $\bm{\alpha} \leq \boldsymbol{1}$ since this reduces the size of the inactive set, and hence the number of unknowns in the linear problem to be solved in \eqref{eq:RS_Newtonstep}.
\end{remark}
In the PETSc implementation \cite{balay_1997,balay_2024b}, the active and inactive DOFs are determined with a tolerance of $10^{-8}$.

\paragraph{Penalization}
With the penalty method, an additional term is added to the energy functional \eqref{eq:energy_functional}, yielding the modified energy functional \cite{gerasimov_penalization_2019}
\begin{equation}\label{eq:energy_fun_penalized}
    \tilde{\mathcal{E}}_n (\boldsymbol{u}, \alpha; \alpha_{n-1}) = \mathcal{E} (\boldsymbol{u}, \alpha) + \frac{\epsilon}{2} \int_\Omega \langle \alpha - \alpha_{n-1} \rangle_-^2 \mathrm{d}\boldsymbol{x}
\end{equation}
with the penalty parameter $\epsilon \gg 1$.
After discretization in space, $\tilde{E}_n(\mathbf{u}, \bm{\alpha}) = \tilde{\mathcal{E}}_n(\boldsymbol{u}_n^h, \bm{\alpha}_n^h)$, the evolution of the system follows a modified form of~\eqref{eq:full_prob}, namely
\begin{equation}\label{eq:full_prob_pen}
    \mathbf{z}_n = [\mathbf{u}_n, \bm{\alpha}_n]^\intercal = \arglocmin_{[\mathbf{u}, \bm{\alpha}]^\intercal \in \tilde{\mathcal{Z}}_n} \tilde{E}_n(\mathbf{u},\bm{\alpha})
    \quad\text{with}\quad
    \tilde{\mathcal{Z}}_n = \left\{ \mathbf{z} = [\mathbf{u}, \bm{\alpha}]^\intercal \in \mathbb{R}^N \equiv \mathbb{R}^{N_u + N_\alpha} \right\}
\end{equation}
which is unconstrained.
Accordingly, the first-order stability condition for the damage now leads to $\tilde{\mathbf{R}}_{\bm{\alpha}}(\mathbf{u}_n,\bm{\alpha}_n) = \boldsymbol{0}$, which can be solved with the standard Newton's method.

\begin{properties}
    The penalized formulation has the following properties:
    \begin{itemize}
        \item since the penalty term added to the energy \eqref{eq:energy_fun_penalized} is convex, the modified energy $\tilde{E}$ retains the convexity properties of the original one;
        \item due to the penalty term, the stiffness matrix is no longer independent of $\bm{\alpha}$ and needs to be re-evaluated at each Newton iteration;
        \item the constraints are fulfilled approximately, and exactly only in the limit $\epsilon \rightarrow \infty$ \cite{gerasimov_penalization_2019}.
    \end{itemize}
\end{properties}
The choice of $\epsilon$ is crucial: a too large value may lead to ill-conditioning and iterative convergence issues, and a too small value may yield an insufficiently accurate fulfillment of the constraints.
In this work, we adopt the lower bound for $\epsilon$ proposed in \cite{gerasimov_penalization_2019}, i.e. $\epsilon = \tfrac{G_{\text{c}}}{\ell} \tfrac{27}{64 \mathtt{TOL}_{\mathtt{ir}}^2}$ (for the \texttt{AT1} model) where $0 < \mathtt{TOL}_{ir} \ll 1$ is an irreversibility tolerance threshold.
We select a quite strict tolerance $\mathtt{TOL}_{\text{ir}} = 10^{-4}$ to facilitate the comparison with the reduced-space strategy (which enforces the constraints with significantly higher precision) and to challenge the convergence of Newton's method for the damage problem.

\subsection{Iterative convergence issues}\label{subsec:convergence_issues}
Having chosen the solution approaches, we can evaluate their robustness.
To this end, we consider a set of five benchmark problems, which we refer to as the \textit{obstacle course} in the following.
The set includes a \textit{nucleation test}, a \textit{sliding test}, a \textit{plate with a hole}, a \textit{plate with an inclined crack}, and a \textit{perforated plate}. The first two tests are intended to demonstrate that, with the standard solution approach in Section \ref{sec:standard}, convergence issues may occur even for seemingly very simple setups. The remaining tests are chosen to capture multi-axial stress states leading to non-trivial crack nucleation and/or propagation.
We compute all these tests with the \texttt{AT1} model and the 'classical' strain energy decompositions, namely the volumetric-deviatoric split, the spectral split, the no-tension split, and the DP-like split with $\gamma=\sqrt{2\mu_0/\kappa_0}$, and additionally with the star-convex formulation using $\gamma^\star=1$ and $\gamma^\star=5$.
A detailed description of the setups and of the results is given in Appendix~\ref{app:obstacle_course}, while all implementation details are reported in Appendix~\ref{app:implementational_details}. In this section, we focus on the numerical convergence behavior for the various benchmarks with the different strain energy decompositions and the two methods to enforce irreversibility.

In Tab.~\ref{tab:convergence_obstacle_course} we report whether convergence was reached or not -- denoted by (\cmark) or (\xmark), respectively.
The subscript of (\xmark) indicates whether failure of iterative convergence occurred in the Newton solver for the mechanical problem (\xmark$_\mathbf{u}$) or for the damage problem (\xmark$_{\bm{\alpha}}$).
Clearly, the standard solution approach exhibits convergence issues in the mechanical problem for most of the considered cases.
For both irreversibility enforcement strategies, in all negative cases, failure to reach iterative convergence in the mechanical problem leads to the overall failure of the alternate minimization scheme.

\begin{table}[H]
    \centering
    \footnotesize
    \caption{Overview of whether the standard solution approach can reach convergence for the obstacle course. We denote success and failure with \cmark and \xmark, respectively.
    With \xmark$_{\mathbf{u}}$ we indicate failure to converge  in the mechanical problem.}
    \label{tab:convergence_obstacle_course}
    \begin{tabularx}{\textwidth}{llccccc}
        \toprule
        &&nucleation test &sliding test &plate with hole &plate w. inclined crack &perforated plate\\
        &&\raisebox{-.5\height}{\includegraphics{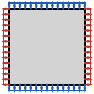}}
        &\raisebox{-.5\height}{\includegraphics{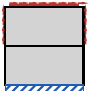}}
        &\raisebox{-.5\height}{\includegraphics{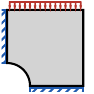}}
        &\raisebox{-.5\height}{\includegraphics{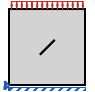}}
        &\raisebox{-.5\height}{\includegraphics{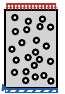}}
        \\
        \midrule
        \parbox[t]{1mm}{\multirow{5}{*}{\rotatebox[origin=c]{90}{red. space$\,\,\,\,$}}}
        &vol-dev                                       &\fbox{\xmark$_{\mathbf{u}}$}           &\cmark                      &\xmark$_{\mathbf{u}}$                  &\xmark$_{\mathbf{u}}$              &\xmark$_{\mathbf{u}}$                  \\
        &star-convex, $\gamma^\star = 1$               &\xmark$_{\mathbf{u}}$                  &\cmark                      &\xmark$_{\mathbf{u}}$                  &\xmark$_{\mathbf{u}}$              &\xmark$_{\mathbf{u}}$                  \\
        &star-convex, $\gamma^\star = 5$               &\xmark$_{\mathbf{u}}$                  &\cmark                      &\xmark$_{\mathbf{u}}$                  &\xmark$_{\mathbf{u}}$              &\xmark$_{\mathbf{u}}$                  \\
        &spectral                                      &\cmark                                 &\cmark                      &\xmark$_{\mathbf{u}}$                  &\cmark                             &\xmark$_{\mathbf{u}}$                  \\
	    &no-tension                                    &\cmark                                 &\xmark$_{\mathbf{u}}$       &\cmark                                 &\cmark                             &\xmark$_{\mathbf{u}}$                  \\
        &DP-like, $\gamma = \sqrt{2\mu_0 / \kappa_0}$      &\xmark$_{\mathbf{u}}$                  &\xmark$_{\mathbf{u}}$       &\xmark$_{\mathbf{u}}$                  &\xmark$_{\mathbf{u}}$              &\xmark$_{\mathbf{u}}$                  \\
        \midrule
        \parbox[t]{1mm}{\multirow{5}{*}{\rotatebox[origin=c]{90}{penalization$\,\,\,\,$}}}
        &vol-dev                                       &\xmark$_{\mathbf{u}}$                  &\cmark                      &\xmark$_{\mathbf{u}}$                  &\xmark$_{\mathbf{u}}$              &\xmark$_{\mathbf{u}}$                  \\
        &star-convex, $\gamma^\star = 1$               &\xmark$_{\mathbf{u}}$                  &\cmark                      &\xmark$_{\mathbf{u}}$                  &\xmark$_{\mathbf{u}}$              &\xmark$_{\mathbf{u}}$                  \\
        &star-convex, $\gamma^\star = 5$               &\xmark$_{\mathbf{u}}$                  &\cmark                      &\xmark$_{\mathbf{u}}$                  &\xmark$_{\mathbf{u}}$              &\xmark$_{\mathbf{u}}$                  \\
        &spectral                                      &\cmark                                 &\cmark                      &\xmark$_{\mathbf{u}}$                  &\cmark                             &\xmark$_{\mathbf{u}}$                  \\
	    &no-tension                                    &\cmark                                 &\xmark$_{\mathbf{u}}$       &\cmark                                 &\cmark                             &\xmark$_{\mathbf{u}}$                  \\
        &DP-like, $\gamma = \sqrt{2\mu_0 / \kappa_0}$      &\xmark$_{\mathbf{u}}$                  &\xmark$_{\mathbf{u}}$       &\xmark$_{\mathbf{u}}$                  &\xmark$_{\mathbf{u}}$              &\xmark$_{\mathbf{u}}$                  \\
        \bottomrule
    \end{tabularx}
\end{table}

\begin{figure}[H]
    \centering
    \begin{subfigure}[t]{0.28\textwidth}
        \centering
        \includegraphics{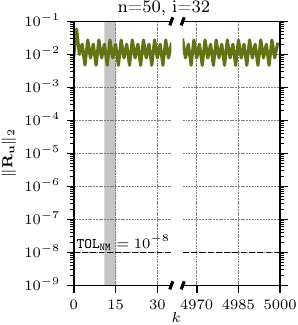}
        \caption{}
        \label{fig:toyproblem_residuals}
    \end{subfigure}
    \hfill
    \begin{subfigure}[t]{0.7\textwidth}
        \centering
        \includegraphics{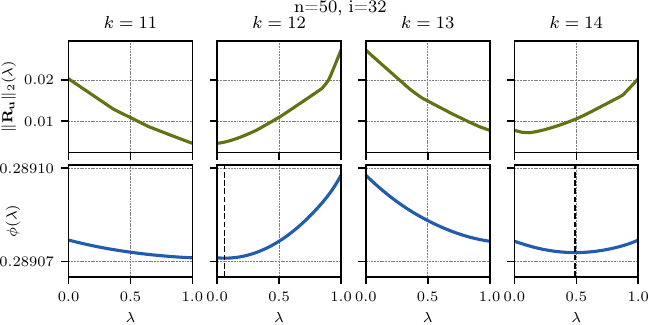}
        \caption{}
        \label{fig:toyproblem_linesearch_sampling}
    \end{subfigure}
    \caption{Oscillation of $\|\mathbf{R}_{\mathbf{u}}\|$ during Newton iterations for the nucleation test with the volumetric-deviatoric split (a), and behavior of $\|\mathbf{R}_{\mathbf{u}}\|_2 (\lambda)$ and $\phi(\lambda)$ between the previous solution iterate and the full Newton step for selected iterations (b). In (b), the value of $\lambda \in [0,1]$ at which the energy is minimized is highlighted with a vertical dashed line.}
    \label{fig:toyproblem}
\end{figure}

To examine these convergence issues further, we select the first case in Tab.~\ref{tab:convergence_obstacle_course}, i.e. the nucleation test with the volumetric-deviatoric split computed with the reduced-space active set strategy (highlighted with a box in the table).
For this example, the residual norm $\|\mathbf{R}_{\mathbf{u}}\|_2$ oscillates between four values until the solver of the mechanical problem reaches the maximum number of allowed iterations, as shown in Fig.~\ref{fig:toyproblem_residuals}.
For these failing iterations, we analyze the change of the residual norm and of the energy in the direction of the Newton step between the solution iterates $\mathbf{u}^{i,k}$ and $\mathbf{u}^{i,k+1}$, i.e.
\begin{equation}
    \|\mathbf{R}_{\mathbf{u}}\|_2 (\lambda) = \|\mathbf{R}_{\mathbf{u}} (\mathbf{u}^{i,k} + \lambda \Delta \mathbf{u}^{i,k}, \bm{\alpha}^{i-1}) \|_2
    \qquad\text{and}\qquad
    \phi (\lambda) = E_n (\mathbf{u}^{i,k} + \lambda \Delta \mathbf{u}^{i,k}, \bm{\alpha}^{i-1})
    \quad\text{,}
\end{equation}
parametrized by $\lambda\in[0,1]$.
Accordingly, $\lambda=0$ corresponds to not updating the solution iterate, and $\lambda=1$ corresponds to the 'full' Newton step~\eqref{eq:Newton_update}, while the intermediate values of $\lambda$ (we sample $100$ values) lead to a 'scaled' Newton step.
Fig.~\ref{fig:toyproblem_linesearch_sampling} shows the behavior of the energy and of the residual norm for the four repeating solution iterates; further examples of the energy along the search direction are reported in Appendix~\ref{app:linesearch_sampling}.
The residual norm displays a strongly non-linear behavior featuring distinct changes in slope along the Newton direction.
Further, while during iterations $k=11$ and $k=13$ the maximum possible decrease of energy within the interval $\lambda \in [0,1]$ is achieved by taking the full Newton step, during iterations $k=12$ and $k=14$ taking the full Newton step leads the energy to either increase or to remain at a similar value as with the previous solution iterate.
These two observations (at least partially) explain the convergence failure of Newton's method and naturally suggest to modify the method by scaling down the full Newton step to seek the minimum of the energy along the Newton step direction.

\section{Exact line search with bisection}\label{sec:bisection_linesearch}
To determine a 'better' step length, either by satisfying suitable conditions \cite{Wolfe_convergence_1969,Armijo_minimization_1966,More_line_1994}, or by (approximately) optimizing an objective function, is precisely the idea of the well-known line-search methods.
With the determined $\lambda^\star$, the modified solution update in Newton's method for the general variable $\mathbf{w} \in \mathbb{R}^{N_w}$ reads
\begin{equation}\label{eq:Newton_ls_udpate}
    \mathbf{w}^{k+1} = \mathbf{w}^k + \lambda^\star \Delta \mathbf{w}^k
\end{equation}
at iteration $k$.
For the reduced-space active set strategy, the solution update reads $\mathbf{w}^{k+1} = \pi(\mathbf{w}^k + \lambda^\star \Delta \mathbf{w}^k)$ to take the bounds into account.
Typically, the problem of determining $\lambda^\star$ is solved only approximately; in this way, not too much computational effort is invested in a possibly 'bad' search direction, since the search direction itself may change drastically from one Newton iteration to the next \cite[p.98]{martins_enginnering_2022}.
Various strategies to determine $\lambda^\star$ can be adopted, see e.g. \cite{martins_enginnering_2022,Nocedal_numerical_2006,asher_first_2011,dennis_numerical_1996}.
Appendix~\ref{app:other_line_search_algorithms} gives a brief overview of some commonly used approaches, which will also be used in Section~\ref{sec:assessment} for performance comparisons to our proposed approach.

\subsection{Finding the minimum of the energy along the Newton direction}
Motivated by the fact that variational phase-field fracture computations are based on the energy minimization \eqref{eq:full_prob} solved by alternate minimization (Section \ref{sec:standard}), and in light of the observations made in Section~\ref{subsec:convergence_issues}, we propose to determine $\lambda^\star$ so as to find the minimum of the energy along the search direction.
This means that, at each Newton iteration of each subproblem, we solve the minimization problem
\begin{equation}\label{eq:ls_minimization_problem}
    \lambda^\star = \argmin_{\lambda \in (0, 1]} \phi (\lambda) \qquad\text{with}\qquad \phi (\lambda) = E (\mathbf{w}^{k} + \lambda \Delta \mathbf{w}^{k}) \qquad\text{.}
\end{equation}
For notational simplicity, we omit the subscript $n$ in the energy and only explicitly denote the argument $\mathbf{w} \in \mathbb{R}^{N_w}$ currently solved for, i.e. either $\mathbf{w}=\mathbf{u}$ ($N_w=N_u$) or $\mathbf{w}=\boldsymbol{\alpha}$ ($N_w=N_{\alpha}$) for the displacement and damage subproblems, respectively. As mentioned in the introduction, classical textbooks such as \cite{Nocedal_numerical_2006} would classify this approach as an \textit{exact line search}. 
\begin{remark}\label{rmk:lambda_max}
    The allowed interval for the step length multiplier $\lambda$ must not necessarily be restricted to $(0, 1]$.
    Depending on the specific problem, $\lambda > 1$ may be beneficial in terms of the overall performance of Newton's method, as reported e.g. in \cite{shea2025greedy}.
    In this study, we mostly keep an upper bound of $\lambda = 1$ to retain a fast \textit{local} convergence of Newton's method close to the solution.
    In Section~\ref{sec:lambda_max_comparison}, we numerically evaluate the influence of the choice of the upper bound of $\lambda$ on the overall performance within an alternate minimization loop.
\end{remark}

\begin{properties}\label{prop:linesearch_objective}
    Due to Properties \ref{prop:R_K}, the line search minimization problem~\eqref{eq:ls_minimization_problem} has the following properties:
    \begin{itemize}
        \item $\phi (\lambda)$ is strictly convex and continuously differentiable in $\lambda$;
        \item hence, and with $\phi^\prime (0) < 0$ by virtue of Newton's method, there is a unique minimum of $\phi(\lambda)$ within the admissible range of $\lambda$.
    \end{itemize}
\end{properties}
Consequently, the energy is guaranteed to decrease, $\phi(\lambda^\star) < \phi(0)$.
The derivative of $\phi(\lambda)$ corresponds to the directional derivative of the energy along the Newton direction,
\begin{equation}\label{eq:dir_deriv}
    \phi^\prime (\lambda) = \frac{\mathrm{d} E (\mathbf{w}^{k} + \lambda \Delta \mathbf{w}^{k})}{\mathrm{d} \lambda} = \mathbf{R}_{\mathbf{w}} (\mathbf{w}^{k} + \lambda \Delta \mathbf{w}^{k}) \cdot \Delta \mathbf{w}^k \qquad\text{,}
\end{equation}
while for the reduced-space strategy outlined in Section~\ref{subsec:treatment_of_constraints}, the directional derivative has to be computed taking the bounds into account, i.e.
\begin{equation}\label{eq:dir_deriv_rs}
    \phi^\prime (\lambda) = \check{\mathbf{R}} (\pi(\mathbf{w}^{k} + \lambda \Delta \mathbf{w}^{k})) \cdot \Delta \mathbf{w}^k
\end{equation}
with $\check{\mathbf{R}}$ as in \eqref{eq:RS_residualconvergence}.

\begin{figure}[H]
    \centering
    \begin{subfigure}[t]{0.475\textwidth}
        \centering
        \includegraphics{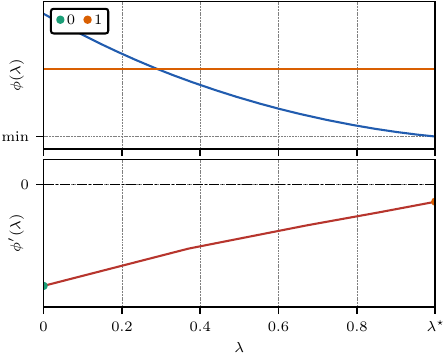}
        \caption{}
        \label{fig:linesearch_bisection_nosignchange}
    \end{subfigure}
    \begin{subfigure}[t]{0.475\textwidth}
        \centering
        \includegraphics{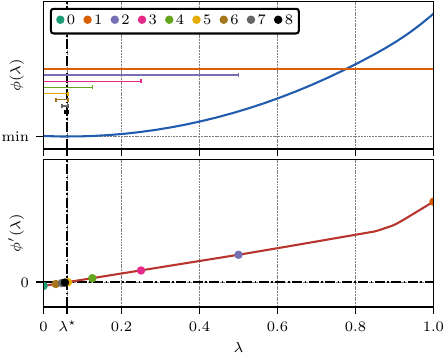}
        \caption{}
        \label{fig:linesearch_bisection_signchange}
    \end{subfigure}
    \caption{Iterations of the bisection line search algorithm for the illustrative example of Fig.~\ref{fig:toyproblem_linesearch_sampling} at Newton iteration $k=11$ where $\phi^\prime (0)\phi^\prime (1)>0$ (a), and at Newton iteration $k=12$ where $\phi^\prime (0)\phi^\prime (1)<0$ (b).
    For each bisection iteration, the iterate $\lambda_l$ is highlighted with a dot, while the bracket $[\lambda^{\text{left}}, \lambda^{\text{right}}]$ is represented by a segment in the upper plot.}
    \label{fig:linesearch_bisection}
\end{figure}

Being $\phi(\lambda)$ strictly convex, $\phi^\prime(\lambda)$ is monotonically increasing.
This allows us to devise a simple yet robust algorithm to solve the line search minimization problem posed in~\eqref{eq:ls_minimization_problem}.
If there is no change of sign from $\phi^\prime (0)$ to $\phi^\prime (1)$, this implies that $\lambda=1$ is the minimizer to~\eqref{eq:ls_minimization_problem}.
Otherwise, $\phi^\prime (\lambda)$ has a root within $(0, 1)$ which corresponds to the minimum of $\phi(\lambda)$.
To find this unique root, we propose to employ the bisection algorithm, as detailed in Algorithm~\ref{alg:ls_bisection}.
We have recently contributed a scalable implementation of this algorithm to the open-source library \texttt{PETSc} \cite{balay_2024a,balay_2024b,balay_1997}, where it is accessible as \texttt{SNESLineSearchBisection}.
Fig.~\ref{fig:linesearch_bisection} visualizes the iterations and bisection intervals $[\lambda^{\text{left}}, \lambda^{\text{right}}]$ for the example of Fig.~\ref{fig:toyproblem_linesearch_sampling}.

\begin{algorithm}[H]
    \DontPrintSemicolon
    \KwIn{$\mathbf{w}^k$, $\Delta \mathbf{w}^k$}
    \KwOut{$\lambda^\star = \argmin_{\lambda \in (0,1]} \phi(\lambda)$}
    initialize $l=1$, $\lambda_0 = \lambda^{\text{left}} = 0$, $\lambda_l = \lambda^{\text{right}} = 1$ \;
    \eIf{$\phi^\prime(\lambda_{0}) \phi^\prime(\lambda_1) \geq 0$}{
        exit with $\lambda_1 = 1$\;
        }{
        \While{$l < l_{\max}$}{
            exit if $\phi^\prime(\lambda_l) / \| \Delta \mathbf{w}^k \|_2 \leq \mathtt{atol}$ or $\phi^\prime(\lambda_l)/\phi^\prime(\lambda_0) \leq \mathtt{rtol}$ or $|\lambda_l - \lambda_{l-1}| \leq \mathtt{ltol}$ \;
            \If{$l>1$}{
                \eIf{$\phi^\prime(\lambda_l) \phi^\prime(\lambda^{\text{left}}) < 0$}{
                    bisect to the left $\lambda^{\text{right}} \gets \lambda_l$ \;
                    }{
                    bisect to the right $\lambda^{\text{left}} \gets \lambda_l$ \;
                    }
                }
            bisect interval $\lambda_{l} = \frac{\lambda^{\text{left}} + \lambda^{\text{right}}}{2}$ \;
            $l \gets l + 1$ \;
            }
        }
    \caption{Bisection line search to find the minimum of the energy along the Newton direction}
    \label{alg:ls_bisection}
\end{algorithm}

A proof of convergence of the bisection algorithm can be found in classical textbooks, e.g. in \cite{Burden_numerical_2010}.
By a trivial modification of \cite[Theorem~2.1]{Burden_numerical_2010} for a strictly convex $\phi(\lambda)$, we obtain the following result.
\begin{proposition}\label{prop:bsct_convergence}
    Let $\phi^\prime (\lambda) \in \mathcal{C} ([0, 1])$ be strictly monotonically increasing and $\phi^\prime (0) \phi^\prime(1) < 0$.
    Then, the bisection algorithm produces a sequence $\{\lambda_l\}_{l=1}^\infty$ that converges to the unique root $\lambda^\star \in (0, 1)$ of $\phi^\prime (\lambda)$ as
    \begin{equation}\label{eq:bsct_convergence}
        |\lambda_l - \lambda^\star| \leq \frac{1}{2^l} \quad \forall \, l \geq 1 \quad\text{.}
    \end{equation}
\end{proposition}
It follows from \eqref{eq:bsct_convergence} that at most $\log_2 (1/\delta)$ bisection iterations are necessary to reach an absolute error of $\delta$ for the root.
For efficiency reasons, in Algorithm~\ref{alg:ls_bisection} we introduce three possible convergence criteria by setting an absolute tolerance $\mathtt{atol}$, a relative tolerance $\mathtt{rtol}$ on the directional derivative, and a tolerance $\mathtt{ltol}$ on the change of $\lambda$ between successive iterations.
In our implementation, we normalize the directional derivative with $\|\Delta \mathbf{w}^k\|_2$ for the check of the absolute tolerance to obtain the correct physical units of the slope \cite[p.84]{martins_enginnering_2022}.
Not normalizing would not change the behavior or properties of the algorithm, only the influence of the numerical tolerance $\mathtt{atol}$.
In this work, we set $\mathtt{atol} = 10^{-12}$ and $\mathtt{ltol} = 10^{-6}$ (corresponding to at most $l_{\max}=20$ iterations and hence an accuracy of at least $9.5 \cdot 10^{-7}$ if $\mathtt{atol}$ is not met before), while we do not use $\mathtt{rtol}$.
These tolerances are selected keeping in mind the objective to find an accurate minimum of the energy along the search direction.

The choice of bisection may seem inefficient, since the algorithm has slow convergence.
However, our aim is to devise a \textit{robust yet simple} algorithm with a global proof of convergence for Newton's method enhanced with the exact line search.
In Section~\ref{sec:assessment}, we show that the proposed algorithm can in fact increase the computational efficiency of Newton's method in comparison to other line search algorithms; we further refer to \cite{Oliveira_enhancement_2020} for an alternative to bisection and a comparison of different strategies.
Another advantage of the bisection algorithm is that it does not involve higher derivatives of $\phi (\lambda)$.
Each bisection iteration only needs the evaluation of the residual $\mathbf{R}_{\mathbf{w}} (\mathbf{w}^k + \lambda_l \Delta \mathbf{w}^k)$ and the dot product with the search direction, operations which scale linearly with the number of DOFs.

\subsection{Global convergence of Newton's method with exact line search}\label{glob_conv}
For Newton's method enhanced with exact line search -- denoted as 'greedy Newton' in \cite{shea2025greedy} -- we can now establish global convergence to the minimum of $E(\mathbf{w})$:

\begin{proposition}\label{prop:convergence_proof_Newtonextls}
    Let $E(\mathbf{w}) \in \mathcal{C}^2(\mathbb{R}^{N_w})$ be strictly convex and coercive.
    Further, let us denote by $\rho_n$, $n \in \mathbb{N}$, $1 \le n \le N_w$, the $n$-th eigenvalue of $\mathbf{K}_{\mathbf{w}\mathbf{w}}(\mathbf{w}) = \nabla^2 E(\mathbf{w})$.
    We assume that there exist lower and upper bounds for the eigenvalues of $\mathbf{K}_{\mathbf{w}\mathbf{w}}(\mathbf{w})$, denoted by $\rho_\mathrm{L}$ and $\rho_\mathrm{U}$, with $\rho_\mathrm{L} \leq \rho_\mathrm{U}$, i.e.,
    \begin{equation}\label{eq:eigval_bounds}
        \rho_n \in [\rho_\mathrm{L}, \rho_\mathrm{U}] \quad \text{for all } n = 1, \dots, N_w \quad\text{.}
    \end{equation}
    Then, the sequence of residual norms $\{\|\mathbf{R}_{\mathbf{w}}(\mathbf{w}^k)\| \}_{k=0}^\infty$ generated by Newton's method with exact line search converges to zero, i.e.,
    \begin{equation}\label{eq:convergence_proof_result}
        \lim_{k \to \infty} \|\mathbf{R}_{\mathbf{w}}(\mathbf{w}^k)\| = 0 \quad\text{.}
    \end{equation}
\end{proposition}
Based on the proof presented in \cite{shea2025greedy}, we prove this proposition in Appendix~\ref{app:convergence_proof}.
As a corollary, by strict convexity and coercivity of the energy functional, we know that \eqref{eq:convergence_proof_result} is sufficient to show that the sequence globally converges to the unique minimizer of $E(\mathbf{w})$.

\begin{remark}\label{rmk:convergence_guarantee}
    Based on the above, if both mechanical and damage subproblems are strictly convex and coercive, alternate minimization with exact line search is guaranteed to converge to a critical point of the energy, since:
    \begin{itemize}
        \item the alternate minimization scheme converges to a critical point of the energy \cite[Theorem~1]{bourdin_numerical_2007}, provided the mechanical and the damage subproblems are successfully solved at each staggered iteration;
        \item for each subproblem, Newton's method with the exact line search converges to the minimizer of the energy, Proposition~\ref{prop:convergence_proof_Newtonextls}, provided the line search minimization problem is successfully solved;
        \item the bisection line search converges to the minimizer of the energy along the Newton direction, Proposition~\ref{prop:bsct_convergence}.
    \end{itemize}
\end{remark}
The above does not generally hold for (I) the case of the reduced-space active set strategy where no convergence proof is available, and (II) the star-convex model with $\gamma^\star > 0$ for which the damage problem may become non-convex (see also Section~\ref{star-convex}).
For these two cases, we will establish the robustness of the overall solution approach empirically with numerical experiments in the following section.

\section{Numerical assessment of the method}\label{sec:assessment}
With the algorithm and its properties being established theoretically, we now shift our attention to the numerical assessment of the proposed solution approach.
We first assess the robustness by revisiting the obstacle course of Section~\ref{subsec:convergence_issues}.
Secondly, we compare different choices for the maximum $\lambda$ for the bisection line search, and then compare the proposed exact line search to other commonly used line search algorithms in terms of computational efficiency.

\subsection{Robustness of the method: Revisiting the obstacle course}\label{subsec:assessment_robustness}
We revisit the obstacle course of Section~\ref{subsec:convergence_issues}, for which the standard solution approach exhibited convergence issues for most cases, and evaluate the convergence behavior with the exact line search.
We employ the bisection line search either for the mechanical problem only, or for both subproblems.
Results are shown in Tab.~\ref{tab:convergence_obstacle_course_linesearch} for all five tests, which we compute as before with the various splits and with the two constraint enforcement methods (reduced-space active set method and penalization).
When irreversibility is enforced with the reduced-space active set method, the exact line search applied only to the mechanical problem is sufficient to achieve convergence in all cases, suggesting that it is not necessary to employ line search also for the damage problem.
When irreversibility is enforced with the penalty method, the exact line search only on the mechanical problem solves the convergence issues of Tab.~\ref{tab:convergence_obstacle_course} in many, but not in all cases.
In the remaining problematic cases, Newton's method now succeeds to converge in the mechanical problem but fails in the damage problem.
By applying the exact line search for the damage problem as well, iterative convergence is achieved in all cases.
See Appendix~\ref{app:obstacle_course} for details on the obtained solutions.

\begin{table}[H]
    \centering
    \footnotesize
    \caption{Overview of whether the proposed solution approach can reach convergence for the obstacle course.
    We denote success and failure with \cmark and \xmark, respectively.
    With \xmark$_{\bm{\alpha}}$, we indicate failure to converge in the damage problem.}
    \label{tab:convergence_obstacle_course_linesearch}
    \begin{tabularx}{\textwidth}{llccccc}
        \toprule
        &&nucleation test &sliding test &plate with hole &plate w. inclined crack &perforated plate\\
        &&\raisebox{-.5\height}{\includegraphics{setup_nucleation_mini.pdf}}
        &\raisebox{-.5\height}{\includegraphics{setup_sliding_mini.pdf}}
        &\raisebox{-.5\height}{\includegraphics{setup_platewithhole_mini.pdf}}
        &\raisebox{-.5\height}{\includegraphics{setup_platewithinclinedcrack_mini.pdf}}
        &\raisebox{-.5\height}{\includegraphics{setup_perforatedplate_mini.pdf}}\\
        \midrule
        &line search employed
        &$\mathbf{u} \,\,\,\, \mathbf{u}\text{,}\bm{\alpha}$ &$\mathbf{u} \,\,\,\, \mathbf{u}\text{,}\bm{\alpha}$ &$\mathbf{u} \,\,\,\, \mathbf{u}\text{,}\bm{\alpha}$ &$\mathbf{u} \,\,\,\, \mathbf{u}\text{,}\bm{\alpha}$ &$\mathbf{u} \,\,\,\, \mathbf{u}\text{,}\bm{\alpha}$\\
        \midrule
        \parbox[t]{1mm}{\multirow{5}{*}{\rotatebox[origin=c]{90}{red. space$\,\,\,\,$}}}
        &vol-dev                                       &\cmark $\,$ \cmark                  &\cmark $\,$ \cmark                     &\cmark $\,$ \cmark                   &\cmark $\,$ \cmark                  &\cmark $\,$ \cmark                  \\
        &star-convex, $\gamma^\star = 1$               &\cmark $\,$ \cmark                  &\cmark $\,$ \cmark                     &\cmark $\,$ \cmark                   &\cmark $\,$ \cmark                  &\cmark $\,$ \cmark                  \\
        &star-convex, $\gamma^\star = 5$               &\cmark $\,$ \cmark                  &\cmark $\,$ \cmark                     &\cmark $\,$ \cmark                   &\cmark $\,$ \cmark                  &\cmark $\,$ \cmark                  \\
        &spectral                                      &\cmark $\,$ \cmark                  &\cmark $\,$ \cmark                     &\cmark $\,$ \cmark                   &\cmark $\,$ \cmark                  &\cmark $\,$ \cmark                  \\
	    &no-tension                                    &\cmark $\,$ \cmark                  &\cmark $\,$ \cmark                     &\cmark $\,$ \cmark                   &\cmark $\,$ \cmark                  &\cmark $\,$ \cmark                  \\
        &DP-like, $\gamma = \sqrt{2\mu_0 / \kappa_0}$  &\cmark $\,$ \cmark                  &\cmark $\,$ \cmark                     &\cmark $\,$ \cmark                   &\cmark $\,$ \cmark                  &\cmark $\,$ \cmark                  \\
        \midrule
        \parbox[t]{1mm}{\multirow{5}{*}{\rotatebox[origin=c]{90}{penalization$\,\,\,\,$}}}
        &vol-dev                                       &\xmark$_{\bm{\alpha}}$ $\,$ \cmark                  &\cmark $\,$ \cmark                     &\xmark$_{\bm{\alpha}}$ $\,$ \cmark                   &\xmark$_{\bm{\alpha}}$ $\,$ \cmark                  &\xmark$_{\bm{\alpha}}$ $\,$ \cmark                  \\
        &star-convex, $\gamma^\star = 1$               &\xmark$_{\bm{\alpha}}$ $\,$ \cmark                  &\cmark $\,$ \cmark                     &\xmark$_{\bm{\alpha}}$ $\,$ \cmark                   &\xmark$_{\bm{\alpha}}$ $\,$ \cmark                  &\xmark$_{\bm{\alpha}}$ $\,$ \cmark                  \\
        &star-convex, $\gamma^\star = 5$               &\xmark$_{\bm{\alpha}}$ $\,$ \cmark                  &\cmark $\,$ \cmark                     &\xmark$_{\bm{\alpha}}$ $\,$ \cmark                   &\xmark$_{\bm{\alpha}}$ $\,$ \cmark                  &\xmark$_{\bm{\alpha}}$ $\,$ \cmark                  \\
        &spectral                                      &\cmark $\,$ \cmark                                  &\cmark $\,$ \cmark                     &\cmark $\,$ \cmark                                   &\cmark $\,$ \cmark                                  &\cmark $\,$ \cmark                                  \\
	    &no-tension                                    &\cmark $\,$ \cmark                                  &\cmark $\,$ \cmark                     &\cmark $\,$ \cmark                                   &\cmark $\,$ \cmark                                  &\cmark $\,$ \cmark                                  \\
        &DP-like, $\gamma = \sqrt{2\mu_0 / \kappa_0}$  &\cmark $\,$ \cmark                                  &\cmark $\,$ \cmark                     &\cmark $\,$ \cmark                                   &\cmark $\,$ \cmark                                  &\cmark $\,$ \cmark                                  \\
        \bottomrule
    \end{tabularx}
\end{table}

\begin{figure}[H]
    \centering
    \begin{subfigure}[t]{0.475\textwidth}
        \centering
        \includegraphics{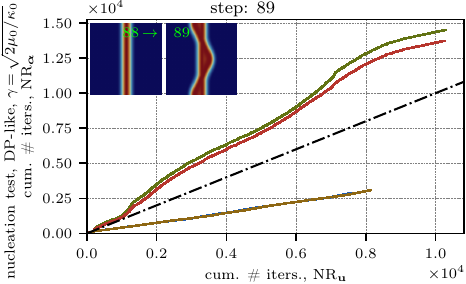}
        \caption{}
        \label{fig:staggeredpath_irreversibility_a}
    \end{subfigure}
    \begin{subfigure}[t]{0.475\textwidth}
        \centering
        \includegraphics{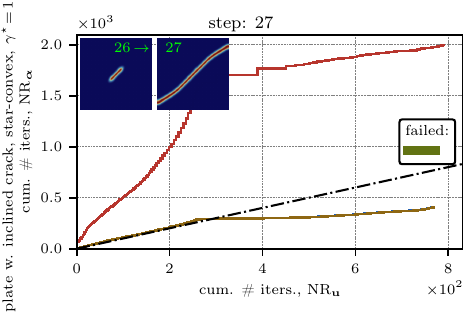}
        \caption{}
        \label{fig:staggeredpath_irreversibility_c}
    \end{subfigure}
    \newline
    \begin{subfigure}[t]{0.475\textwidth}
        \centering
        \includegraphics{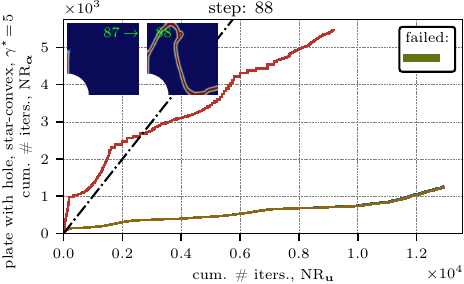}
        \caption{}
        \label{fig:staggeredpath_irreversibility_b}
    \end{subfigure}
    \begin{subfigure}[t]{0.475\textwidth}
        \centering
        \includegraphics{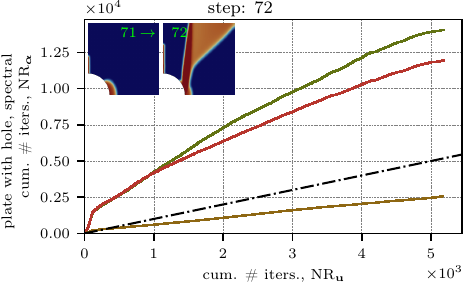}
        \caption{}
        \label{fig:staggeredpath_irreversibility_f}
    \end{subfigure}
    \newline
    \begin{subfigure}[t]{0.475\textwidth}
        \centering
        \includegraphics{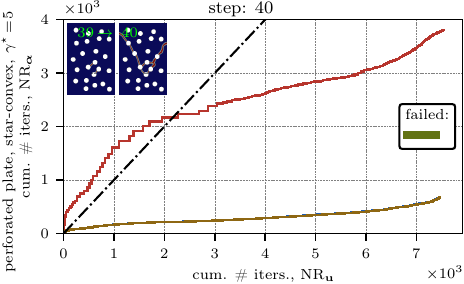}
        \caption{}
        \label{fig:staggeredpath_irreversibility_d}
    \end{subfigure}
    \begin{subfigure}[t]{0.475\textwidth}
        \centering
        \includegraphics{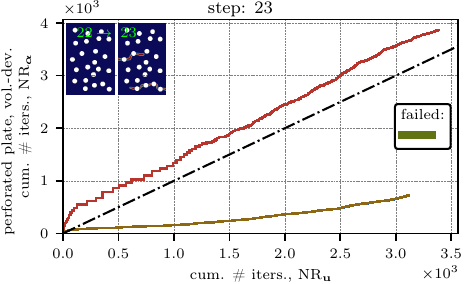}
        \caption{}
        \label{fig:staggeredpath_irreversibility_h}
    \end{subfigure}
    \newline
    \includegraphics{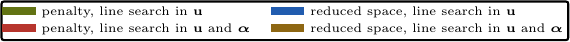}
    \caption{Staggered paths for various cases from the obstacle course at selected load steps.
    For each constraint enforcement method, we compare the iterative solution performance using exact line search only for the mechanical problem and for both subproblems.}
    \label{fig:staggeredpath_irreversibility}
\end{figure}

To gain further insights on the effect of the exact line search based on bisection in combination with different approaches for constraint enforcement, we depict in Fig.~\ref{fig:staggeredpath_irreversibility} the 'staggered paths' for a few examples, each at a selected load step where significant phase-field evolution takes place.
This means that for each of the staggered iterations for a selected load step,
\begin{itemize}
    \item we move one step right for each Newton iteration of the mechanical problem until convergence is reached, and then from this point,
    \item we move one step up for each Newton iteration of the damage problem until convergence is reached. Then again from this point, we move right for each Newton iteration in the mechanical problem, etc.
\end{itemize}
Accordingly, the abscissa represents the cumulative number of Newton iterations in the mechanical problem, while the ordinate represents the cumulative number of Newton iterations in the damage problem.
Clearly, the shorter the path, the less Newton iterations are necessary in total to reach convergence in the selected time step.
Thus, the length of the path is an indicator for the computational cost.
For reference, we plot a straight line for the hypothetical case where convergence is reached within the same number of iterations for both the mechanical and the damage problem.
Staggered paths above or below this line denote that the damage problem requires more or less total Newton iterations than the displacement problem, respectively.

Some trends emerge from these plots. For the solution of the damage problem, the penalty method leads to a significantly larger number of Newton iterations to reach convergence than the reduced-space method (partly due to the strict tolerance $\mathtt{TOL}_{\text{ir}} = 10^{-4}$).
The reduced-space solver does not benefit from the exact line search, since in this case the damage problem is linear (Properties~\ref{prop:vectorE}) and the iterations are only needed to find the correct active and inactive sets.
In fact, the line search is even detrimental as it slightly increases the computational effort.
Conversely, the penalized phase-field problem is strongly non-linear, as evident when sampling the energy and the directional derivative in the Newton direction (see Appendix~\ref{app:linesearch_sampling}).
This explains the need for the exact line search to overcome convergence issues, and its ability to reduce the total number of necessary Newton iterations in the damage problem.
For the cases of Fig.~\ref{fig:staggeredpath_irreversibility} where the pure Newton solver achieves convergence with the penalized formulation, the exact line search leads to a small, but not negligible reduction in the total number of Newton iterations.

Overall, the successful solution of the challenging problems in the obstacle course confirms the theoretical findings from Section~\ref{sec:bisection_linesearch} and empirically extends them to the two cases where no proofs are available.
In particular, with the reduced-space active set strategy for the solution of the damage problem, we observe convergence in all tested cases even without the exact line search. We report specific considerations and results on the star-convex model in Section~\ref{star-convex}.
In summary, alternate minimization with exact line search seems to be 'all one needs' for robust computations of phase-field brittle fracture.

\subsection{Robustness of the method: Some observations on the star-convex energy decomposition}\label{star-convex}
\begin{figure}[H]
    \centering
    \begin{subfigure}[t]{0.475\textwidth}
        \centering
        \includegraphics[height=4cm]{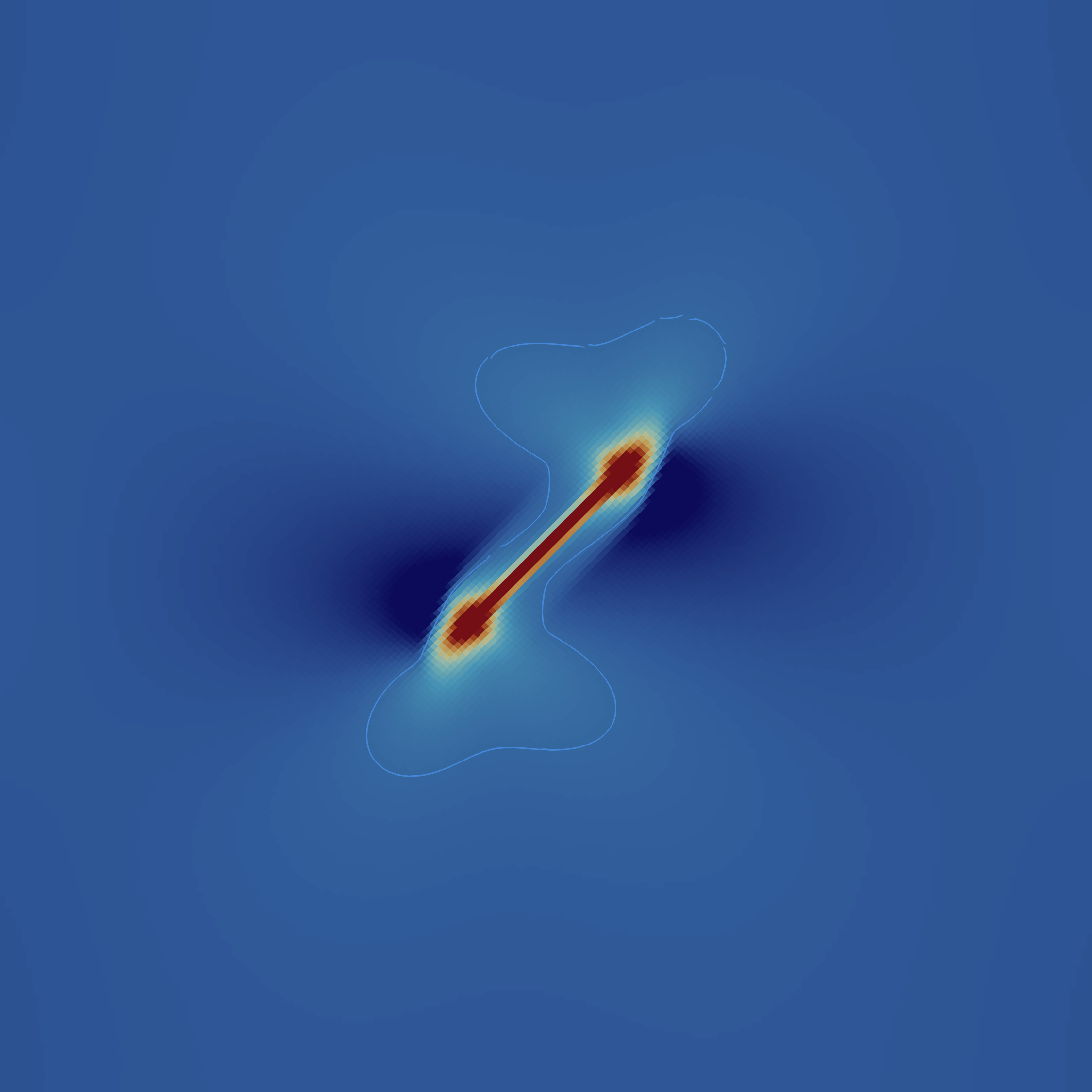}
        \includegraphics{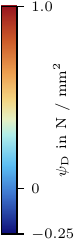}
        \caption{}
        \label{fig:psi_D_platewithinclinedcrack_starconvex5}
    \end{subfigure}
    \begin{subfigure}[t]{0.475\textwidth}
        \centering
        \includegraphics[height=4cm]{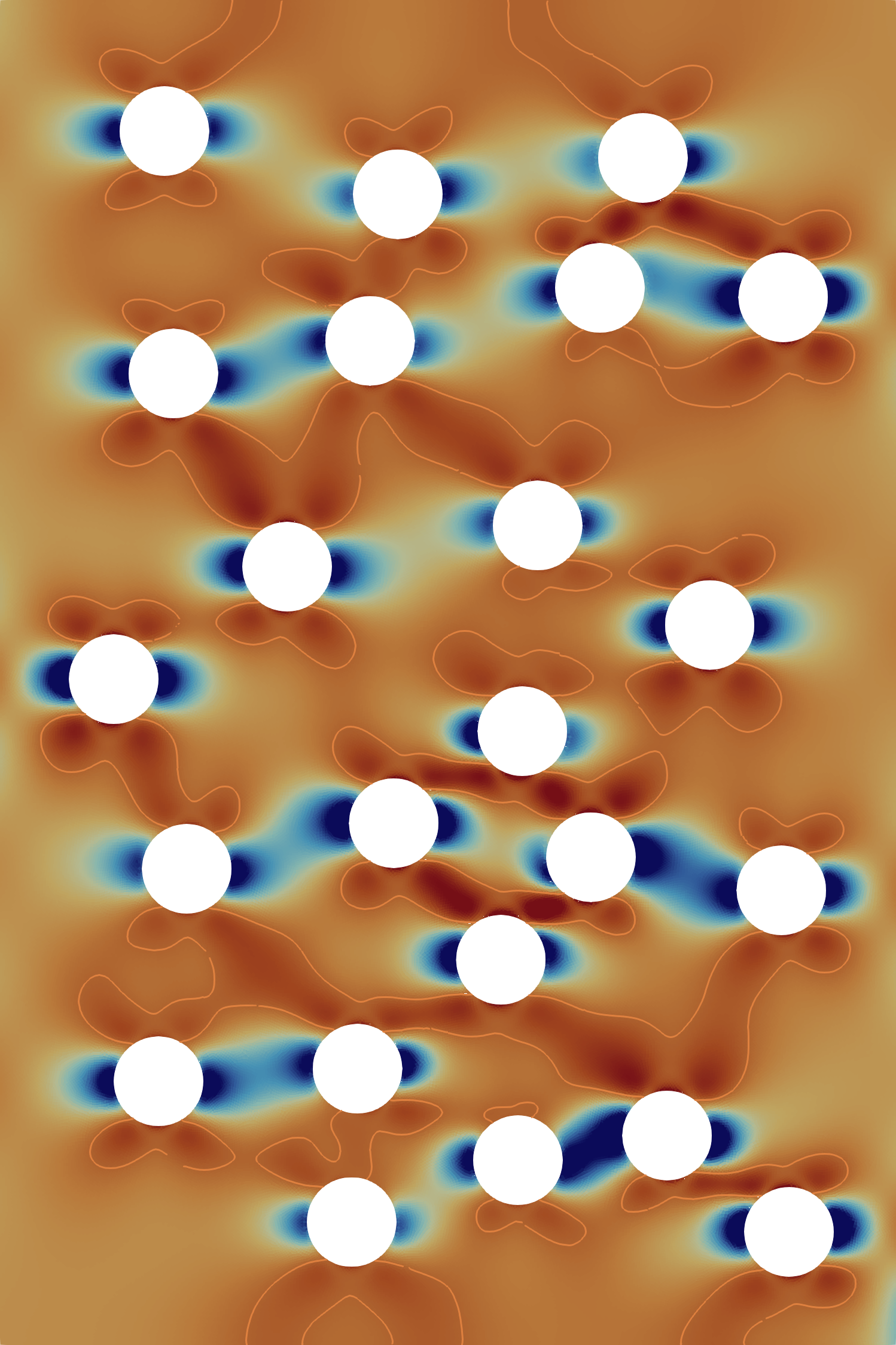}
        \includegraphics{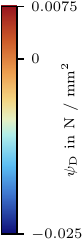}
        \caption{}
        \label{fig:psi_D_perforatedplate_starconvex5}
    \end{subfigure}
    \caption{Crack-driving contribution of the strain energy density $\psi_{\text{D}}$ with negative values in significant parts of the domain, for (a) the plate with an inclined crack at load step $n=27$ with $\gamma^\star = 5$, and (b) the perforated plate at load step $n=35$ with $\gamma^\star = 5$.
    Both time steps are prior to crack nucleation. The plots include the isolines with $\psi_{\text{D}}=0$.}
    \label{fig:psi_D}
\end{figure}

For the star-convex model with $\gamma^\star > 0$, the elastic energy density $\psi(\boldsymbol{\varepsilon},\alpha)$ is strictly convex with respect to $\boldsymbol{\alpha}$ only for $\psi_\text{D}(\boldsymbol{\varepsilon})>0$.
Negative values for $\psi_{\text{D}}$ can indeed be observed for predominantly compressive deformations, as we show in Fig.~\ref{fig:psi_D} with two examples from the obstacle course.
As a result, the strict convexity of $E_n(\mathbf{u},\bm{\alpha})$ with respect to $\bm{\alpha}$ and the positive definiteness of $\mathbf{K}_{\bm{\alpha}\bm{\alpha}}(\mathbf{u})$ depend on $\mathbf{u}$, and the proof of global convergence in Section \ref{glob_conv} for Newton's method with the exact line search no longer holds in general. 

While we are not able to formally prove the strict convexity of $E_n(\mathbf{u},\bm{\alpha})$ with respect to $\bm{\alpha}$ on the inactive set, nor the strict convexity with respect to $\bm{\alpha}$ of the modified energy $\tilde{E}_n(\mathbf{u},\bm{\alpha})$ for the penalized formulation, we can empirically evaluate convexity by numerically sampling the energy and its directional derivative with respect to damage along the Newton direction.
For all tests in the obstacle course, we use the star-convex model with $\gamma^\star = 1$ and $\gamma^\star = 5$ and sample the energy and its directional derivative with $100$ points for each Newton iteration.
On this rather large data set, of which we report selected examples in Appendix~\ref{app:linesearch_sampling}, we checked the monotonicity of the directional derivative along the line search direction and could not find any evidence of non-convexity up to numerical tolerances.
Correspondingly, we observe that when using the star-convex model with $\gamma^\star>0$ Newton's method with the exact line search performs as robustly as with the 'classical' energy decompositions, as evident from Tables \ref{tab:convergence_obstacle_course} and \ref{tab:convergence_obstacle_course_linesearch} and already noted in Section \ref{subsec:assessment_robustness}.

\subsection{Influence of the maximum step length}\label{sec:lambda_max_comparison}
\begin{figure}[H]
    \centering
    \hspace*{-0.2cm}
    \begin{subfigure}[t]{0.475\textwidth}
        \centering
        \includegraphics{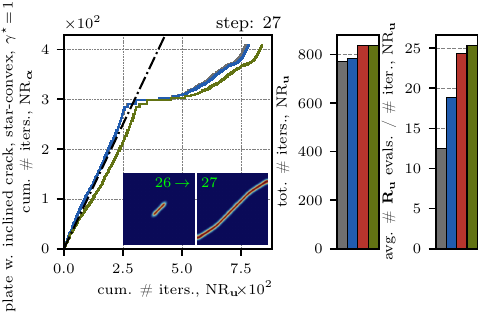}
        \caption{}
        \label{fig:performance_lambdamax_a}
    \end{subfigure}
    \hspace{0.2cm}
    \begin{subfigure}[t]{0.475\textwidth}
        \centering
        \includegraphics{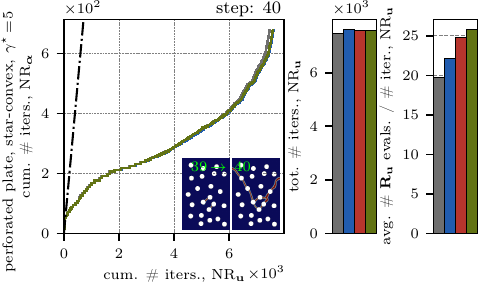}
        \caption{}
        \label{fig:performance_lambdamax_b}
    \end{subfigure}
    \newline
    \hspace*{-0.2cm}
    \begin{subfigure}[t]{0.475\textwidth}
        \centering
        \includegraphics{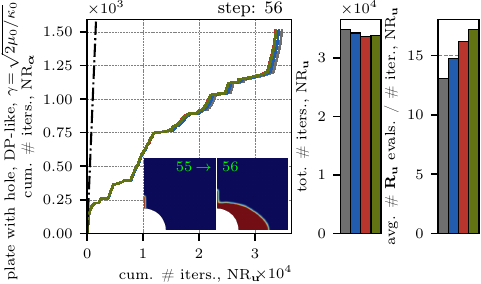}
        \caption{}
        \label{fig:performance_lambdamax_c}
    \end{subfigure}
    \hspace{0.2cm}
    \begin{subfigure}[t]{0.475\textwidth}
        \centering
        \includegraphics{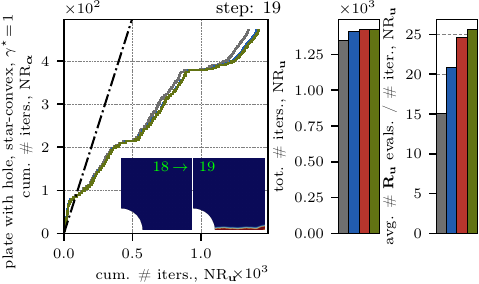}
        \caption{}
        \label{fig:performance_lambdamax_d}
    \end{subfigure}
    \newline
    \includegraphics{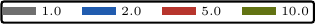}
    \caption{Performance metrics for various tests comparing $\lambda_1 \in (1.0, 2.0, 5.0, 10.0)$: the plate with an inclined crack with the star-convex split, $\gamma^\star=1$ (a), the perforated plate with the star-convex split, $\gamma^\star=5$ (b), and the plate with a hole with the DP-like split, $\gamma = \sqrt{\tfrac{2\mu_0}{\kappa_0}}$ (c), and with the star-convex split, $\gamma^\star = 1.0$ (d).}
    \label{fig:performance_lambdamax}
\end{figure}

As anticipated in Remark~\ref{rmk:lambda_max}, we now evaluate the influence of the upper bound of $\lambda$ on the overall performance within an alternate minimization loop.
In particular, we set the starting point (corresponding to the maximum attainable $\lambda^\star$) for the bisection line search algorithm to $\lambda_1 \in (1.0, 2.0, 5.0, 10.0)$ for selected tests of the obstacle course.
We use the bisection line search only for the solution of the mechanical problem and adopt the reduced-space active set strategy to enforce irreversibility in the damage subproblem.
In this way, we focus only on the influence of the non-linearity of the mechanical problem induced by the strain energy decomposition.
Fig.~\ref{fig:performance_lambdamax} shows computational performance metrics for the plate with an inclined crack, the perforated plate, and the plate with a hole, computed with different splits:
on the left-hand side of each subfigure we show the staggered paths, and on the right-hand side the total number of Newton iterations as well as the average number of residual evaluations per Newton iteration for the mechanical problem.
We deliberately refrain from comparing the total computational time, since this strongly depends on problem, setup, and implementation.
Instead, we consider the total number of Newton iterations to be the strongest indicator of computational efficiency.

As evident in these figures, choosing a larger $\lambda_1$ does not have a significant impact on the overall performance of Newton's method within the alternate minimization.
For three of the four chosen cases, setting $\lambda_1=1$ gives the overall shortest path, i.e., the overall lowest number of Newton iterations to reach convergence.
In these cases, choosing a larger $\lambda_1$ leads to up to $9.0$~\% more Newton iterations, and up to $103$~\% more residual evaluations per Newton iteration (for the case of the plate with an inclined crack with the star-convex model, $\gamma^\star = 1$, Fig.~\ref{fig:performance_lambdamax_a}, and $\lambda_1 = 10$).
The additional residual evaluations are directly linked to the bisection algorithm needing more iterations to reach a smaller optimal $\lambda$.
Only in one of the cases (the plate with a hole with the DP-like model, $\gamma = \sqrt{\tfrac{2\mu_0}{\kappa_0}}$, Fig.~\ref{fig:performance_lambdamax_c}), setting a larger $\lambda_1$ lead to slightly less Newton iterations.
Namely, $\lambda_1=5.0$ leads to $-3.3$~\% Newton iterations in total, however with $24.1$~\% more residual evaluations per Newton iteration in comparison to $\lambda_1=1.0$.
Based on these results, since $\lambda_1 = 1$ works best in most cases within the alternate minimization loop, we adopt this choice in the remainder of this work.

\subsection{Computational efficiency of the method: Comparison to other line searches}\label{sec:performance_comparison}
To further assess the performance of the proposed exact line search algorithm based on bisection, we compare it to other commonly used line search algorithms.
The first comparison is with a \textit{backtracking} line search which halves the step size until \textit{sufficient decrease (Armijo) conditions} are fulfilled, in one variant for the residual norm, or in another variant for the energy.
The second comparison is with a \textit{secant method} that tries to minimize the energy or the residual norm along the search direction, or that seeks to find the roots of the directional derivative.
We give a more detailed description of these algorithms in Appendix~\ref{app:other_line_search_algorithms}.
For the strictly convex cases, if successful, Newton's method theoretically converges to the same solution regardless of the specific line search algorithm being used.
However, since different line searches find different step size multipliers $\lambda$ at each Newton iteration, the number of Newton iterations necessary to find the solution and correspondingly the overall computational cost may vary significantly.
As follows, we assess the overall computational efficiency in terms of the number of Newton iterations necessary to reach convergence.

\begin{figure}[p]
    \centering
    \begin{subfigure}[t]{\textwidth}
        \centering
        \includegraphics{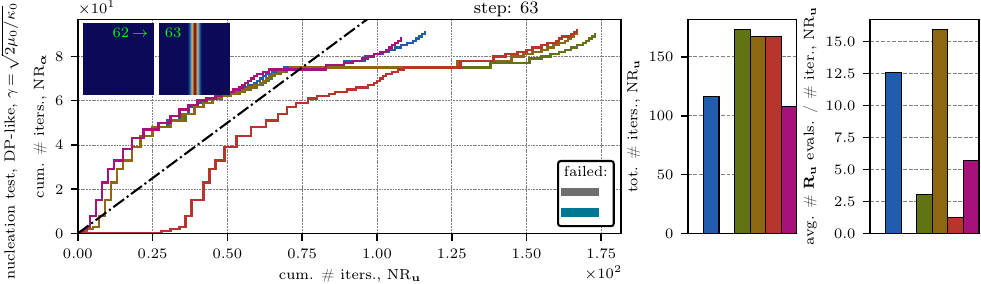}\\[-0.5cm]
        \caption{}
        \label{fig:performance_a}
    \end{subfigure}
    \newline
    \begin{subfigure}[t]{\textwidth}
        \centering
        \includegraphics{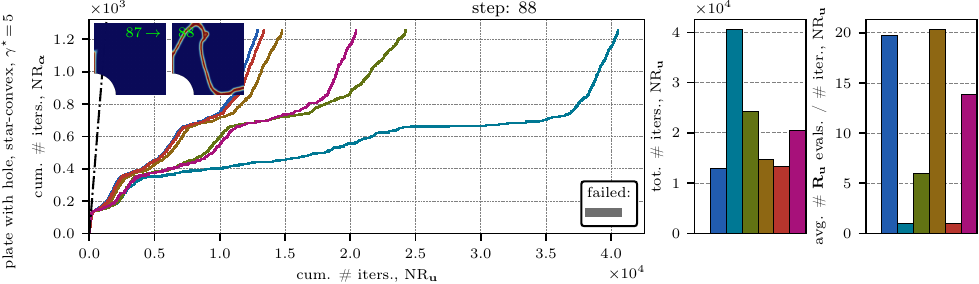}\\[-0.5cm]
        \caption{}
        \label{fig:performance_b}
    \end{subfigure}
    \newline
    \begin{subfigure}[t]{\textwidth}
        \centering
        \includegraphics{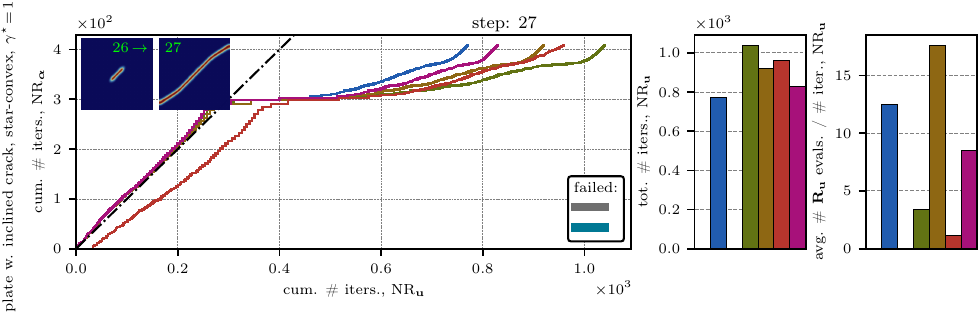}\\[-0.8cm]
        \caption{}
        \label{fig:performance_c}
    \end{subfigure}
    \newline
    \begin{subfigure}[t]{\textwidth}
        \centering
        \includegraphics{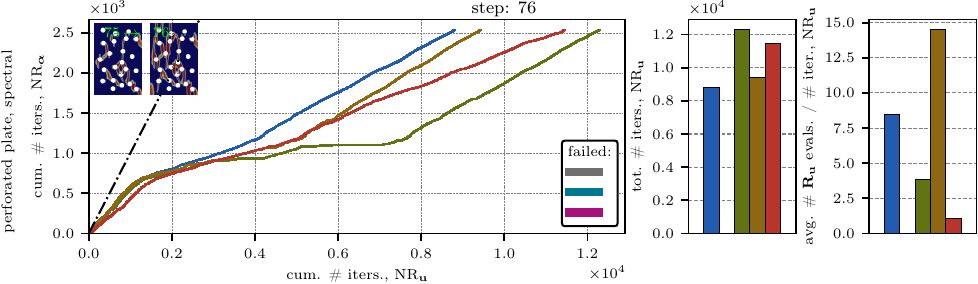}\\[-0.5cm]
        \caption{}
        \label{fig:performance_d}
    \end{subfigure}
    \newline
    \includegraphics{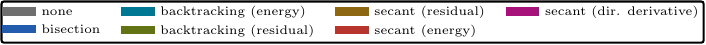}
    \caption{Performance metrics for the nucleation test with the DP-like split, $\gamma = \sqrt{\tfrac{2\mu_0}{\kappa_0}}$ (a), the plate with a hole with the star-convex split, $\gamma^\star=5$ (b), the plate with an inclined crack with the star-convex split, $\gamma^\star=1$ (c), and the perforated plate with the spectral split (d).}
    \label{fig:performance}
\end{figure}

As before, we employ the line search algorithms only for the solution of the mechanical problem; for the linear complementarity damage problem, we use Newton's method with the reduced-space active set strategy to enforce irreversibility.
We select four test cases from the obstacle course, and for each of them we analyze a particularly challenging load step (Fig.~\ref{fig:performance}), i.e. the load step at nucleation (Fig.~\ref{fig:performance_a}), or at abrupt propagation and failure (Fig.~\ref{fig:performance_b},~\ref{fig:performance_c}, \ref{fig:performance_d}).
For each case, we plot on the left-hand side of the figures the staggered paths.
Since we employ no line search in the damage subproblem, the paths are stretched along the horizontal axis and all methods require the same number of total Newton iterations for the damage problem.
On the right-hand side of the figures, we visualize further metrics to evaluate the computational efficiency, namely, the total number of Newton iterations and the average number of residual evaluations per Newton iteration for the mechanical problem.
Note that the secant method and the backtracking line search evaluate the energy and not the residual, hence they need one residual evaluation per Newton iteration.

We observe significant differences in performance between the tested line search algorithms.
In general, the proposed exact line search based on bisection achieves the shortest staggered path (hence the lowest total number of Newton iterations), or the second shortest one.
Only in one case the secant method, in its version which seeks roots in the directional derivative, achieves a slightly shorter staggered path.
However, the secant method performs less consistently, as sometimes it is close to the performance of the proposed line search, but sometimes it needs up to twice as many total Newton iterations, or it even fails.
A similar observation holds for the secant method which seeks the minimum of the residual or of the energy, whereby the former performs slightly better.
This can be explained with the complicated behavior of the energy in the search direction, illustrated in Appendix~\ref{app:linesearch_sampling}, which causes the gradient-based secant method to struggle to converge to the minimum (or the root of the directional derivative).
The backtracking algorithm shows the worst overall performance, leading to significantly longer staggered paths. This again can be attributed to the complicated behavior of the respective objective in the search direction, where the algorithm can exit with non-optimal step size multipliers.
Further, the energy-based backtracking is able to reach convergence only in one of the assessed cases and with a significantly longer staggered path in comparison to the other algorithms.
In contrast, the bisection algorithm which seeks the roots of the directional derivative is not influenced by the specific energy landscape along the Newton direction, and it always obtains the step size which minimizes the energy along the search direction.

From these observations, we conclude that Newton's method enhanced with the proposed exact line search based on bisection is not only robust, but also the one that converges with the lowest total number of Newton iterations.
Its only drawback is a slightly higher cost of the line search procedure, as evidenced by the larger average number of residual evaluations per Newton iteration.
However, this cost becomes less significant for larger problems where the linear solve dominates the computational cost; here the reduction of the total number of Newton iterations can lead to significant computational savings.

\section{Putting the method through its paces: the Brazilian test}\label{sec:brazilian3D}
To showcase the performance of the proposed exact line search algorithm, we test it on a large and particularly challenging three-dimensional problem, the so-called \textit{Brazilian test} proposed independently by Carneiro and Akazawa in 1943 \cite{Fairbarin_tribute_2002}.
This test features a cylindrical disk of radius $R$ and thickness $T$ placed between two loading platens, which compress the disk orthogonally to its cylindrical axis until failure.
The fracture is typically observed in a vertical plane connecting the lines of contact for concrete or rock-like materials, see e.g. \cite{Jin_quasistatic_2017,Erarslan_experimental_2012,Khosravani_fracture_2018} for experimental investigations.
According to testing standards, the loading platens can be either flat (with optional bearing strips) as in \cite{ASTM_D3927_2023}, or curved with a radius of $1.5R$ as in \cite{ISRM_suggested_1978}.
In this work, we adopt the latter option, while idealizing the contact between the specimen and the curved loading plates by means of the following condition
\begin{equation}\label{eq:brazilian_contact}
    u_y (\boldsymbol{x}) \leq g_0 (\boldsymbol{x}) - \bar{u}_y \qquad \forall \boldsymbol{x} \in \left\{ \boldsymbol{x}=(x,y,z)^{\mathsf{T}} : \sqrt{x^2 + y^2} = R \right\}
\end{equation}
with $\bar{u}_y$ as the prescribed displacement of one of the loading platens.
The initial gap function $g_0 (\boldsymbol{x})$ is given by
\begin{equation}
    g_0 (\boldsymbol{x}) = 1.5 R \left[ \cos \left( \arcsin \frac{x}{1.5R} \right) - 1 \right] - R \left[ \cos \left( \arcsin \frac{x}{R} \right) - 1 \right] \qquad \forall \boldsymbol{x} \in \left\{ \boldsymbol{x} : \sqrt{x^2 + y^2} = R \right\} \quad\text{.}
\end{equation}
With this contact idealization, we treat the loading platens as rigid bodies and assume no friction.
This allows us to use the reduced-space active set strategy of Section~\ref{subsec:treatment_of_constraints} for the now constrained minimization problem posed in the displacement field.
This constraint introduces another source of non-linearity in the mechanical problem, while retaining the convexity properties of the energy functional.

Similarly to \cite{Kumar_strength_2024,Bilgen_phasefield_2019}, we set the radius to $R=25$~mm; we choose $T=10$~mm yielding $T/D = 1/5$ which lies within the recommended range of $T/D \in [1/5, 3/4]$ \cite{ASTM_D3927_2023}.
We adopt a Young's modulus $E_0=10\,000$~MPa, a Poisson's ratio $\nu_0=0.15$, $G_{\text{c}}=0.15$~N$/$mm and $\ell=0.5$~mm.
Finally, we adopt the star-convex energy decomposition and use the parameter $\gamma^\star$ to tune the ratio of tensile and compressive strength such that failure occurs in the vertical plane.
With our chosen $\gamma^\star = 5.5$, we obtain tensile, compressive, and shear strengths of
\begin{equation}
    \sigma_{\text{e}}^+ = \sqrt{\frac{E_0 G_{\text{c}}}{8/3 \ell}} = 34 \text{~MPa}
    \qquad\text{,}\qquad
    \sigma_{\text{e}}^- = - \infty
    \qquad\text{and}\qquad
    \tau_{\text{c}} = \sqrt{\frac{\mu_0 G_{\text{c}}}{8/3 \ell}} = 22 \text{~MPa}
\end{equation}
according to \cite{Vicentini_energy_2024}.
We remark that our focus here is to study the convergence behavior and robustness of the proposed solution scheme, not to calibrate the parameters to experimental data.

\begin{figure}[H]
    \centering
    \begin{subfigure}[t]{0.3\textwidth}
        \centering
        \includegraphics{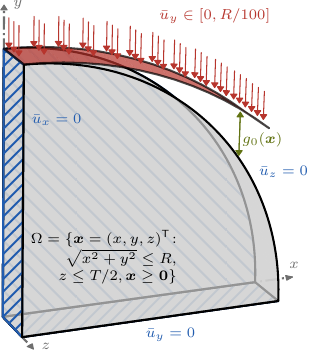}
        \caption{}
        \label{fig:brazilian_setup}
    \end{subfigure}
    \hspace{0.25cm}
    \begin{subfigure}[t]{0.3\textwidth}
        \centering
        \includegraphics{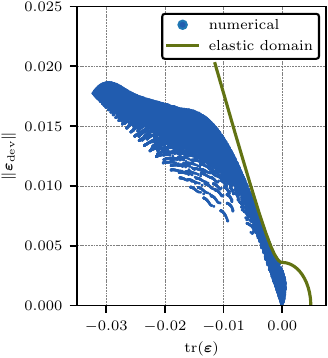}
        \caption{}
        \label{fig:brazilian_strainspace}
    \end{subfigure}
    \hspace{0.25cm}
    \begin{subfigure}[t]{0.3\textwidth}
        \centering
        \includegraphics{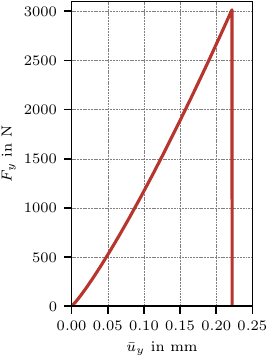}
        \caption{}
        \label{fig:brazilian_reaction}
    \end{subfigure}
    \caption{Brazilian test setup (a), strain states in the volumetric-deviatoric strain space (b), and reaction force (c).}
    \label{fig:brazilian_setup_reaction}
\end{figure}

As depicted in Fig. \ref{fig:brazilian_setup}, we exploit symmetry to compute on one eighth of the domain enforcing $u_z = 0$ in the $xy$-plane (referred to as the 'back' face in the following), $u_y = 0$ in the $xz$-plane (the 'bottom' face), and $u_x=0$ in the $yz$-plane (the 'left' face).
The downward displacement of the rigid platen $\bar{u}_y$ is incrementally applied in $1\,000$ uniform load steps from $0$ to $R/100$, and the computation is stopped once $\max \bm{\alpha} \geq 0.99$ is obtained in the domain for the first time.
We discretize the $xy$-plane with $h\approx \ell/5$ in the area of the expected crack, with a continuously coarser discretization far from the expected crack.
This mesh in the $xy$-plane is extruded in the $z$-direction with $h\approx \ell/2.5$.
As a result, our mesh features $602\,880$ trilinear hexahedral elements and $630\,478$ nodes, yielding $1\,891\,434$ displacement and $630\,478$ phase-field DOFs.
Based on the previous findings, we employ the reduced-space active set strategy with no line search for the damage problem, and the exact line search based on bisection for the mechanical problem.

The numerically determined strain states at the last load step prior to the first phase-field evolution are reported in the $\operatorname{tr}(\boldsymbol{\varepsilon})$-$\| \boldsymbol{\varepsilon}_{\text{dev}}\|$ plane in Fig.~\ref{fig:brazilian_strainspace} along with the theoretical boundary of the elastic domain for the star-convex model with the chosen $\gamma^\star=5.5$.
Clearly, most of the domain sustains a compressive state.
Prior to one critical load step, the material behaves linear elastically with no damage, and the phase-field evolution starts as soon as the first point reaches the boundary of the elastic domain.
As also visible from the abrupt drop of the reaction force in the force-displacement curve of Fig.~\ref{fig:brazilian_reaction}, at this moment sudden failure of the specimen takes place within a single load step.

\begin{figure}[H]
    \centering
    \begin{subfigure}[t]{0.275\textwidth}
        \centering
        \includegraphics[height=4.25cm]{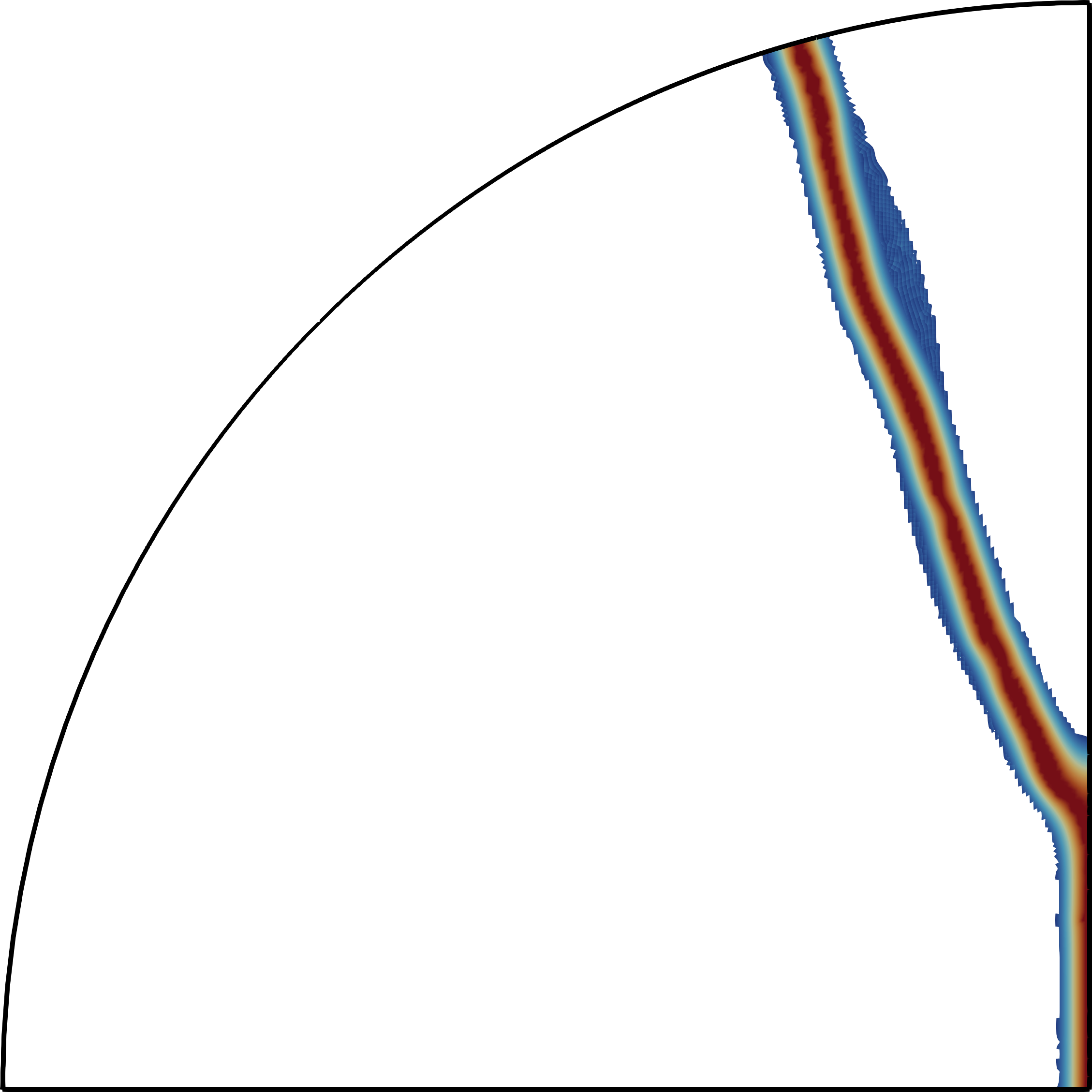}
        \caption{}
        \label{fig:phasefield_brazilian3D_back}
    \end{subfigure}
    \begin{subfigure}[t]{0.05\textwidth}
        \centering
        \includegraphics[height=4.25cm]{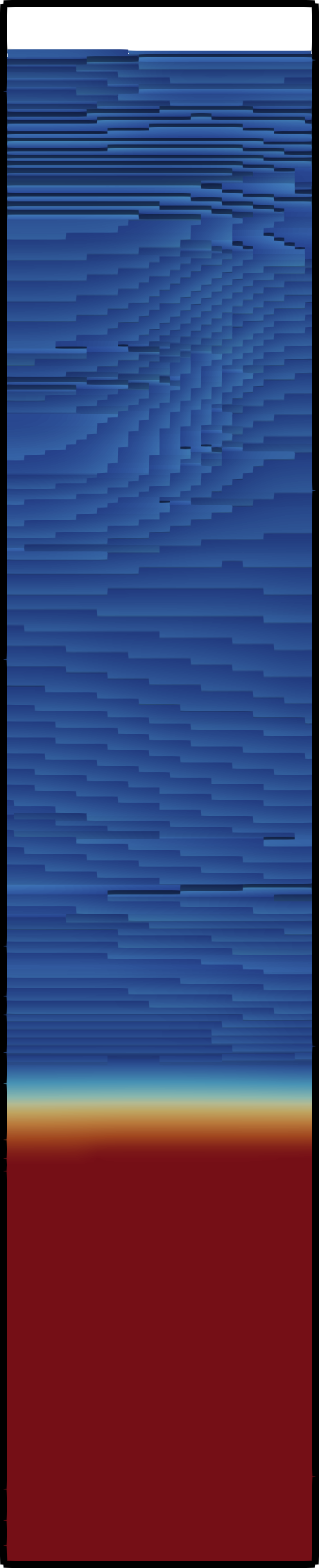}
        \caption{}
        \label{fig:phasefield_brazilian3D_left}
    \end{subfigure}
    \begin{subfigure}[t]{0.275\textwidth}
        \centering
        \includegraphics[height=4.25cm]{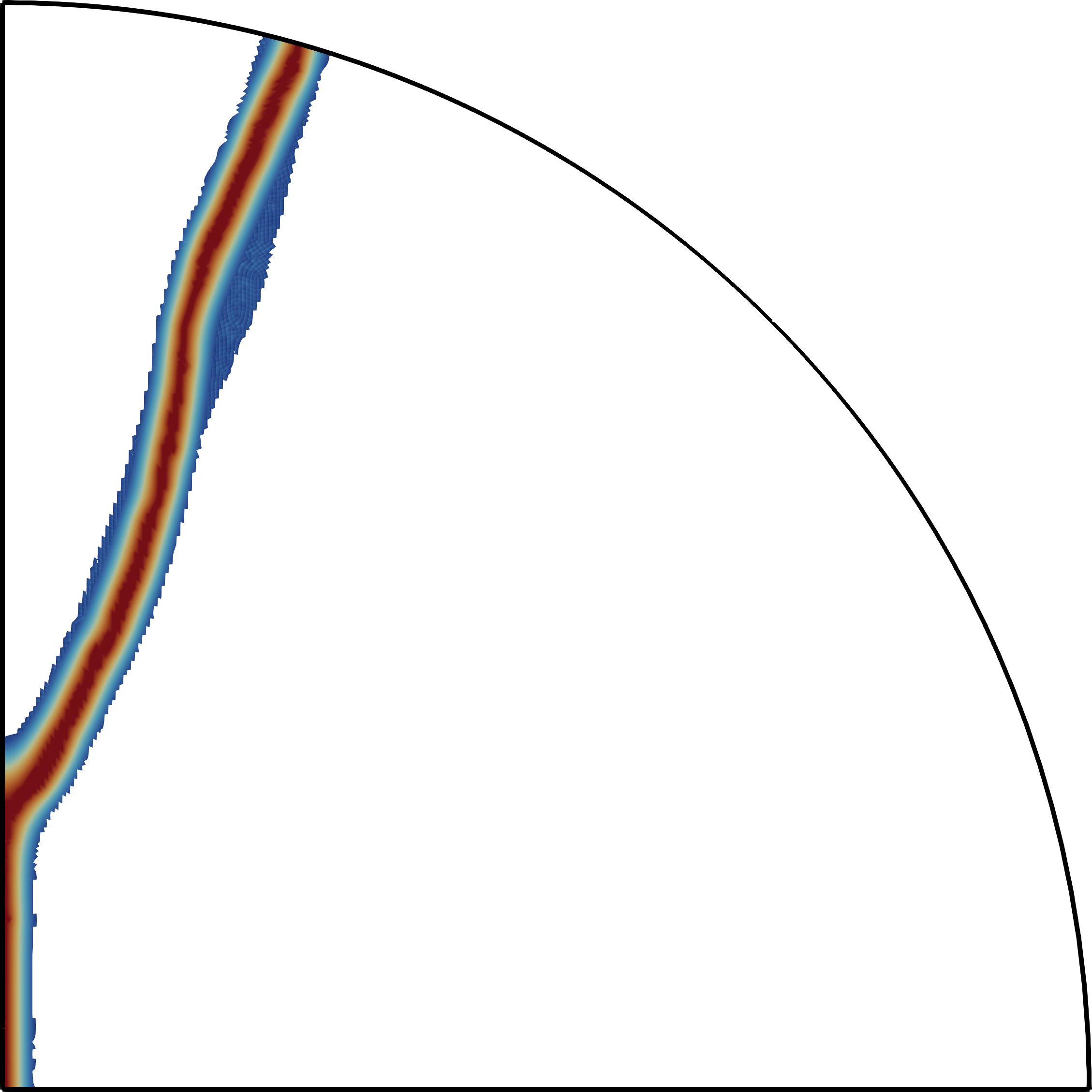}
        \caption{}
        \label{fig:phasefield_brazilian3D_front}
    \end{subfigure}
    \begin{subfigure}[t]{0.225\textwidth}
        \centering
        \includegraphics[height=4.25cm]{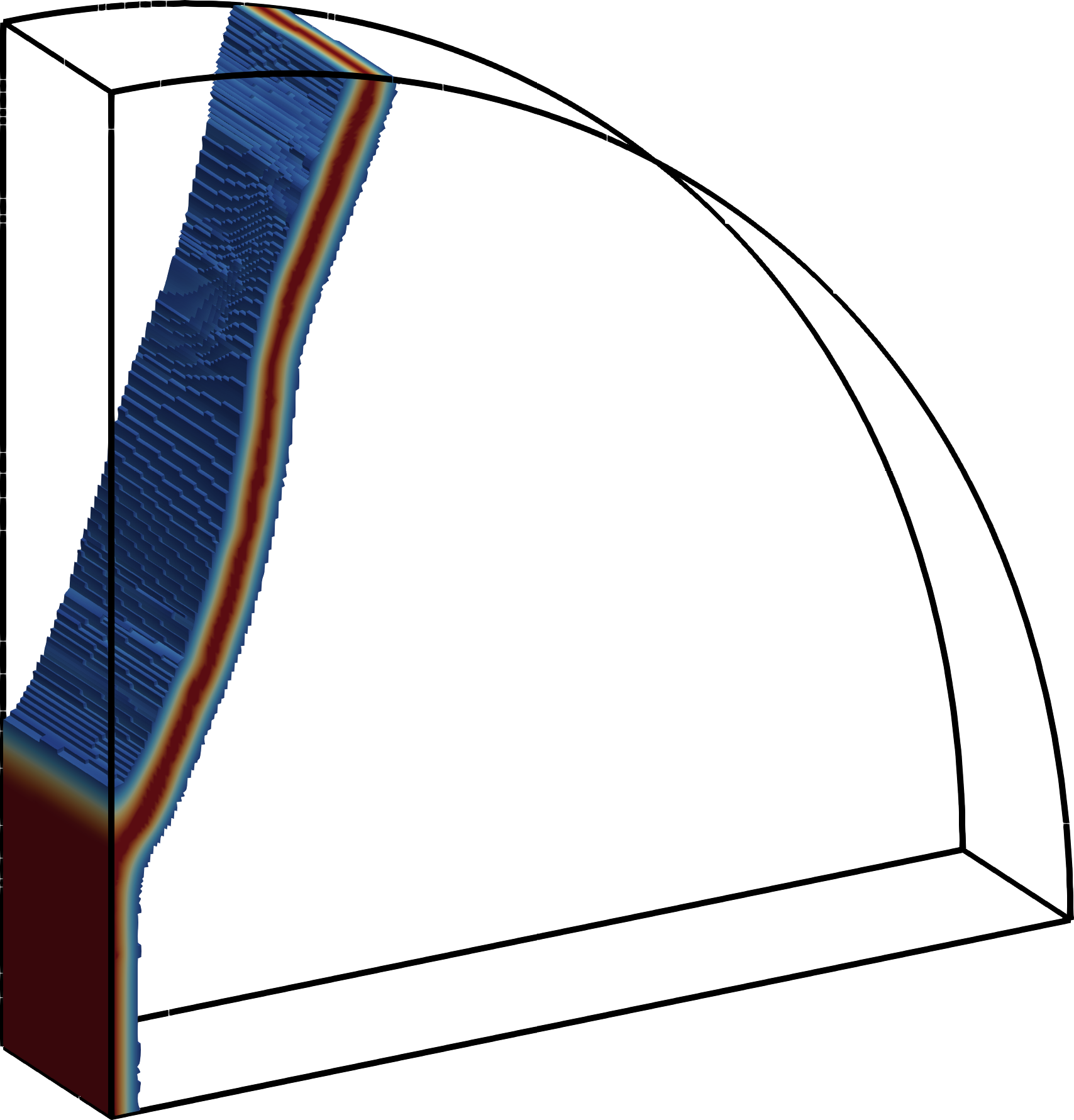}
        \caption{}
        \label{fig:phasefield_brazilian3D_perspective}
    \end{subfigure}
    \hspace{0.325cm}
    \includegraphics{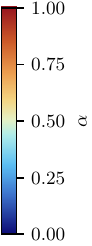}
    \caption{Final phase-field state on the back (a), the left (b), the front (c) sides, and from a 3D perspective (d). For clarity, only elements with $\alpha \geq 0.1$ are depicted.}
    \label{fig:phasefield_brazilian3D}
\end{figure}

The final phase-field solution at this critical load step $888$ is shown in Fig.~\ref{fig:phasefield_brazilian3D}, where all elements with $\alpha < 0.1$ are hidden.
From the center of the specimen, a vertical crack ranges up to roughly one fourth of the radius, after which the crack branches into two slightly curved cracks, reaching the outer radius of the domain at an angle of roughly $20^\circ$.
There is only little variation of the phase-field in the $z$-direction, justifying the coarser discretization in the thickness direction.
However, the crack is slightly straighter in the back face (the $xz$-symmetry plane), and more curved on the front surface.

\begin{figure}[H]
    \centering
    \includegraphics{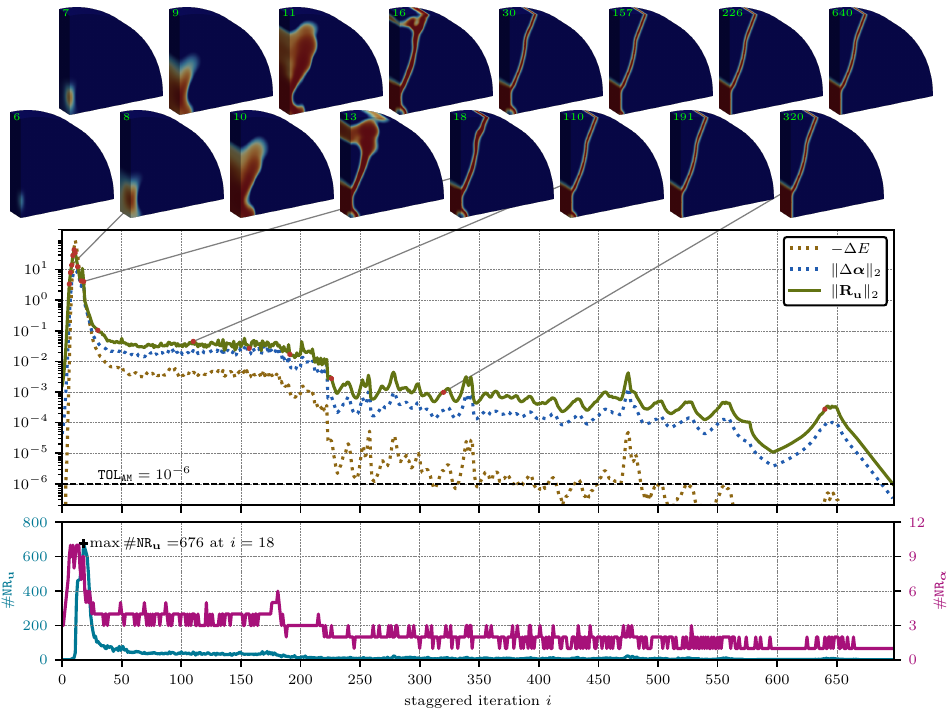}
    \caption{Staggered convergence behavior for the Brazilian test during critical step $888$}
    \label{fig:brazilian_convergence}
\end{figure}

This complicated crack pattern forming within a single load step poses a significant numerical challenge in light of the multi-axial, predominantly compressive stress states and contact conditions, proving the robustness of the proposed solution algorithm.
The convergence behavior during the alternate minimization loop at load step $888$ is illustrated in Fig.~\ref{fig:brazilian_convergence} in terms of
\begin{itemize}
    \item the updated residual of the displacement field with the updated phase-field state, $\|\mathbf{R}_{\mathbf{u}}\|_2 = \| \mathbf{R}_{\mathbf{u}} (\mathbf{u}^{i}, \bm{\alpha}^{i}) \|_2$ (based on which we determine convergence of the alternate minimization algorithm),
    \item the change of the phase-field solution iterate from one staggered iteration to the next $\|\Delta \bm{\alpha} \|_2 = \sqrt{\int_{\Omega} | \alpha^{i} - \alpha^{i-1} |^2 \mathrm{d}\boldsymbol{x}}$,
    \item the decrease of the energy from one alternate minimization loop to the next, $-\Delta E = E(\mathbf{u}^{i-1}, \bm{\alpha}^{i-1}) - E(\mathbf{u}^{i}, \bm{\alpha}^{i})$, and
    \item the number of necessary Newton iterations to reach convergence in the mechanical and damage problems.
\end{itemize}
Further, the intermediate damage solution fields are shown for selected staggered iterations.
During the first iterations, there is barely any evolution of the phase field; a value of $\alpha \geq 0.25$ is obtained only after the sixth staggered iteration slightly upwards from the center on the front face, where the phase-field value quickly evolves to $1$.
The damaged region grows from this position; the nearly final branching location and angle of the crack reaching the outer boundary are found after roughly $20$ staggered iterations.
From this point on, the crack becomes thinner.
At the center of the outer surface, a small hemispherical region remains undamaged during the first staggered iterations and shrinks during the subsequent iterations until only a hemispherical damaged region remains, which disappears after iteration $190$.
These changes of the phase-field pattern are reflected in the value of $\|\Delta \bm{\alpha} \|_2$, which is in the order of $10^{-3}$ after iteration $250$ and drops below $10^{-4}$ after staggered iteration $550$.
In general, the evolutions of $\|\mathbf{R}_{\mathbf{u}}\|_2$, $\|\Delta \bm{\alpha} \|_2$ and the energy decrease $-\Delta E$ follow very similar patterns, albeit in different orders of magnitude.
The energy decrease is generally two to three orders below the residual norm, while the change of the phase-field vector is at most one order of magnitude below the residual in this case.

The number of necessary Newton iterations is also directly related to the evolution of the fields:
Newton's method needs up to $676$ iterations for the mechanical problem in the $18$th alternate minimization loop, while this number quickly drops to around $40$ between staggered iterations $50$ and $200$, around $10$ between iterations $200$ and $550$, and mostly between one and three afterwards.
While these numbers may seem excessively large, we believe they are reasonable due to several factors.
First, they mostly occur around the time when the crack first spans the whole specimen diameter, implying a full loss of load-bearing capacity.
Accordingly, the initial guess of Newton's method is very far from the solution.
Secondly, the active and inactive sets of the reduced space strategy, $\mathcal{A}^{i,k}$ and $\mathcal{I}^{i,k}$, may significantly change among iterations based on the deformation state.
This is further complicated by the overlap of the region where the specimen is contact with the loading platens and the region where the crack appears.
Further, during the staggered iterations with a large number of Newton iterations in the mechanical problem, the phase field transitions from a diffuse pattern (i.e., very large regions of significantly reduced stiffness) to a localized crack.
A different initialization strategy or the adoption of a quasi-Newton method might be beneficial, however further investigations would be needed to guarantee that no energy barriers are crossed and to study the implications for an exact line search.
The damage problem instead needs at most $10$ iterations to converge when there are large changes in the phase-field state in the beginning, and at most three iterations after staggered iteration $250$.
This example clearly shows that stopping the alternate minimization after the first staggered iteration, as suggested in \cite{miehe2010thermodynamically,chafia_massively_2025}, would not lead to accurate results.

Most of the computational time is spent in the Newton iterations of the mechanical problem ($17\,985$ in total for this time step).
As concluded in Section~\ref{sec:assessment}, the exact line search needs a smaller total number of Newton iterations in comparison to other line searches, which plays an important role for large computations.
In this case, for load step $888$, the line search for the mechanical problem amounts to ca. $6.9$~\% of the total computational time (at this stage $5.3$~\% for the $350\,166$ displacement field residual assemblies), and the linear solves to ca. $92.6$~\%.

\section{Conclusions}\label{sec:conclusions}
We addressed the iterative convergence issues of Newton's method within an alternate minimization procedure for computations with variational phase-field models for brittle fracture.
These issues stem from the strong non-linearities induced in the displacement subproblem by the strain energy decomposition, and in the phase-field subproblem by the penalty-based enforcement of damage irreversibility.
As a remedy, we proposed to enhance Newton's method with an exact line search globalization strategy which minimizes the energy along the Newton update direction through a simple yet robust bisection algorithm.
In the specific context of variational phase-field models for brittle fracture, this allowed us to establish a global convergence property of the line-search enhanced Newton's method for strictly convex problems, which in turn guarantees the convergence of the alternate minimization algorithm to critical points of the energy.

With an obstacle course comprising several benchmark tests for brittle fracture which we compute with various commonly adopted strain energy decompositions, using both the penalty method and the reduced-space active set strategy to enforce irreversibility of damage, we demonstrated the robustness of the resulting solution approach and could draw the following additional conclusions:
\begin{itemize}
    \item the limited computational overhead of the proposed exact line search based on bisection is compensated by less iterations to reach convergence in comparison to other line search algorithms.
        The reduction in the number of iterations significantly enhances efficiency, especially for large problems.
    \item For the damage problem, the exact line search does not offer advantages when the reduced-space active set strategy is used to enforce irreversibility; conversely, it is important to overcome convergence issues when using the penalty method.
    \item While the global convergence proof does not cover the reduced-space active set strategy and the star-convex strain energy decomposition with $\gamma^\star > 0$, empirical evidence shows that the algorithm maintains the same level of robustness and efficiency also under these conditions.
\end{itemize}
The adopted line search algorithm is minimally invasive and straightforward to implement, and we have recently contributed it to the \texttt{PETSc} library (all our implementations are publicly available at \url{https://github.com/jonas-heinzmann/phase_field_exact_linesearch}).
Based on the results in this paper, we believe that alternate minimization with an exact line search is 'all one needs' for robust computations of phase-field brittle fracture.

\section*{Acknowledgements}
We gratefully acknowledge funding from the Swiss National Science Foundation (SNF) through Grant No. 200021-219407 'Phase-field modeling of fracture and fatigue: from rigorous theory to fast predictive simulations'.

\section*{Code availability}
The implementations used in this work are publicly available at \url{https://github.com/jonas-heinzmann/phase_field_exact_linesearch}.

\bibliographystyle{elsarticle-num}
\bibliography{references}

\addcontentsline{toc}{section}{Appendix}
\renewcommand{\thesubsection}{\Alph{subsection}}
\section*{Appendix}

\subsection{Strain energy splits}\label{app:splits}
In the following, we report the degradable and residual parts of the strain energy density for the different splits used in this work.
The volumetric-deviatoric split proposed by \cite{amor_regularized_2009} reads
\begin{equation}\label{eq:psi_voldev}
    \psi_{\text{D}} (\boldsymbol{\varepsilon}) = \mu_0 \|\boldsymbol{\varepsilon}_{\text{dev}}\|^2 + \frac{\kappa_0}{2} \langle \text{tr} (\boldsymbol{\varepsilon}) \rangle_+^2 
    \quad\text{,}\quad
    \psi_{\text{R}} (\boldsymbol{\varepsilon}) = \frac{\kappa_0}{2} \langle \text{tr} (\boldsymbol{\varepsilon}) \rangle_-^2 \quad\text{.}
\end{equation}
For the spectral split proposed in \cite{miehe2010thermodynamically}, $\psi_{\text{D}}$ and $\psi_{\text{R}}$ read
\begin{equation}
    \psi_{\text{D}} (\boldsymbol{\varepsilon}) = \frac{1}{2}\lambda_0 \langle \operatorname{tr}(\boldsymbol{\varepsilon}) \rangle_+^2 + \mu_0 \boldsymbol{\varepsilon}_+ \cdot \boldsymbol{\varepsilon}_+
    \quad\text{,}\quad
    \psi_{\text{R}} (\boldsymbol{\varepsilon}) = \frac{1}{2}\lambda_0 \langle \operatorname{tr}(\boldsymbol{\varepsilon}) \rangle_-^2 + \mu_0 \boldsymbol{\varepsilon}_- \cdot \boldsymbol{\varepsilon}_-\\
\end{equation}
with $\boldsymbol{\varepsilon}_\pm = \sum_i \langle \varepsilon_i \rangle_\pm \boldsymbol{e}_i \otimes \boldsymbol{e}_i$, where $\varepsilon_i$ are the eigenvalues of the strain tensor and $\boldsymbol{e}_i$ are the corresponding eigenvectors.
Here, $\lambda_0$ is the first Lamé parameter.
The no-tension model proposed in \cite{freddi_regularized_2010} works with structured deformations, denoted as $\boldsymbol{\eta}$, which are constrained within a convex set $\mathcal{K}_{\boldsymbol{\varepsilon}} = \text{Sym}_+$ (that is the set of all symmetric positive semidefinite tensors) defining the admissible 'structure' of micro-cracks.
The residual energy is then given as the solution of the following minimization problem
\begin{equation}\label{eq:notensionsplit}
    \psi_{\text{R}} (\boldsymbol{\varepsilon}) = \min_{\boldsymbol{\eta} \in \mathcal{K}_{\boldsymbol{\varepsilon}}} \psi_0 (\boldsymbol{\varepsilon} - \boldsymbol{\eta}) \quad\text{with} \quad\psi_0 (\boldsymbol{\varepsilon}) = \frac{1}{2}\lambda_0 \operatorname{tr}(\boldsymbol{\varepsilon})^2 + \mu_0 \boldsymbol{\varepsilon} \cdot \boldsymbol{\varepsilon} \quad\text{,}
\end{equation}
while the degradable part is given by $\psi_{\text{D}} (\boldsymbol{\varepsilon}) = \psi_0 (\boldsymbol{\varepsilon}) - \psi_{\text{R}} (\boldsymbol{\varepsilon})$.
For brevity, the result of the minimization is not reported here, instead we refer to \cite{Vicentini_energy_2024,de_lorenzis_nucleation_2022}.
The DP-like split proposed in \cite{de_lorenzis_nucleation_2022} extends this idea, defining
\begin{equation}
    \mathcal{K}_{\boldsymbol{\varepsilon}} = \{ \boldsymbol{\eta} \in \text{Sym} : \operatorname{tr}(\boldsymbol{\eta}) \geq \gamma \| \operatorname{dev} (\boldsymbol{\eta}) \| \}
\end{equation}
with the parameter $\gamma \geq 0$ giving some flexibility to tune the shape of the elastic domain.
The minimization problem~\eqref{eq:notensionsplit} remains unchanged, and for its solution again we refer to \cite{Vicentini_energy_2024,de_lorenzis_nucleation_2022}.
Finally, the star-convex formulation proposed in \cite{Vicentini_energy_2024} reads
\begin{equation}\label{eq:psi_starconvex}
    \psi_{\text{D}} (\boldsymbol{\varepsilon}) = \mu_0 \|\boldsymbol{\varepsilon}_{\text{dev}}\|^2 
    + \frac{\kappa_0}{2} \Big( \langle \text{tr} (\boldsymbol{\varepsilon}) \rangle_+^2 
    - \gamma^\star \langle \text{tr} (\boldsymbol{\varepsilon}) \rangle_-^2 \Big)
    \quad\text{,}\quad
    \psi_{\text{R}} (\boldsymbol{\varepsilon}) = (1+\gamma^\star) \frac{\kappa_0}{2} \langle \text{tr} (\boldsymbol{\varepsilon}) \rangle_-^2 \quad\text{,}
\end{equation}
where the parameter $\gamma^\star\geq-1$ controls the ratio between compressive and tensile strengths \cite{Vicentini_energy_2024}.
In particular, for $\gamma^\star=-1$ the model reduces to the standard phase-field formulation without energy decomposition, as originally proposed in \cite{bourdin_numerical_2000}, whereas for $\gamma^\star=0$ it recovers the volumetric--deviatoric decomposition \eqref{eq:psi_voldev}.
A detailed explanation and comparison of the various models is given in \cite{Vicentini_energy_2024}.

\subsection{Obstacle course}\label{app:obstacle_course}
In the following, we report the setups for the numerical tests used in the obstacle course, whose domains and boundary conditions are illustrated in the first row of Tab.~\ref{tab:obstacle_course}.
We consider plane-strain conditions and discretize the domain with bilinear quadrilateral Lagrange elements with an element size $h$ such that $\ell/h \approx 5$ in the whole domain.
The Dirichlet boundary conditions are applied in 100 uniform steps, for which we plot the reaction forces for the different splits in the second row of Tab.~\ref{tab:obstacle_course}.
The remaining rows of Tab.~\ref{tab:obstacle_course} depict the phase-field solution at the last load step for the various splits.
These results were obtained with the reduced-space active set strategy with no line search for the damage problem, and the exact line search based on bisection for the mechanical problem.

\begin{figure}[t]
    \centering
    \begin{subfigure}[c]{0.125\textwidth}
        \centering
        \includegraphics[width=2.25cm]{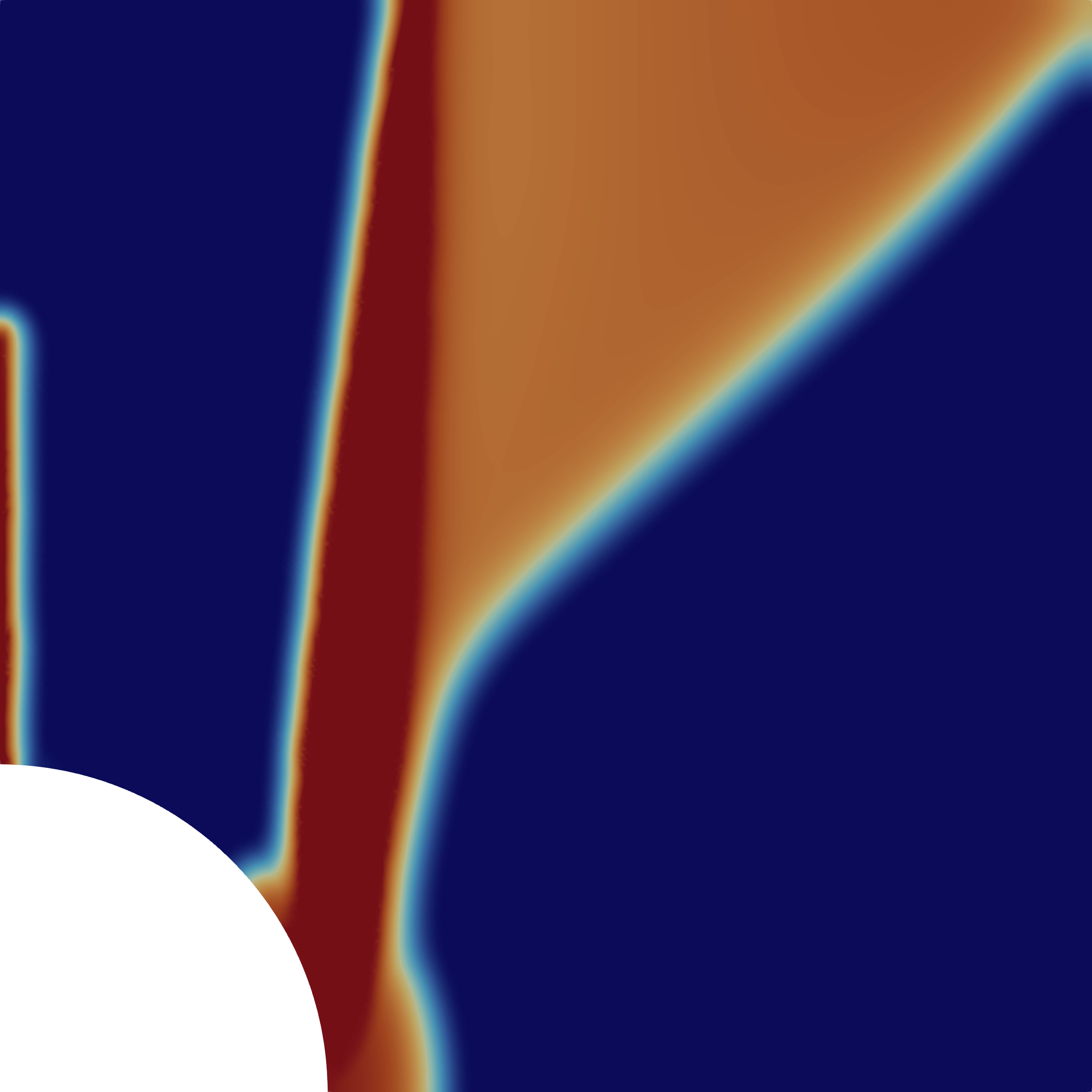}
        \caption{}
        \label{fig:phasefield_platewithhole_spectral_72}
    \end{subfigure}
    \hfill
    \begin{subfigure}[c]{0.125\textwidth}
        \centering
        \includegraphics[width=2.25cm]{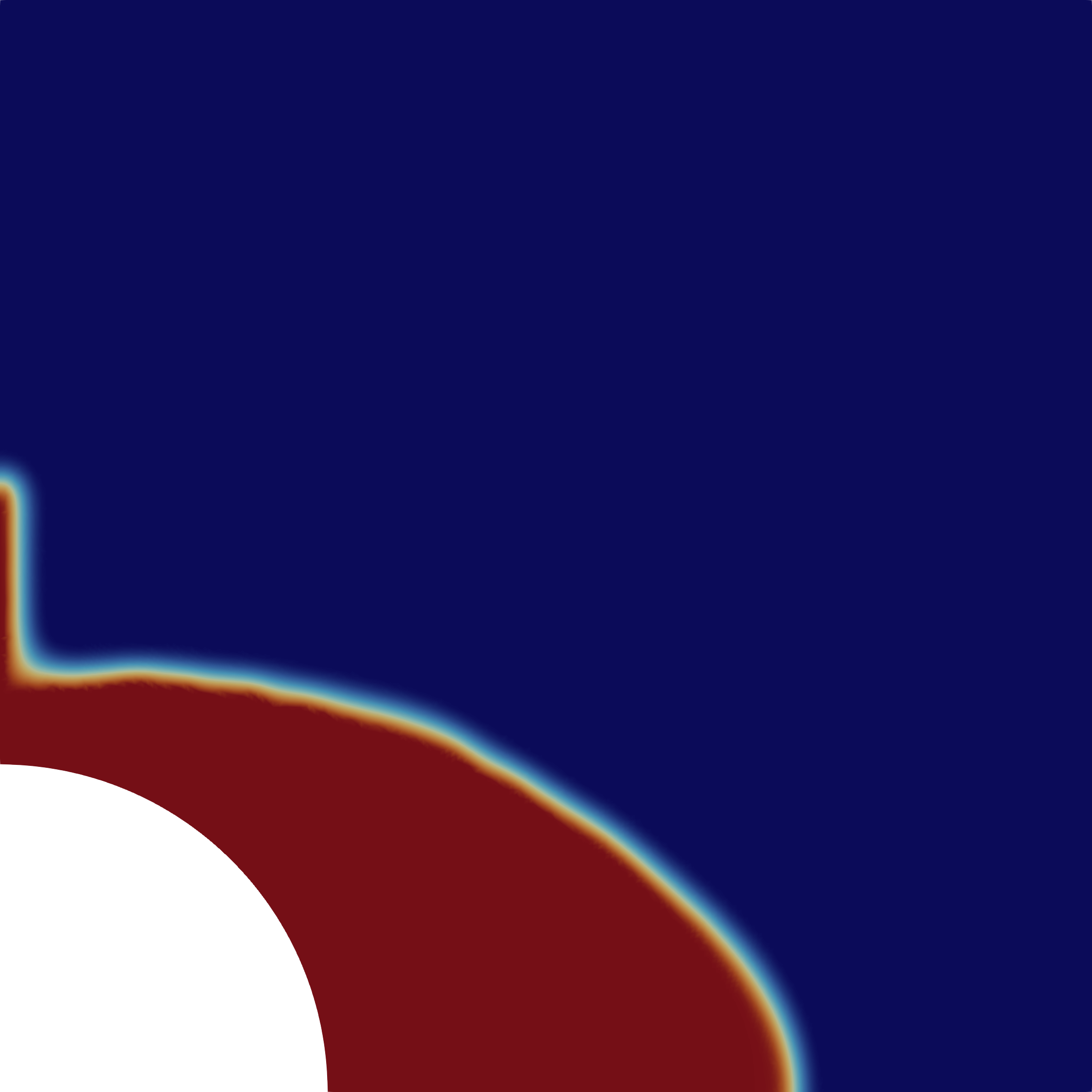}
        \caption{}
        \label{fig:phasefield_platewithhole_DPlike_56}
    \end{subfigure}
    \hfill
    \begin{subfigure}[c]{0.125\textwidth}
        \centering
        \includegraphics[width=2.25cm]{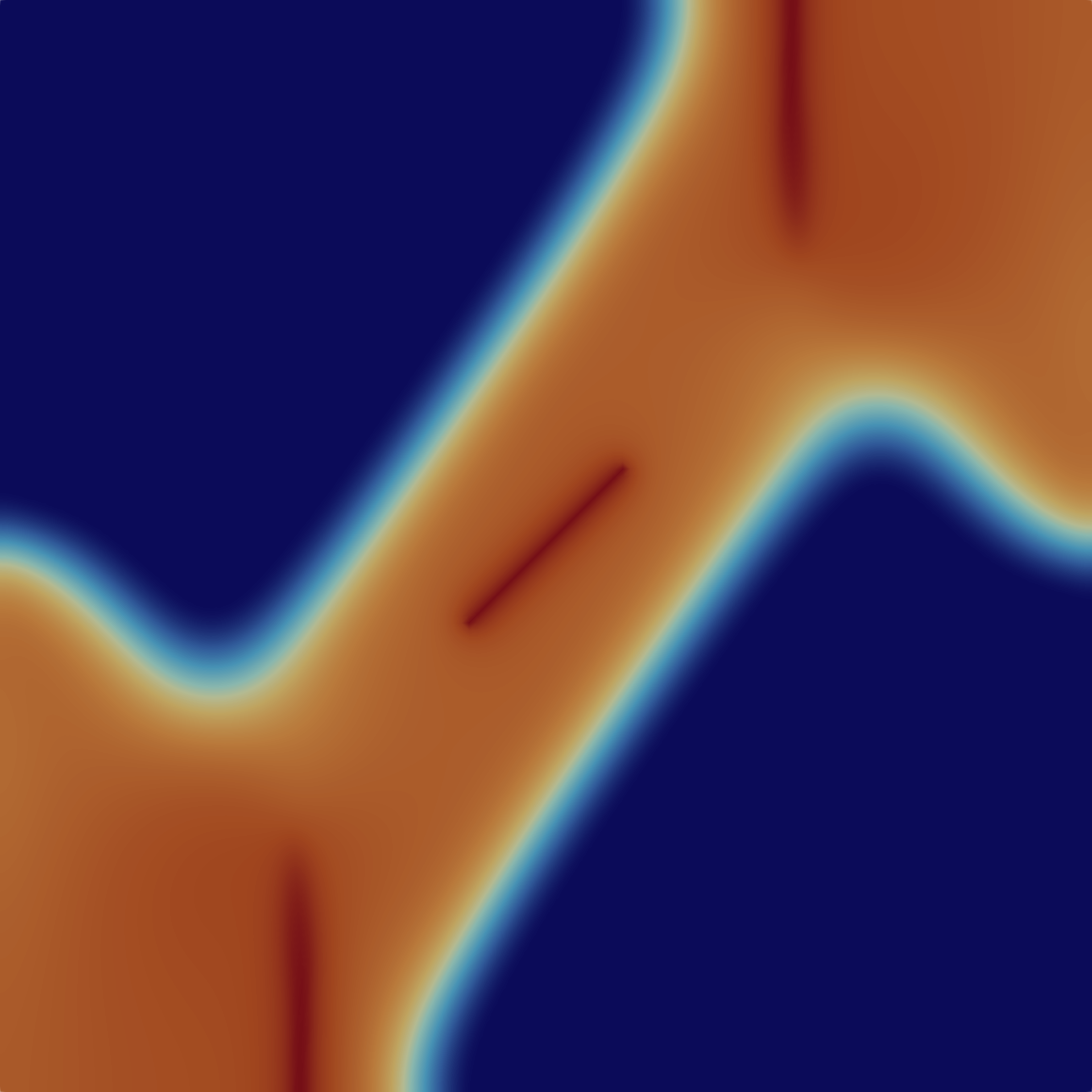}
        \caption{}
        \label{fig:phasefield_platewithinclinedcrack_spectral_91}
    \end{subfigure}
    \hfill
    \begin{subfigure}[c]{0.125\textwidth}
        \centering
        \includegraphics[width=2.25cm]{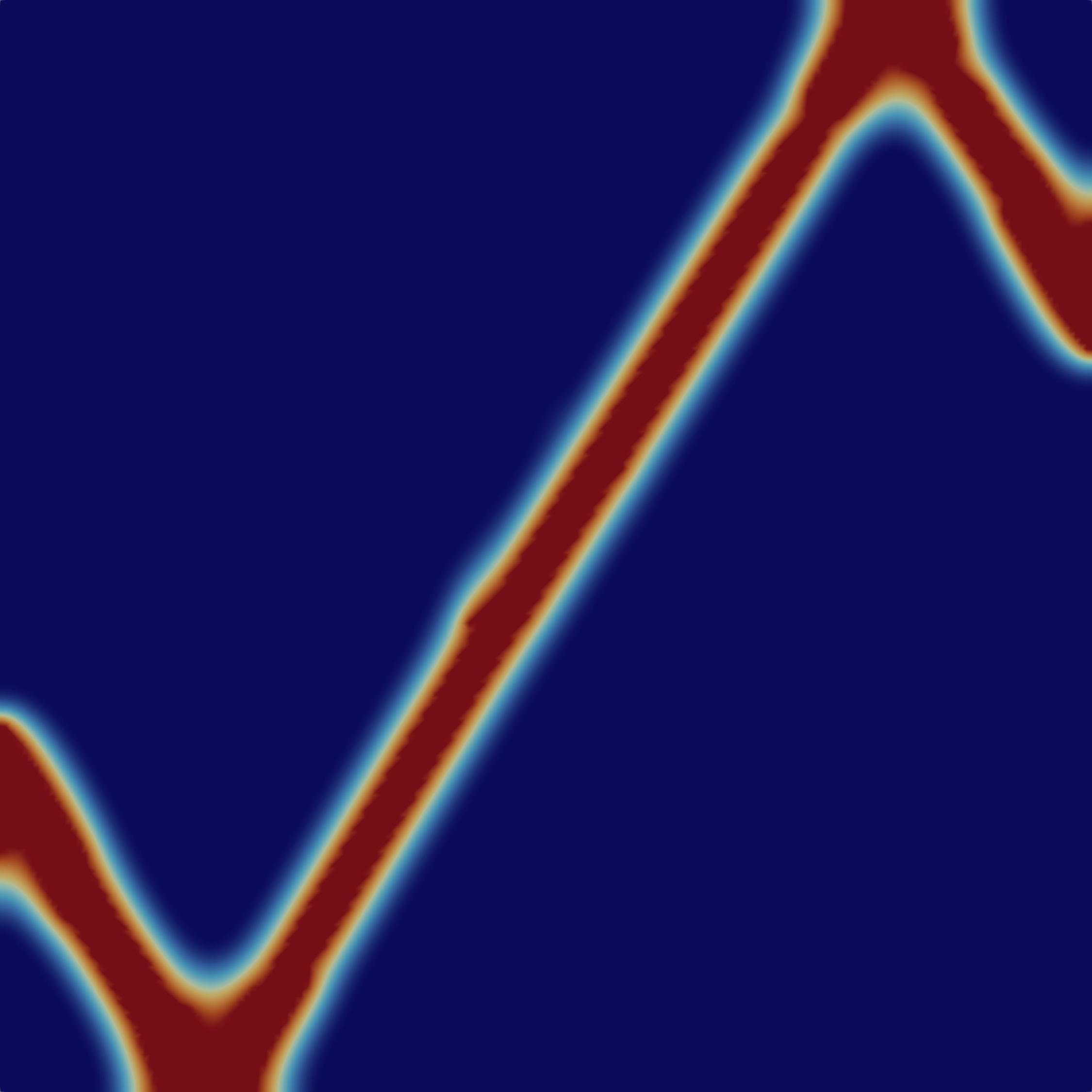}
        \caption{}
        \label{fig:phasefield_platewithinclinedcrack_DPlike_94}
    \end{subfigure}
    \hfill
    \begin{subfigure}[c]{0.125\textwidth}
        \centering
        \includegraphics[width=2.25cm]{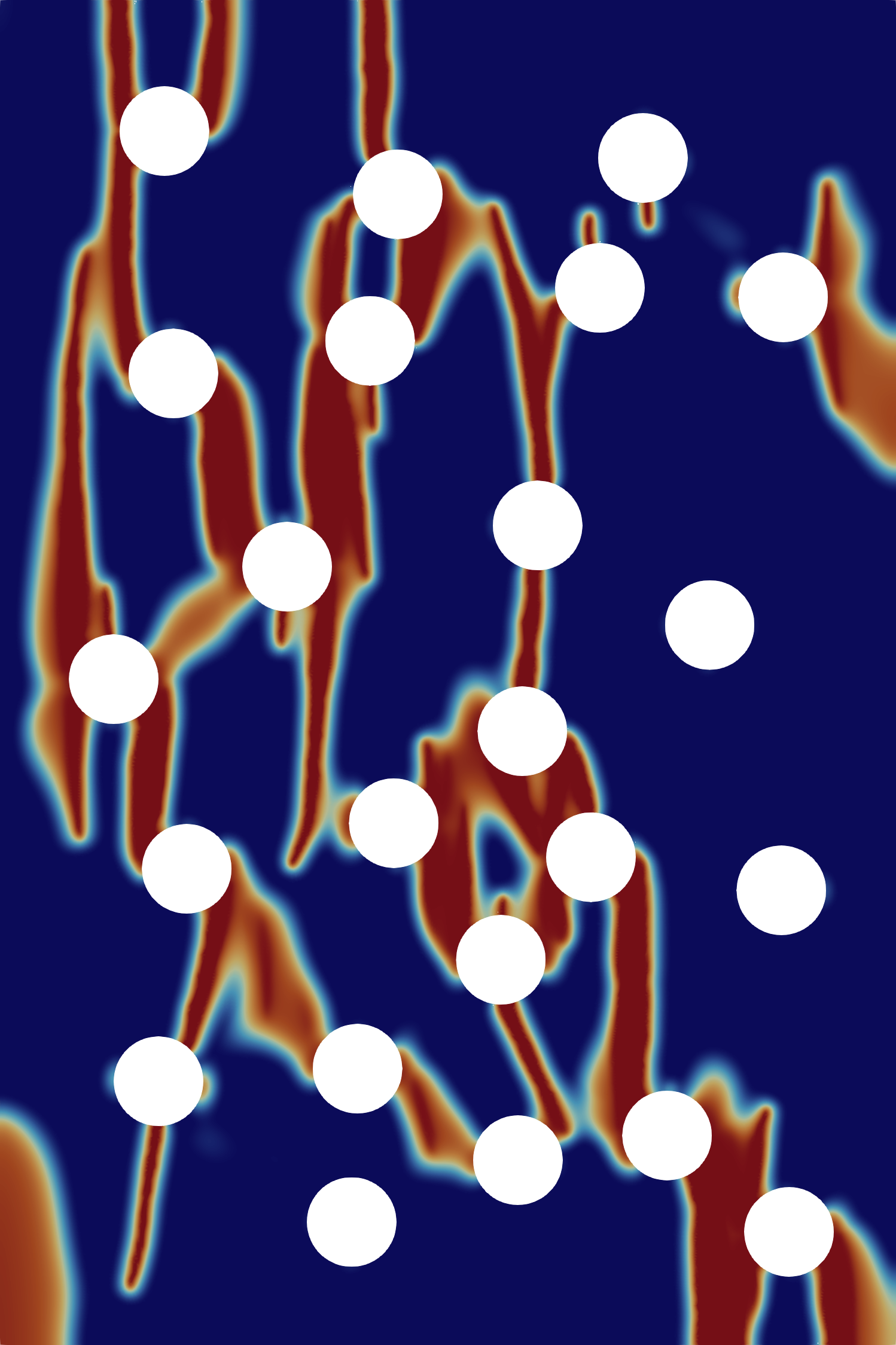}
        \caption{}
        \label{fig:phasefield_perforatedplate_spectral_76}
    \end{subfigure}
    \hfill
    \begin{subfigure}[c]{0.125\textwidth}
        \centering
        \includegraphics[width=2.25cm]{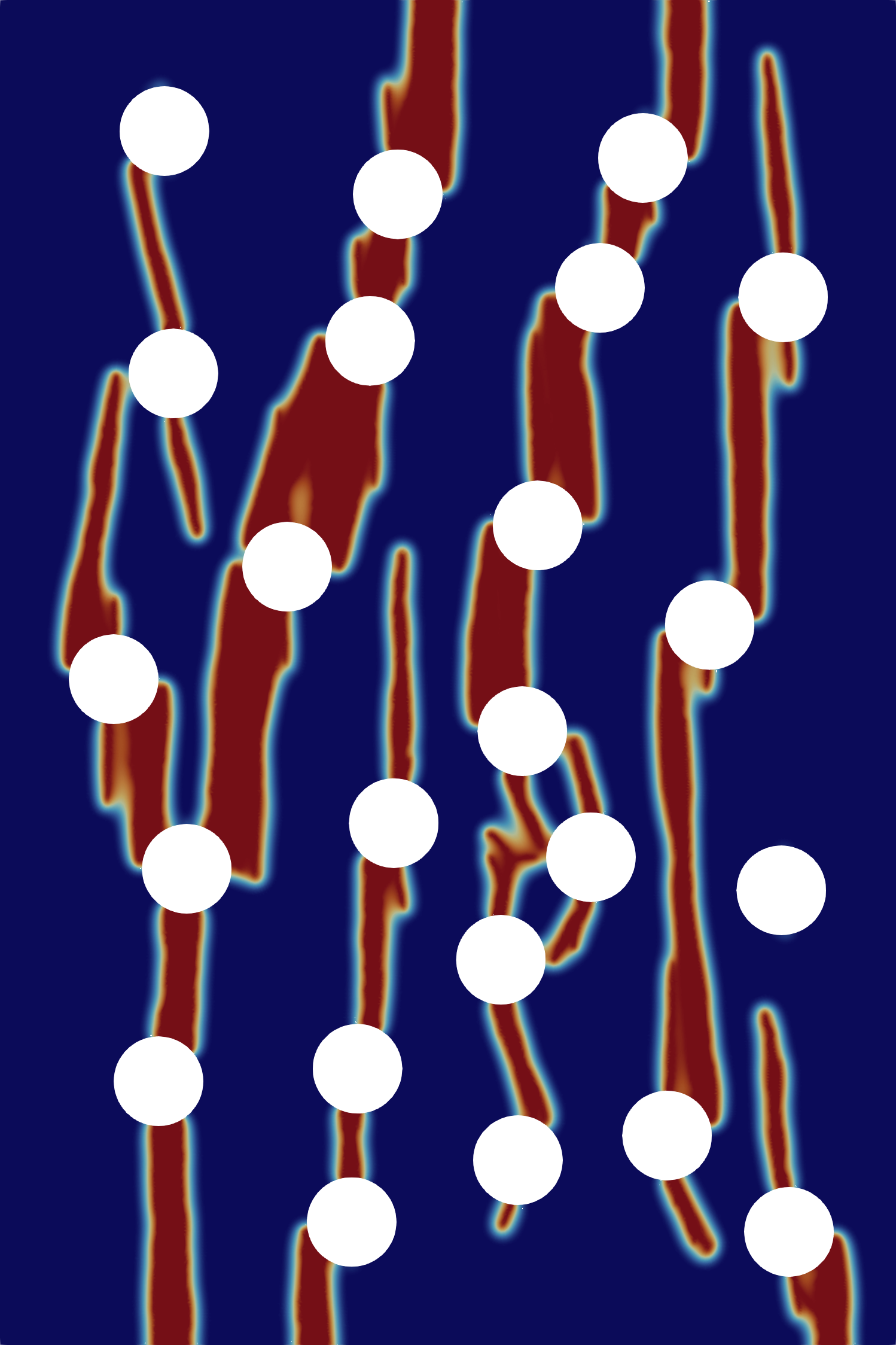}
        \caption{}
        \label{fig:phasefield_perforatedplate_no-tension_90}
    \end{subfigure}
    \hfill
    \begin{subfigure}[c]{0.125\textwidth}
        \centering
        \includegraphics[width=2.25cm]{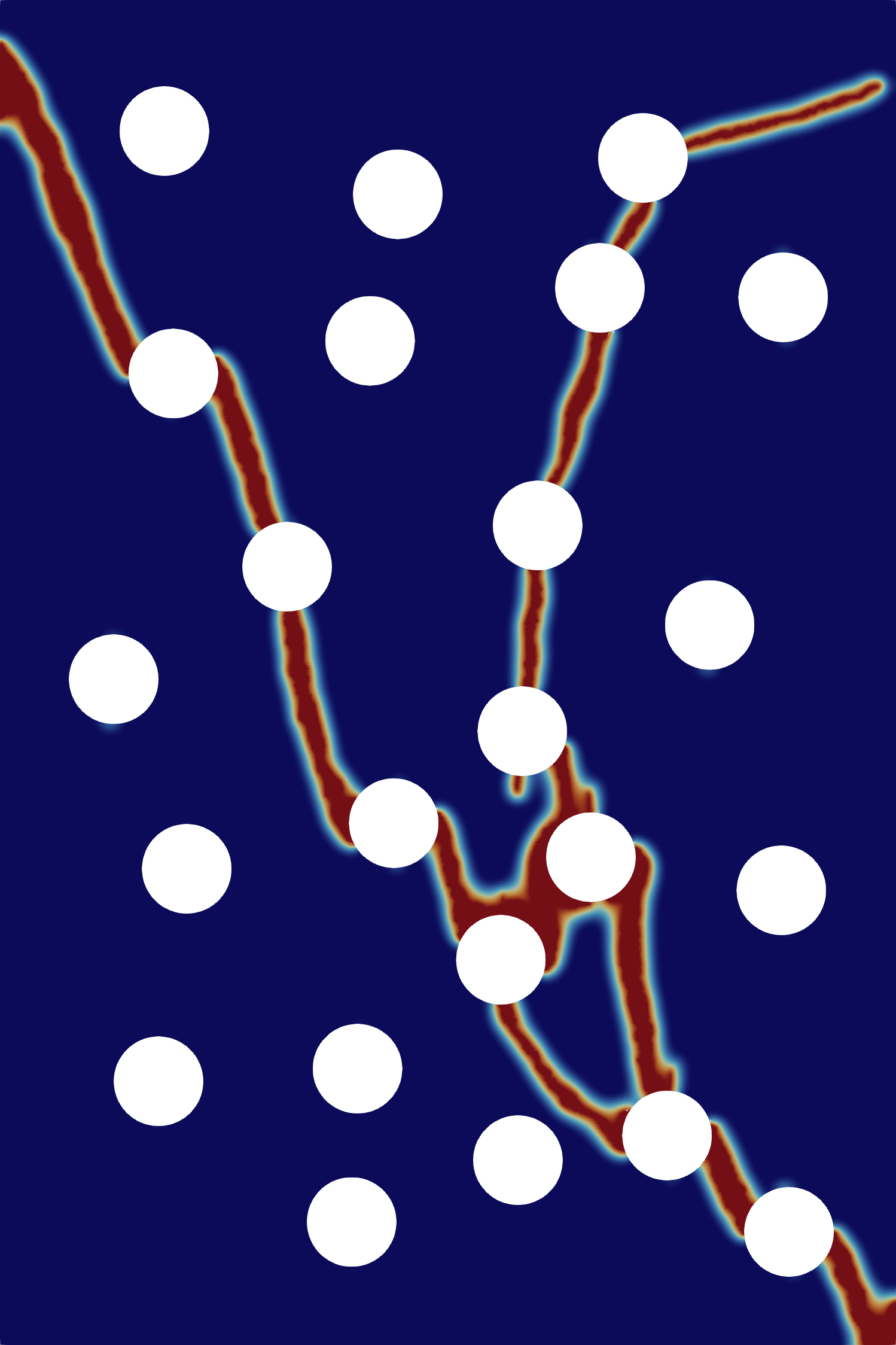}
        \caption{}
        \label{fig:phasefield_perforatedplate_DPlike_66}
    \end{subfigure}
    \caption{Unrealistic phase-field solutions obtained for the plate with hole test with the spectral (a) and DP-like (b) splits, the plate with inclined crack with the spectral (c) and DP-like (d) splits, and the perforated plate with the spectral (e), no-tension (f), and DP-like (g) splits}
    \label{fig:obstaclecourse_notmeaningful}
\end{figure}

We remark that some of the obtained phase-field solutions appear unrealistic, e.g. those obtained with the spectral and DP-like splits for the plate with a hole, the plate with an inclined crack and the perforated plate, where large regions of diffuse damage can be observed.
Since the objective of this paper is not to assess the predictive ability of different models, but rather to prove the numerical robustness of our proposed approach, we consider it valuable to be able to converge to these solutions.
Nevertheless, we stop the computations after the related load step (marked with a cross in the force-displacement plots in Tab.~\ref{tab:obstacle_course}) and report in Tab.~\ref{tab:obstacle_course} the last physically realistic phase-field solutions, while the unrealistic ones obtained at the subsequent load step are depicted in Fig.~\ref{fig:obstaclecourse_notmeaningful}.
For the plate with an inclined crack and the perforated plate with the DP-like split, the tolerance of Newton's method is slightly relaxed to $\mathtt{TOL}_{\mathtt{NM}}=10^{-7}$, while the maximum number of Newton iterations is increased to $50\,000$.
Further, we set $\mathtt{atol}=10^{-16}$ and $\mathtt{ltol}=10^{-10}$ for the exact line search in these two cases to be able to obtain also the physically unrealistic solutions.

\subsubsection{Nucleation test}
The first test is similar to the nucleation test used in \cite[Section~5.3]{vicentini_variational_2025}, i.e. a rectangular domain of edge length $L=1$~mm for which we impose a homogeneous strain state.
To this end, we prescribe a displacement of same magnitude, but opposite directions on opposing edges of the domain, while we impose $\bar{\alpha}=0$ on the four corner nodes to prevent a phase-field crack on the boundary.
Specifically, we impose a homogeneous stress state by setting $\bar{u}_x = \cos(320^\circ) L/10$ and $\bar{u}_y = \sin(320^\circ) L/10$.
We adopt a Young's modulus $E_0=100$~MPa, a Poisson's ratio $\nu_0=0.3$, $G_{\text{c}}=0.1$~N$/$mm and $\ell=0.05$~mm.
We report only the horizontal reaction force at the left edge of the domain.
The regular mesh has $10\,000$ elements and $10\,201$ nodes.

\subsubsection{Sliding test}
We adopt the sliding test from \cite[Section~5.3]{Vicentini_energy_2024}.
The material has Young's modulus $E_0 = 100$~MPa, Poisson's ratio $\nu_0 = 0.3$ and fracture toughness $G_{\text{c}} = 2/15$~N$/$mm, and we adopt $\ell=0.05$~mm.
The initial crack is modeled by means of an initial phase field, and the regular mesh has $10\,000$ elements and $10\,201$ nodes.

\subsubsection{Plate with a hole}
The second test is the plate with a hole, adopted from \cite[Section~5.2]{Vicentini_energy_2024}, i.e. a square plate with edge length $L=2.0$~mm and a central hole with radius $R=0.3$~mm under uniaxial compression.
We impose Dirichlet boundary conditions on the top edge, and symmetry boundary conditions on the left and bottom edges.
The material has Young's modulus $E_0 = 100$~MPa, Poisson's ratio $\nu_0 = 0.3$ and fracture toughness $G_{\text{c}} = 4/75$~N$/$mm, and we adopt $\ell=0.02$~mm.
The mesh features $45\,870$ elements and $46\,277$ nodes.

\subsubsection{Plate with an inclined crack}
Next, we study a plate featuring a pre-existing crack at an angle, similar to the test used in \cite{Ingraffea_finite_1980}.
The plate has length $L=100$~mm and has a central crack of length $a=L/5$ at an angle of $\theta=45^\circ$.
The initial crack is modeled by means of an initial phase field.
As material parameters, we consider $E_0=210\,000$~MPa, $\nu_0=0.3$, $G_{\text{c}}=2.7$~N$/$mm and $\ell=2.5$~mm.
The mesh features $41\,224$ elements and $41\,625$ nodes.

\subsubsection{Perforated plate}
The last test is the \textit{perforated plate} test similar to the one in \cite{nguyen_phase_2015}, where a plate with dimensions $W=2$~mm and $H=3$~mm with $23$ randomly positioned holes of radius $R=0.1$~mm is uniaxially compressed.
Due to the random position of the holes, cracks nucleate and propagate at different loading steps, making convergence particularly challenging especially for models with a strain energy decomposition.
For the material parameters, we set $E_0=12\,000$~MPa, $\nu_0=0.3$, $G_{\text{c}}=0.0014$~N$/$mm and $\ell=0.025$~mm.
The mesh features $191\,296$ elements and $193\,452$ nodes.

\begin{table}[p]
    \vspace{-1cm}
    \caption{Setups, reaction forces, and phase-field solutions for the tests of the obstacle course}
    \label{tab:obstacle_course}
    \footnotesize
    \begin{tabularx}{\textwidth}{p{0.7cm}ccccc}
        \toprule
        &nucleation test &sliding test &plate with hole &plate w. inclined crack &perforated plate\\
        \midrule
        &\raisebox{-.5\height}{\includegraphics{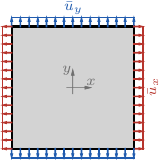}}
        &\raisebox{-.5\height}{\includegraphics{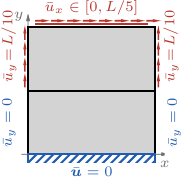}}
        &\raisebox{-.5\height}{\includegraphics{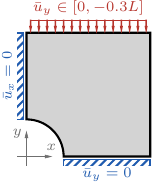}}
        &\raisebox{-.5\height}{\includegraphics{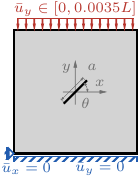}}
        &\raisebox{-.5\height}{\includegraphics{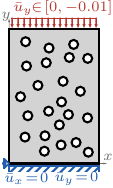}}
        \\
        \midrule
        \multicolumn{6}{c}{\raisebox{-.5\height}{\includegraphics{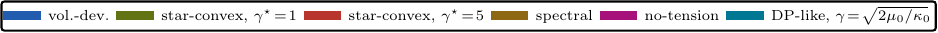}}}
        \\
        &\raisebox{-.5\height}{\includegraphics{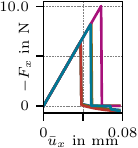}}
        &\raisebox{-.5\height}{\includegraphics{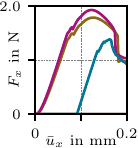}}
        &\raisebox{-.5\height}{\includegraphics{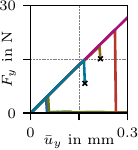}}
        &\raisebox{-.5\height}{\includegraphics{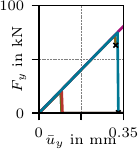}}
        &\raisebox{-.5\height}{\includegraphics{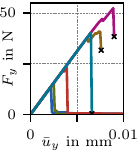}}
        \\
        \midrule
        vol-dev.
        &\raisebox{-.5\height}{\includegraphics[height=2.0cm]{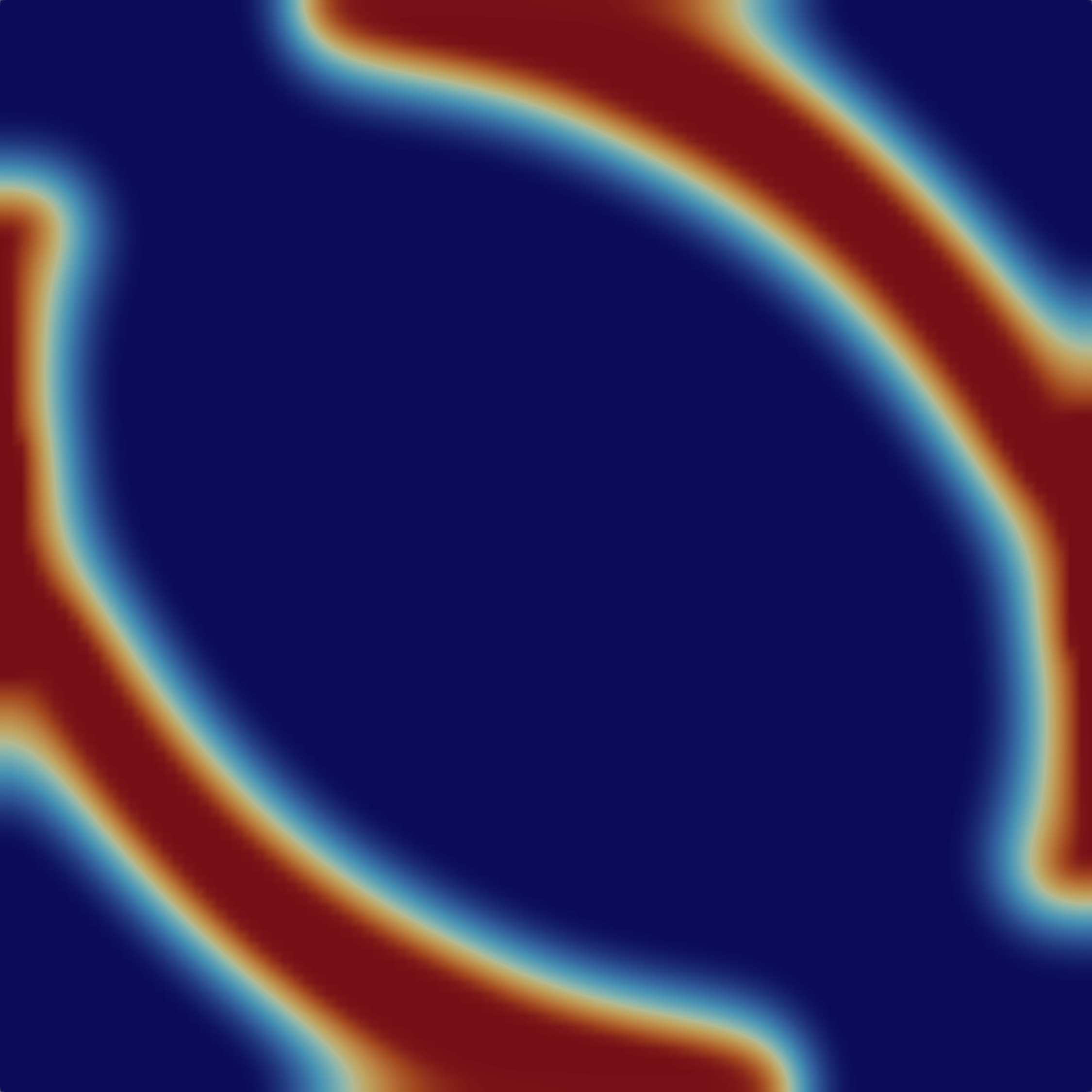}}
        &\raisebox{-.5\height}{\includegraphics[height=2.0cm]{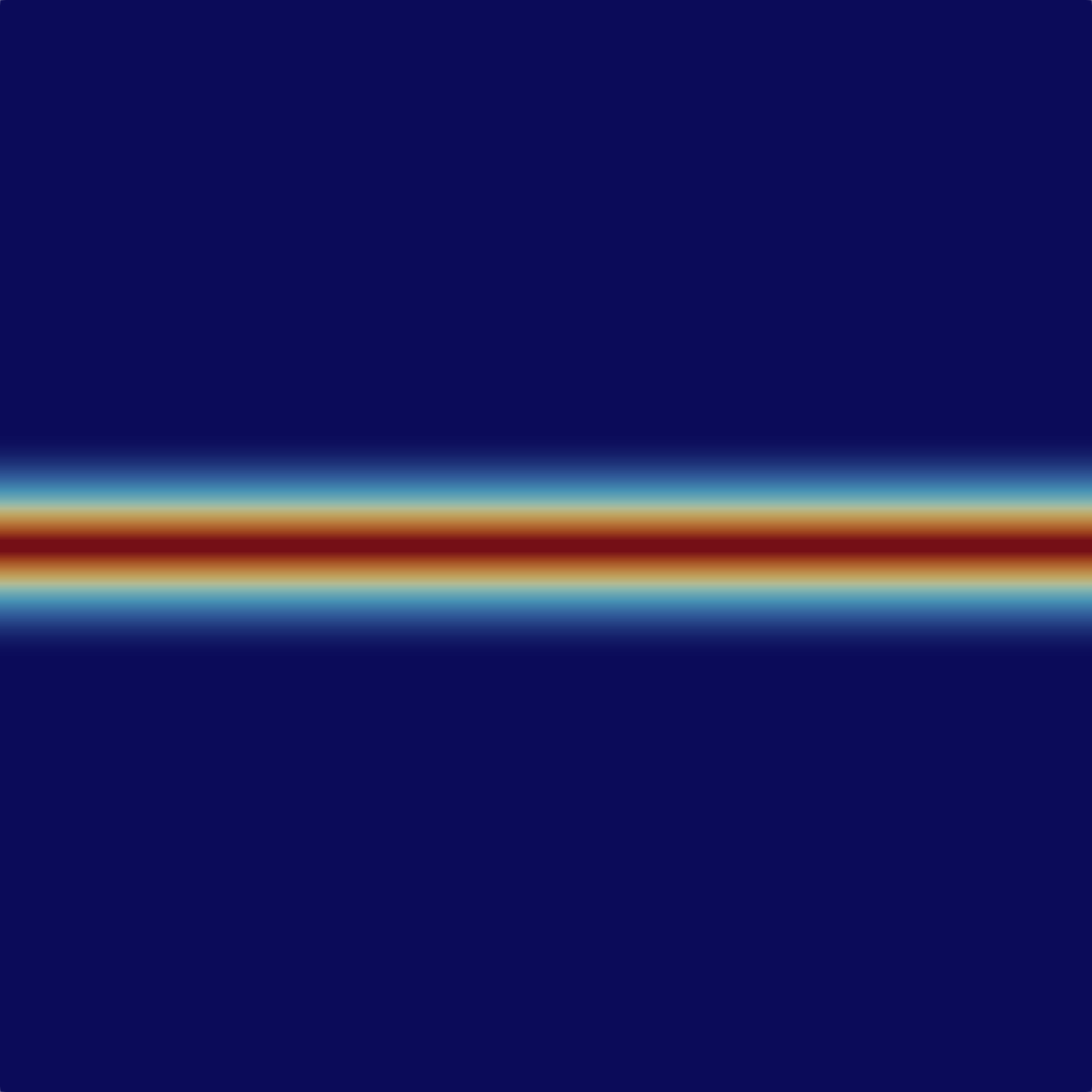}}
        &\raisebox{-.5\height}{\includegraphics[height=2.0cm]{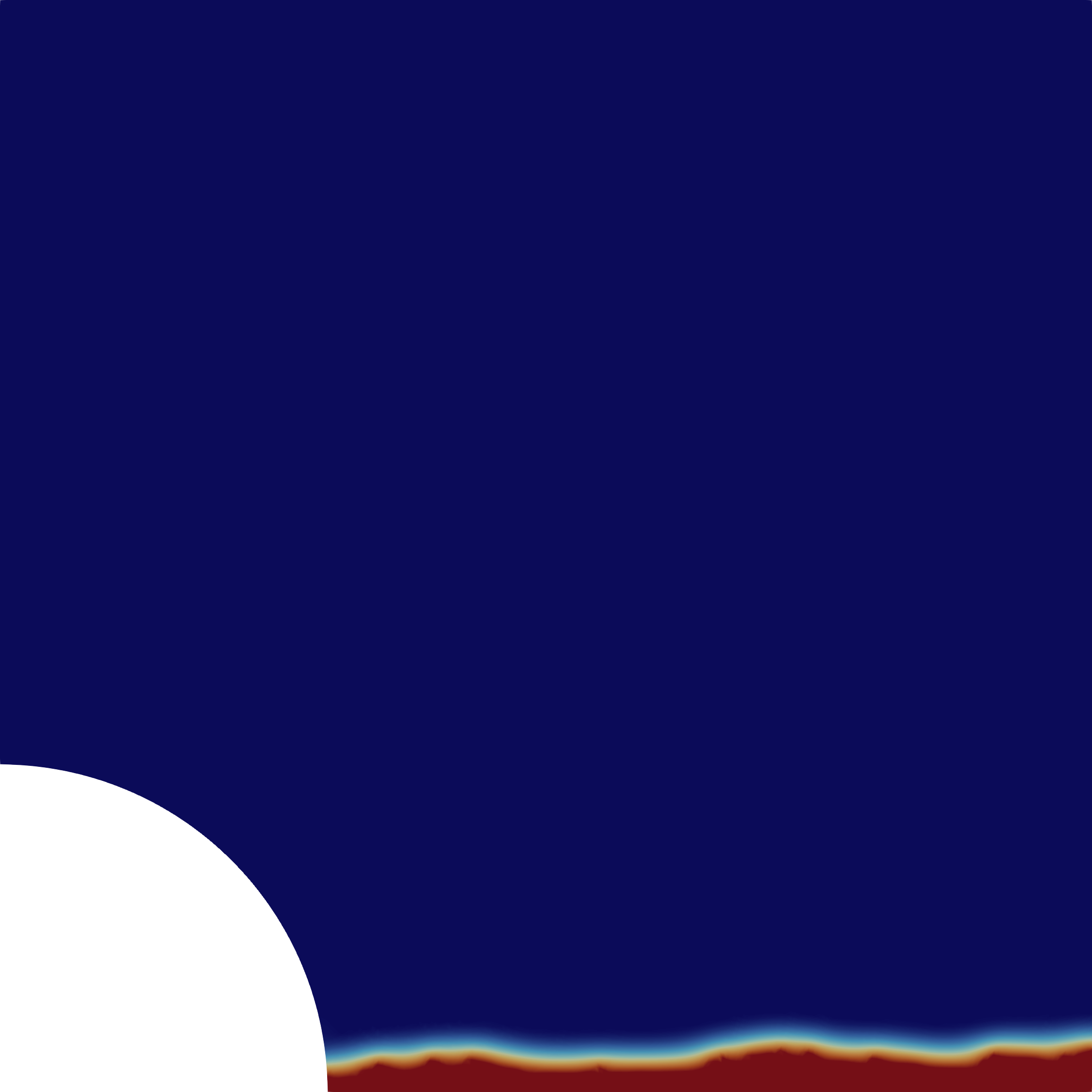}}
        &\raisebox{-.5\height}{\includegraphics[height=2.0cm]{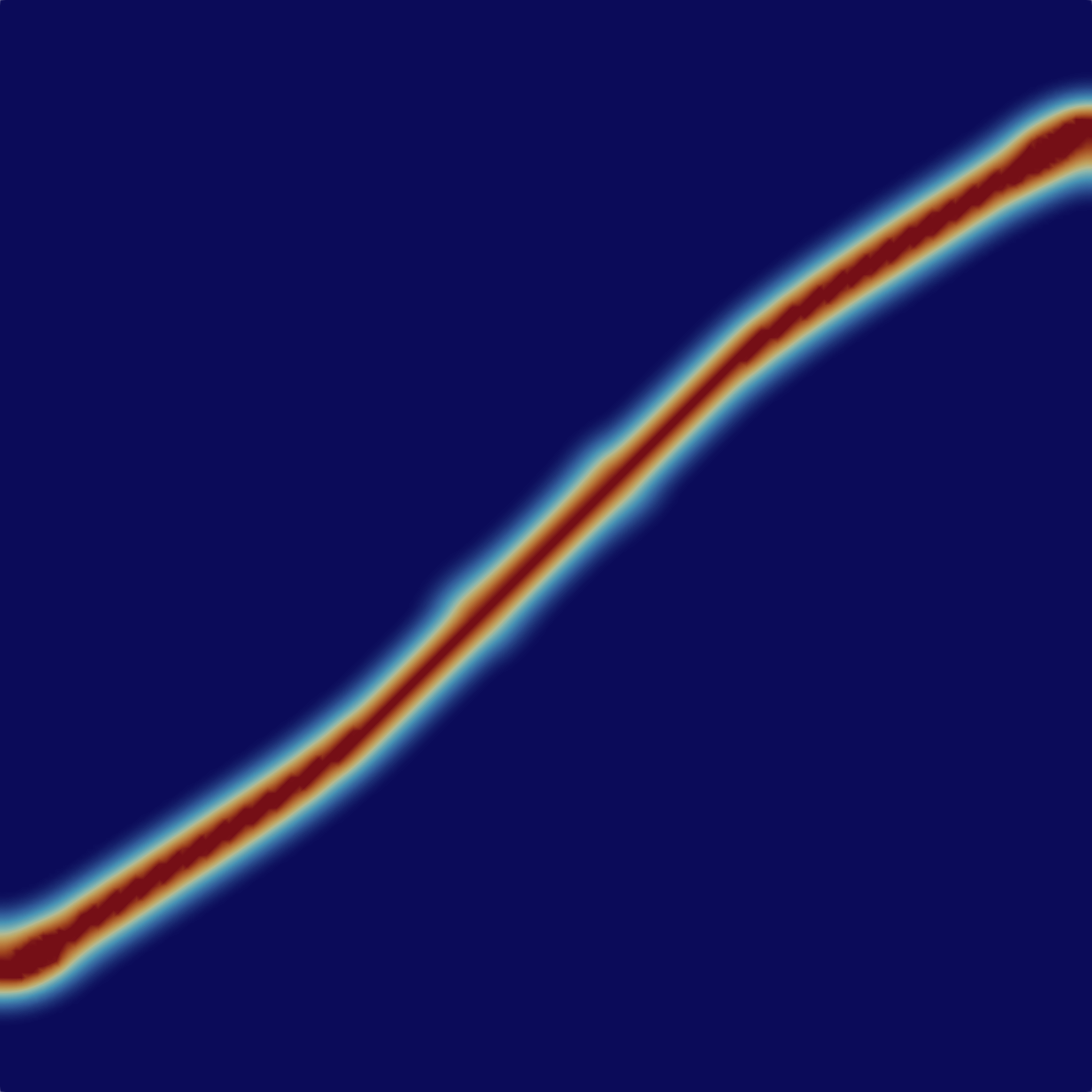}}
        &\raisebox{-.5\height}{\includegraphics[height=2.0cm]{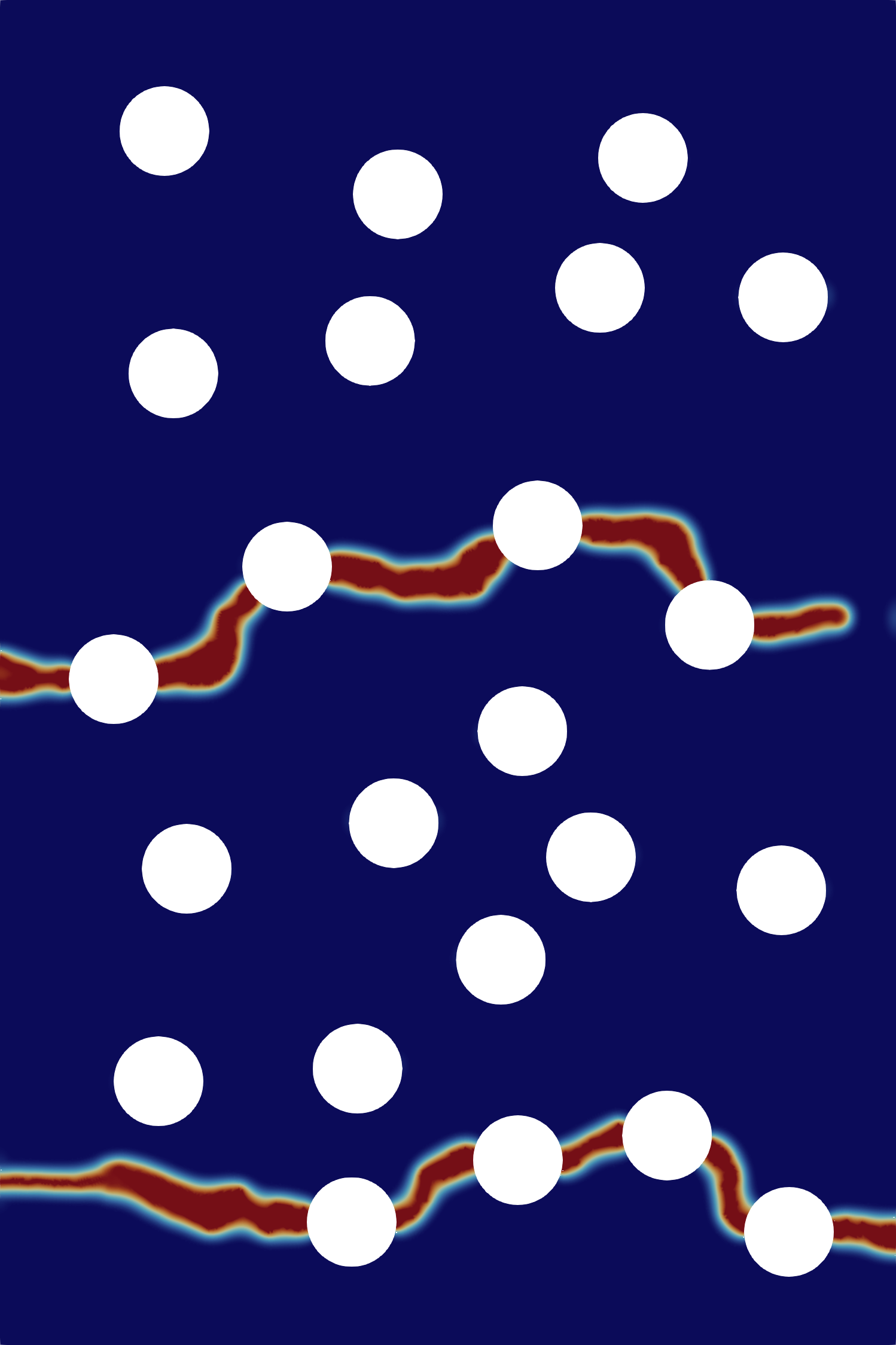}}
        \\[1.25cm]
        star-convex $\gamma^\star\!=\!1\!\!\!$
        &\raisebox{-.5\height}{\includegraphics[height=2.0cm]{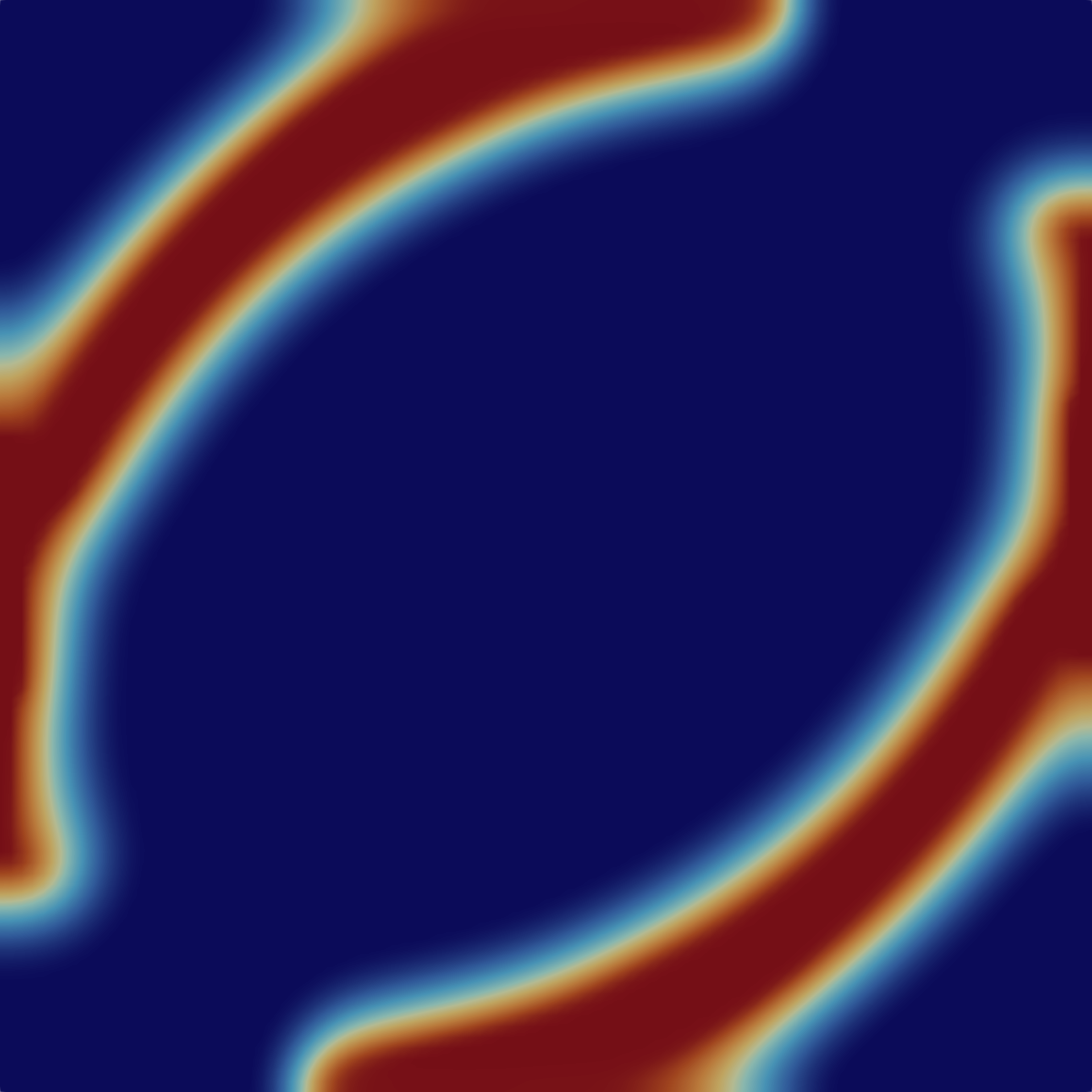}}
        &\raisebox{-.5\height}{\includegraphics[height=2.0cm]{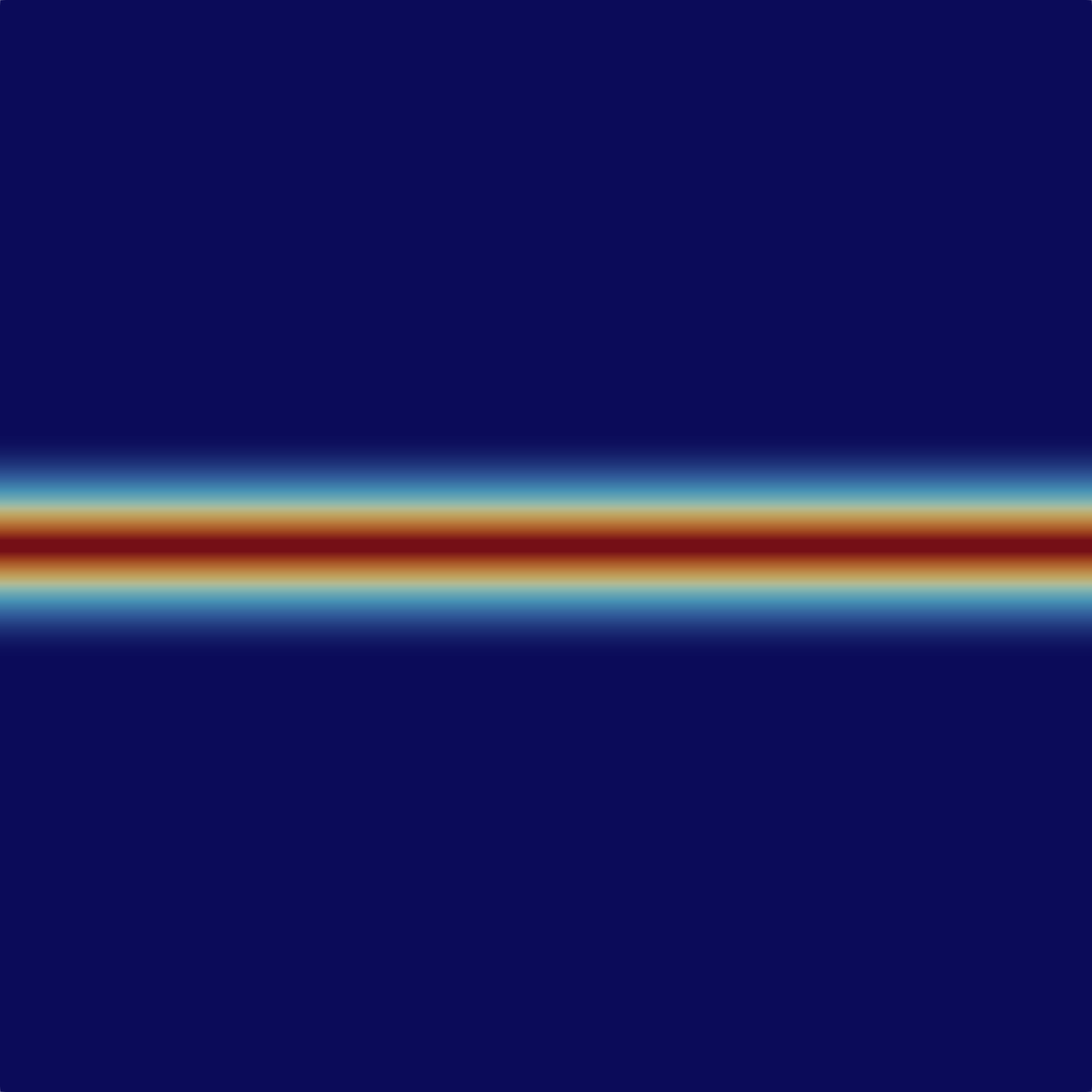}}
        &\raisebox{-.5\height}{\includegraphics[height=2.0cm]{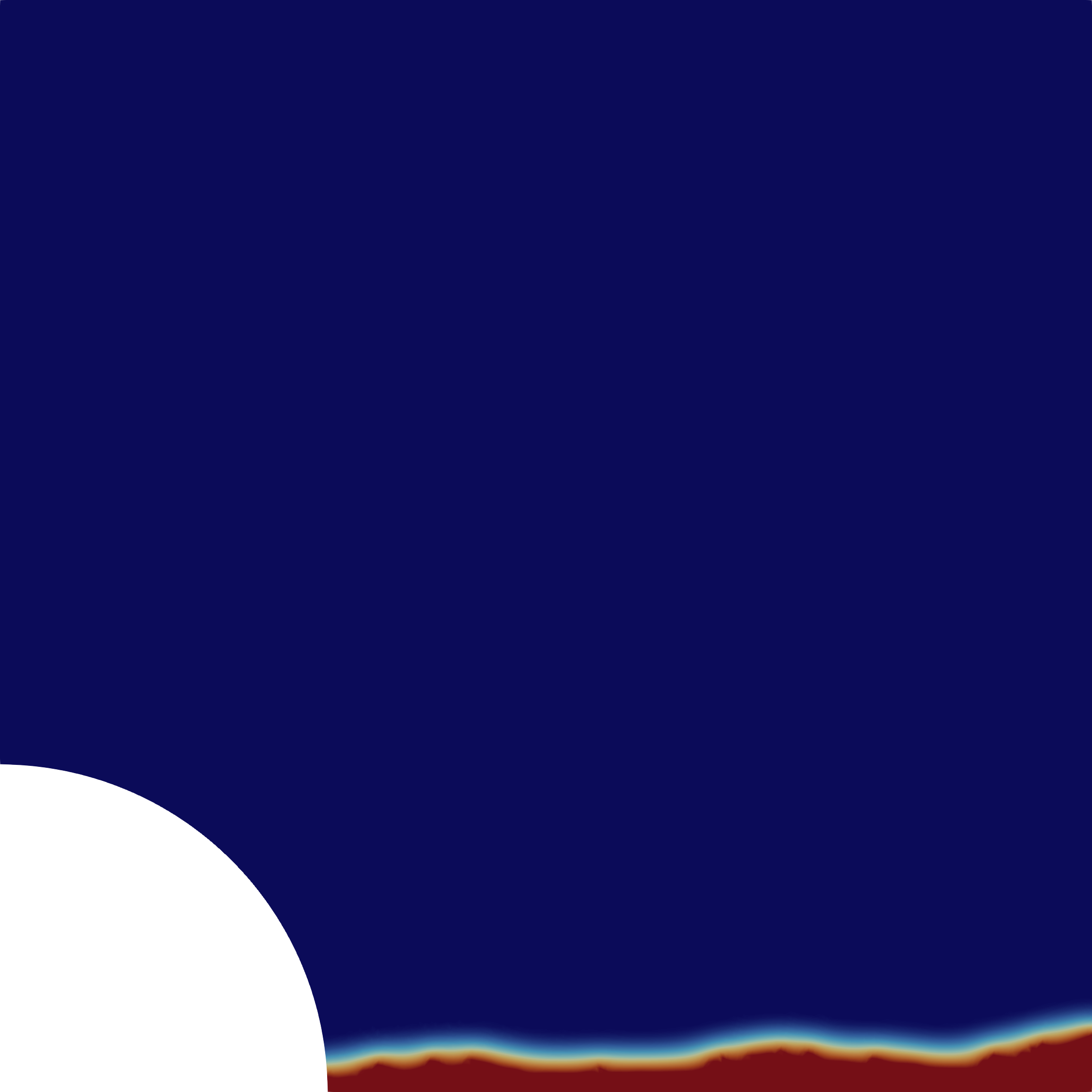}}
        &\raisebox{-.5\height}{\includegraphics[height=2.0cm]{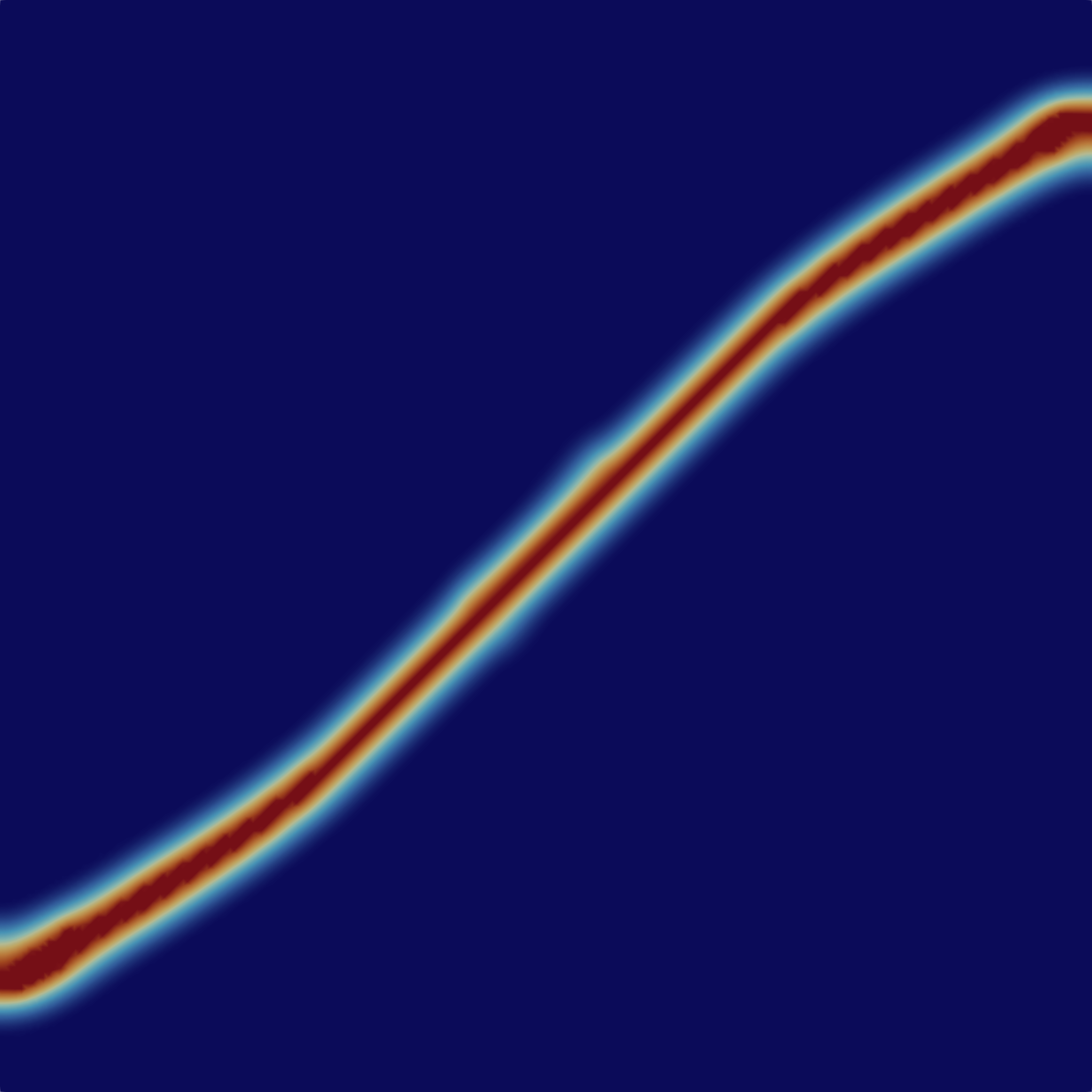}}
        &\raisebox{-.5\height}{\includegraphics[height=2.0cm]{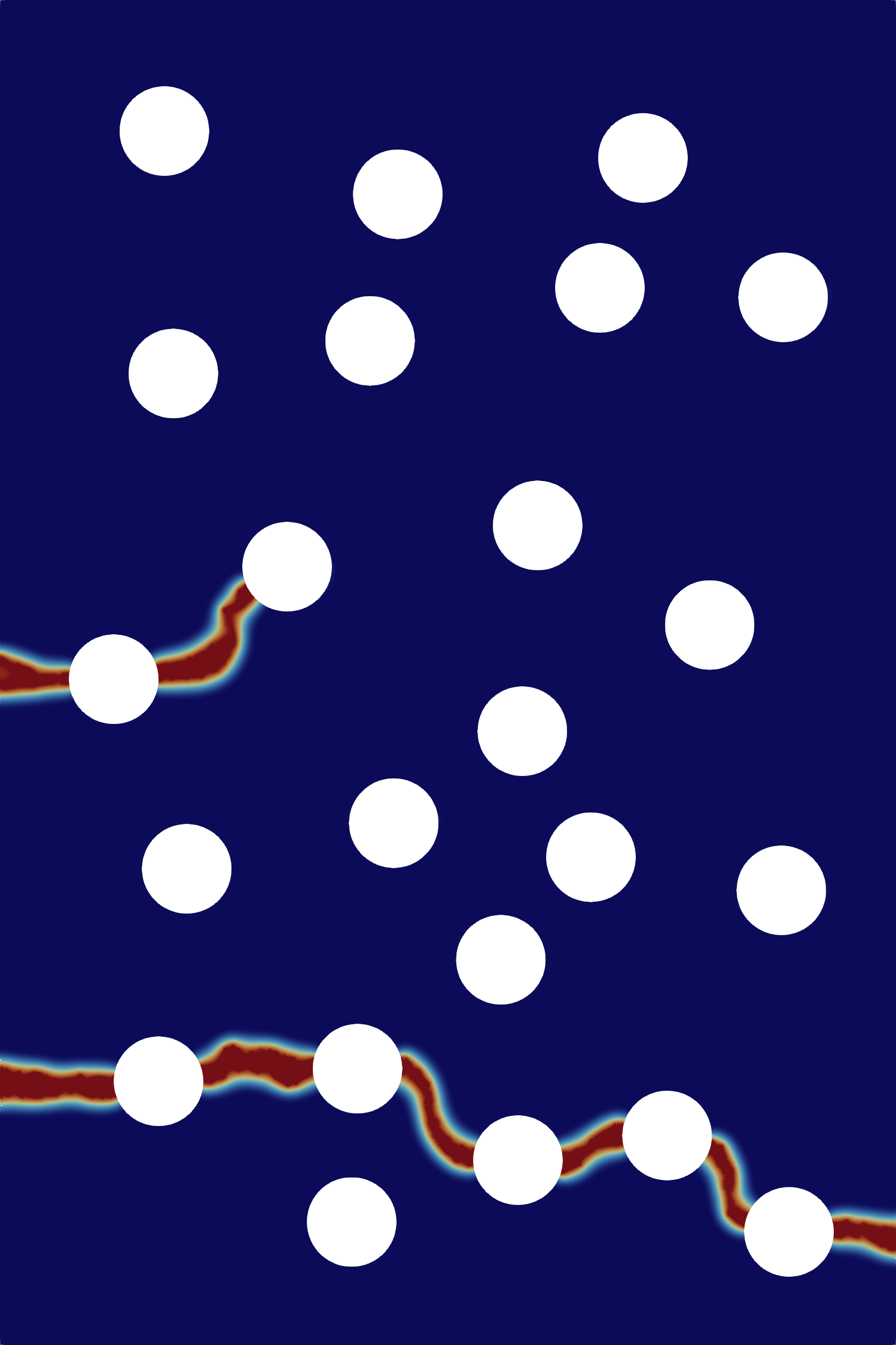}}
        \\[1.25cm]
        star-convex $\gamma^\star\!=\!5\!\!\!$
        &\raisebox{-.5\height}{\includegraphics[height=2.0cm]{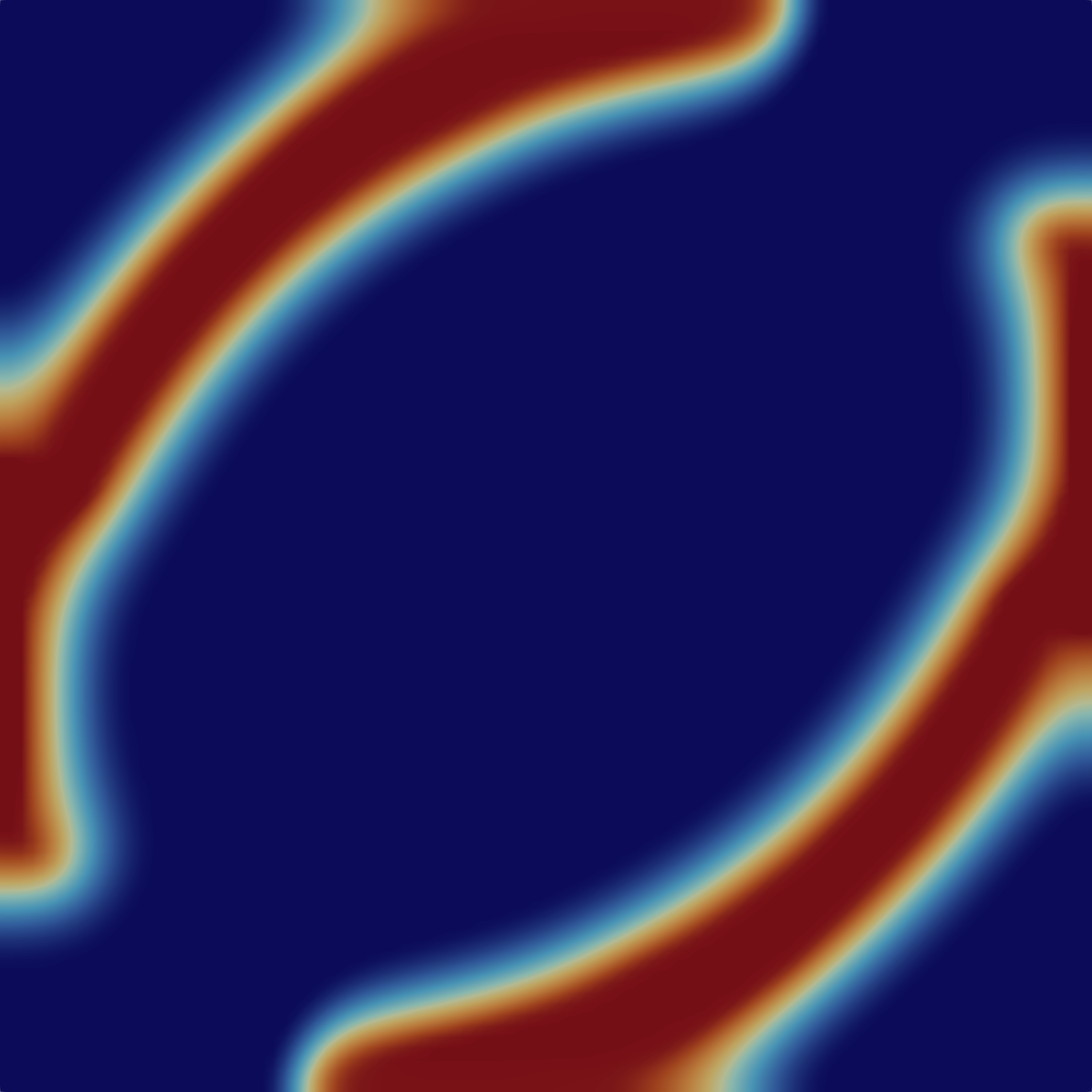}}
        &\raisebox{-.5\height}{\includegraphics[height=2.0cm]{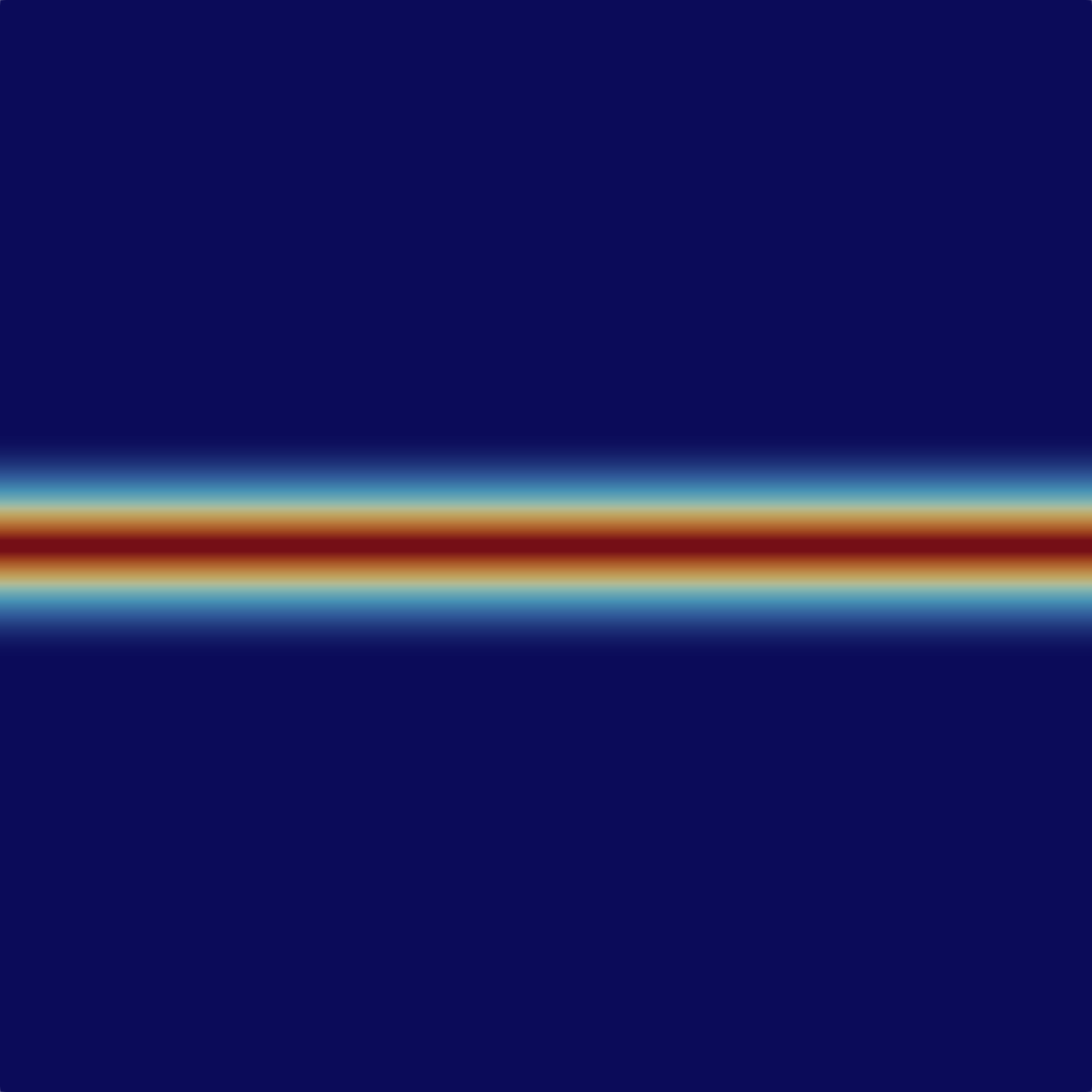}}
        &\raisebox{-.5\height}{\includegraphics[height=2.0cm]{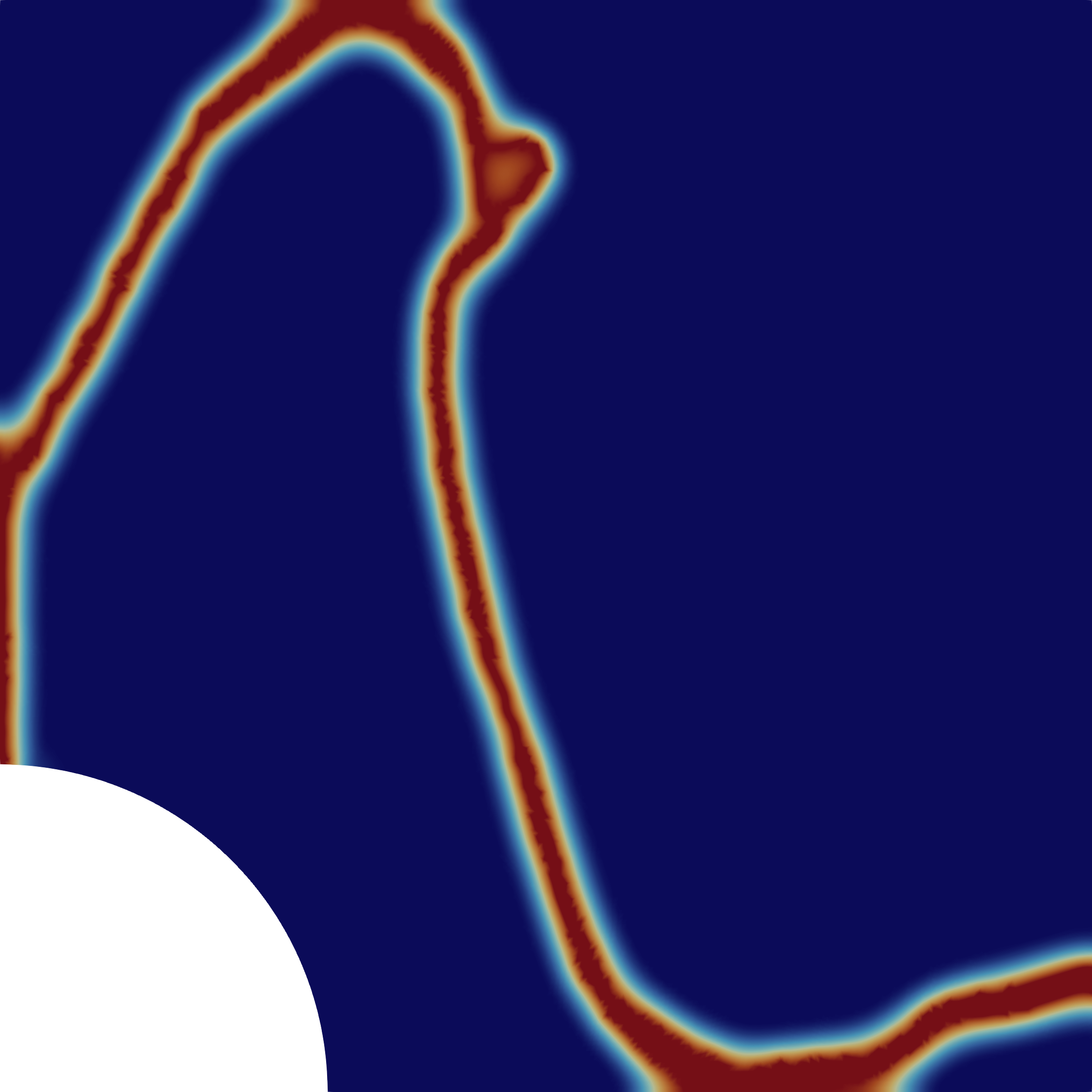}}
        &\raisebox{-.5\height}{\includegraphics[height=2.0cm]{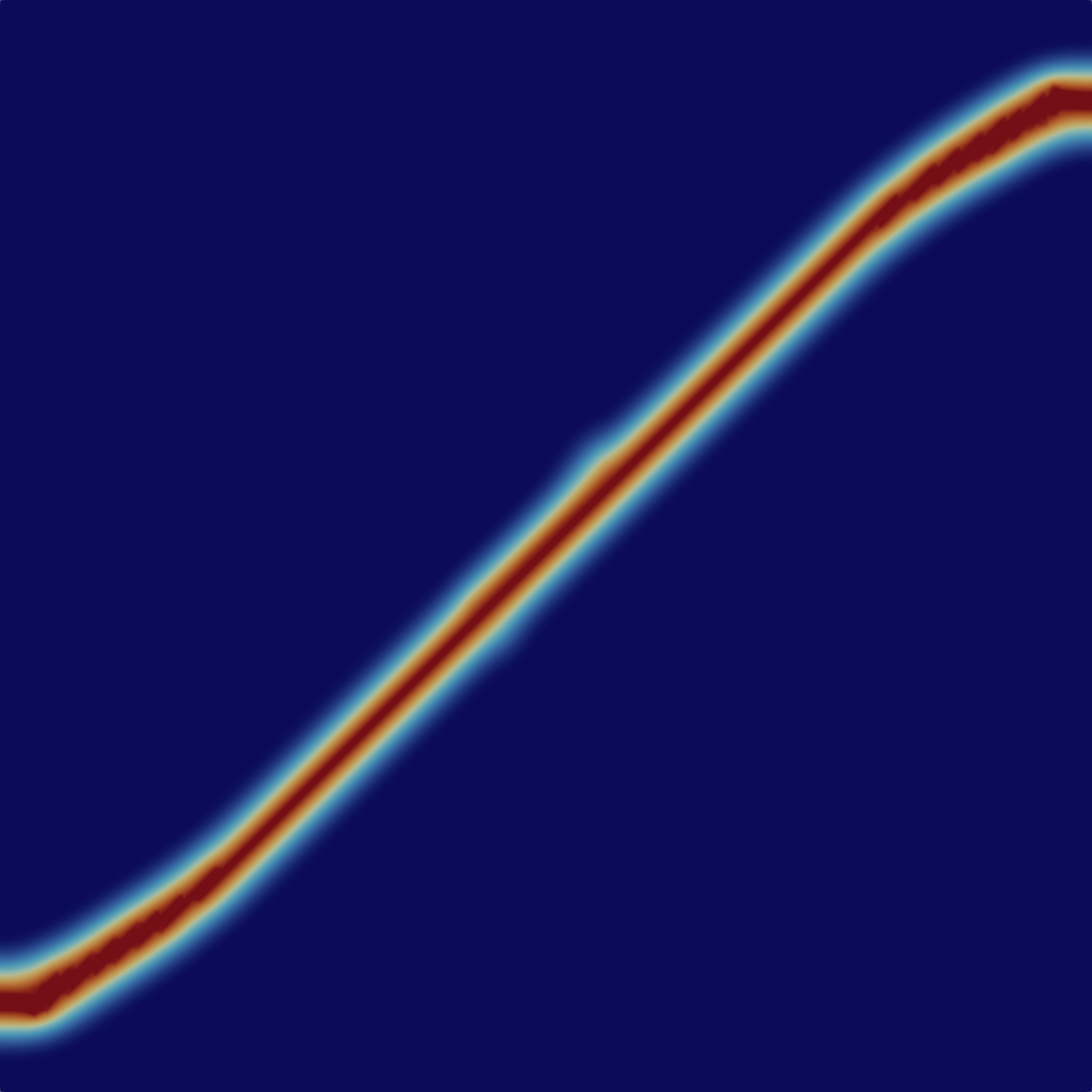}}
        &\raisebox{-.5\height}{\includegraphics[height=2.0cm]{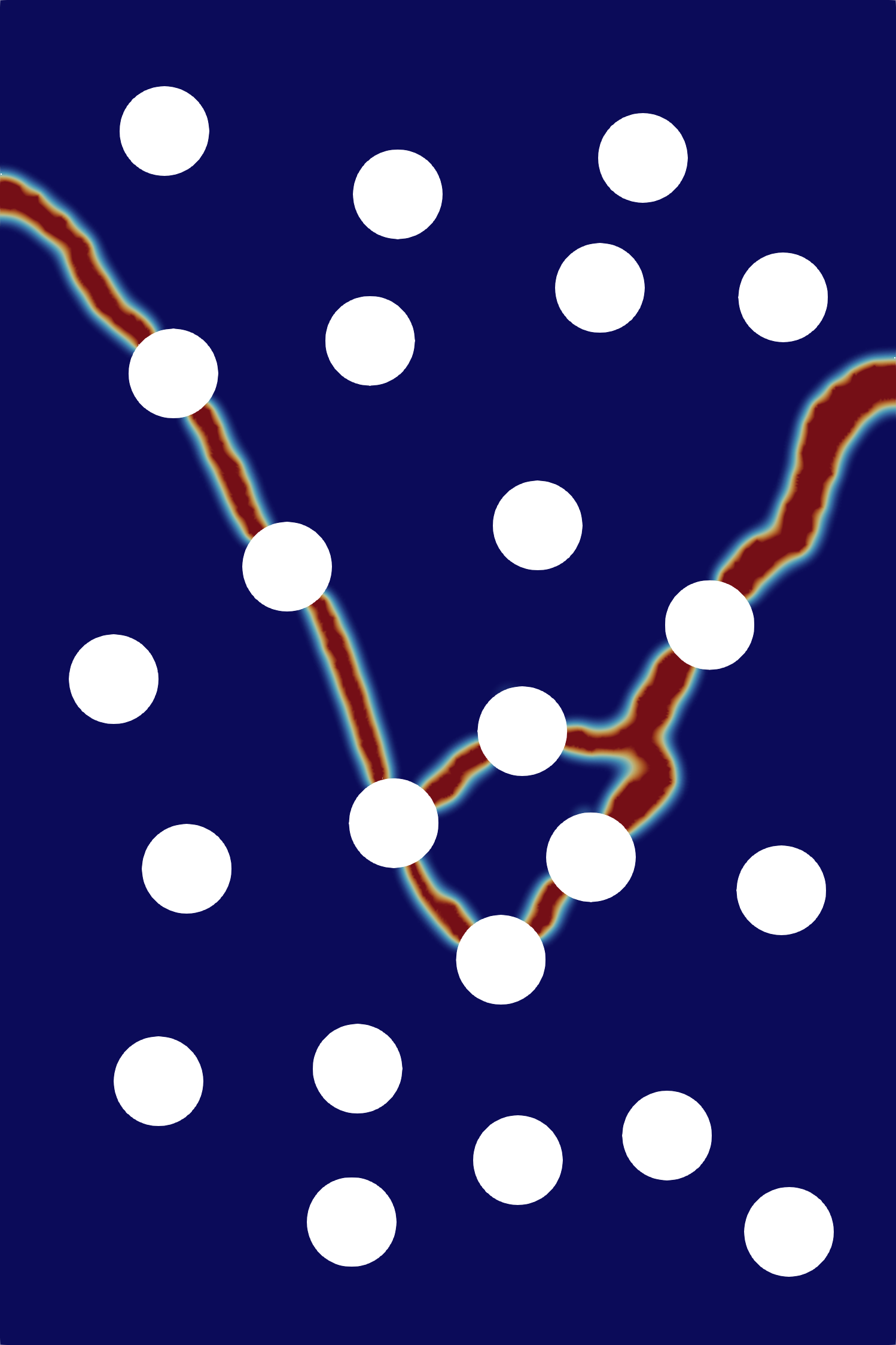}}
        \\[1.25cm]
        spectral
        &\raisebox{-.5\height}{\includegraphics[height=2.0cm]{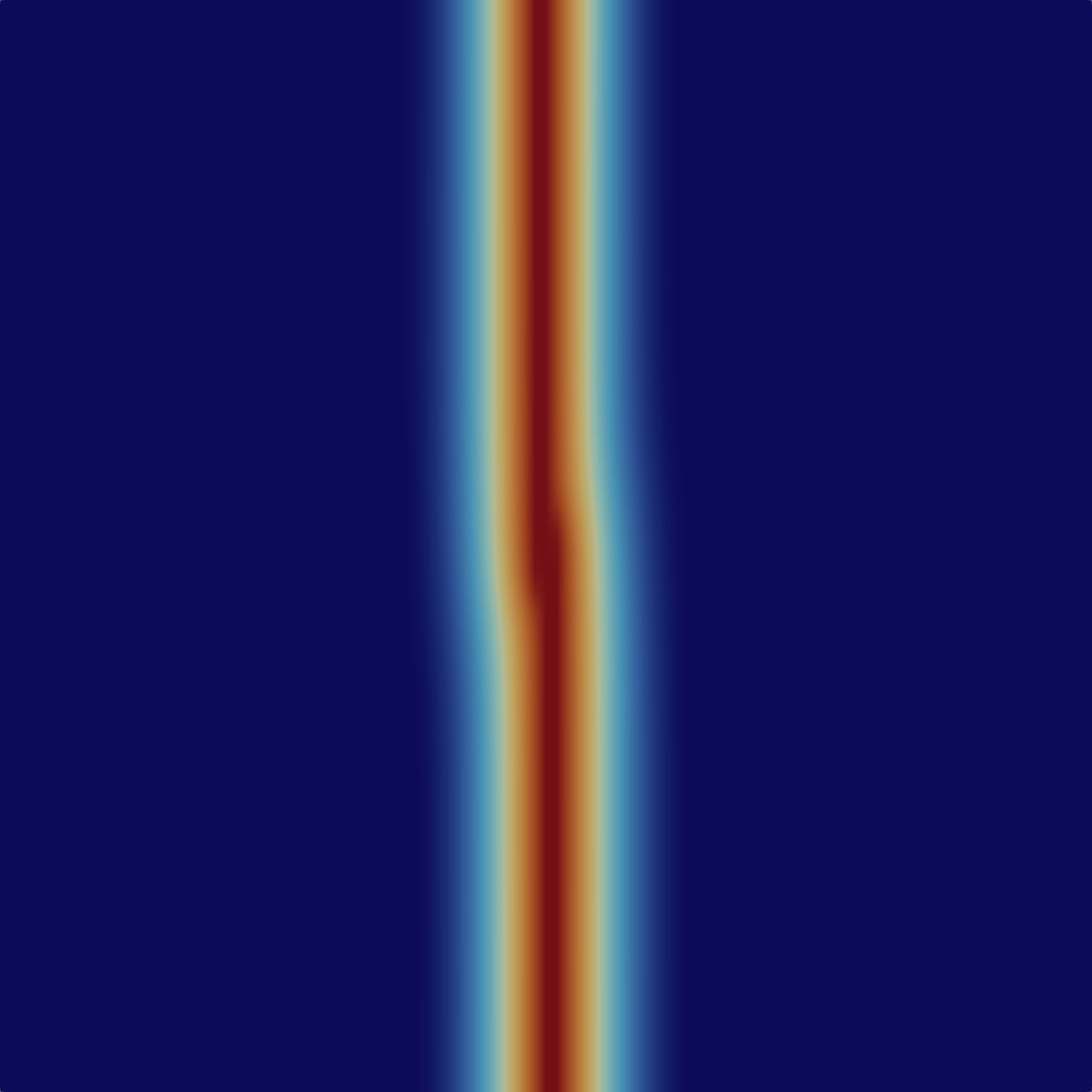}}
        &\raisebox{-.5\height}{\includegraphics[height=2.0cm]{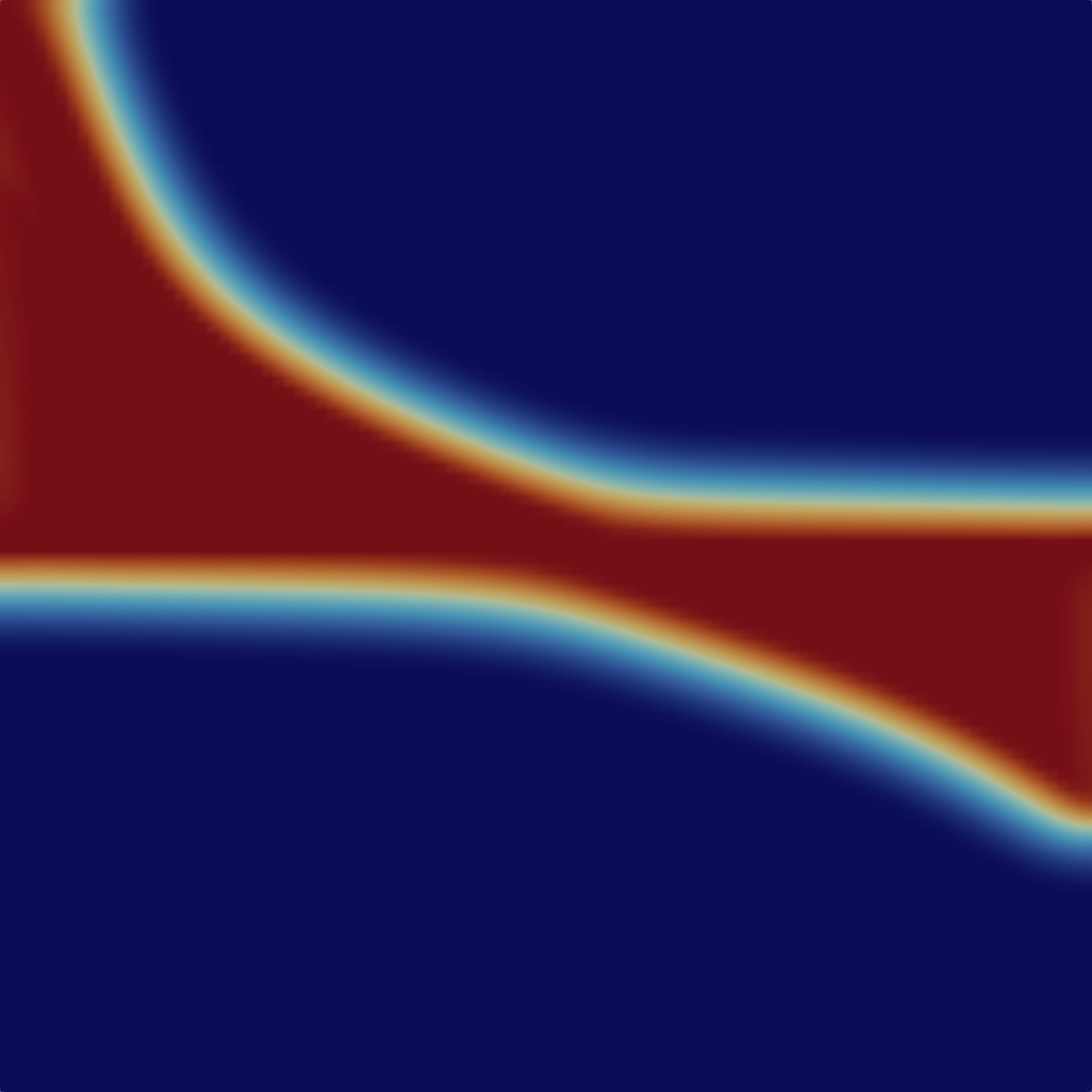}}
        &\raisebox{-.5\height}{\includegraphics[height=2.0cm]{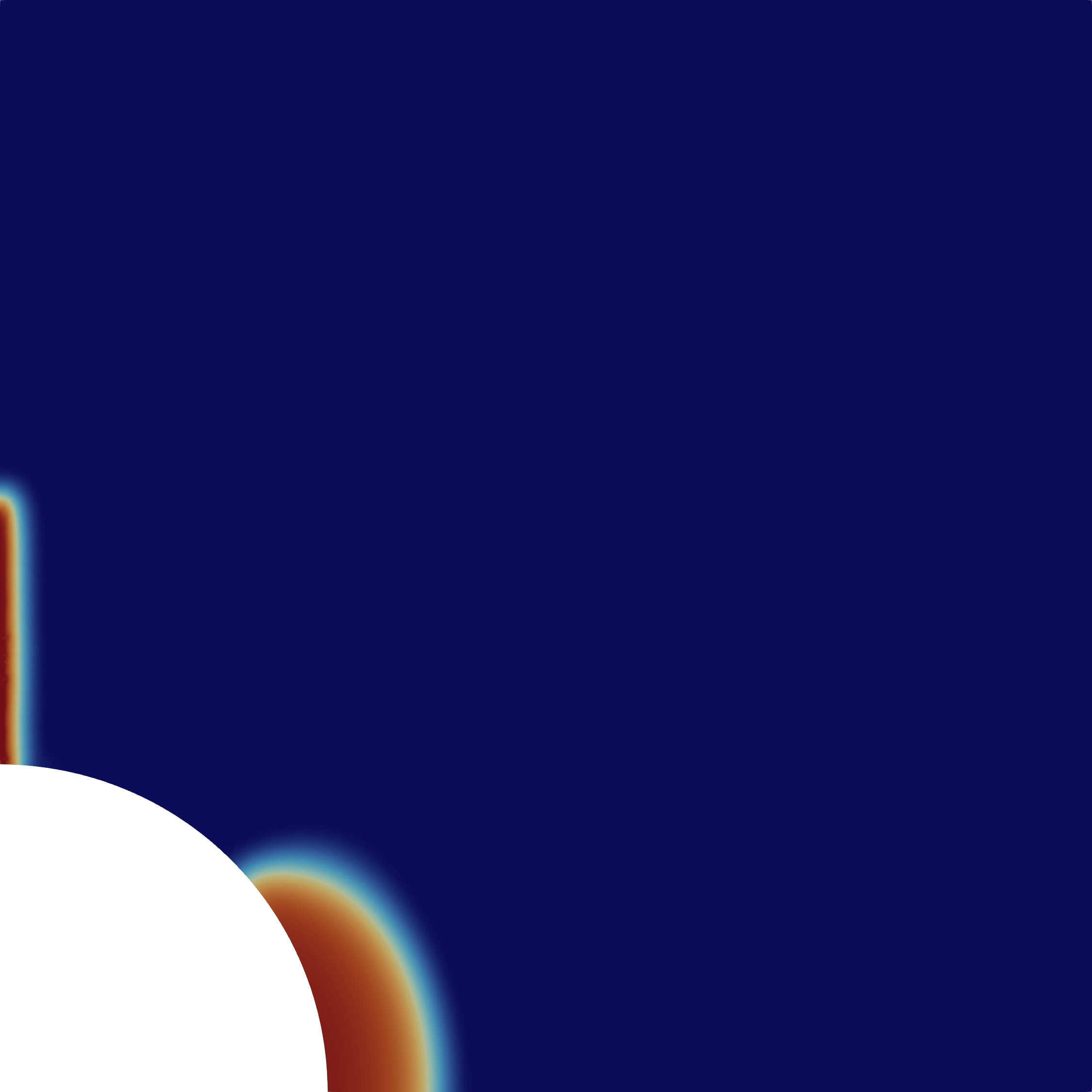}}
        &\raisebox{-.5\height}{\includegraphics[height=2.0cm]{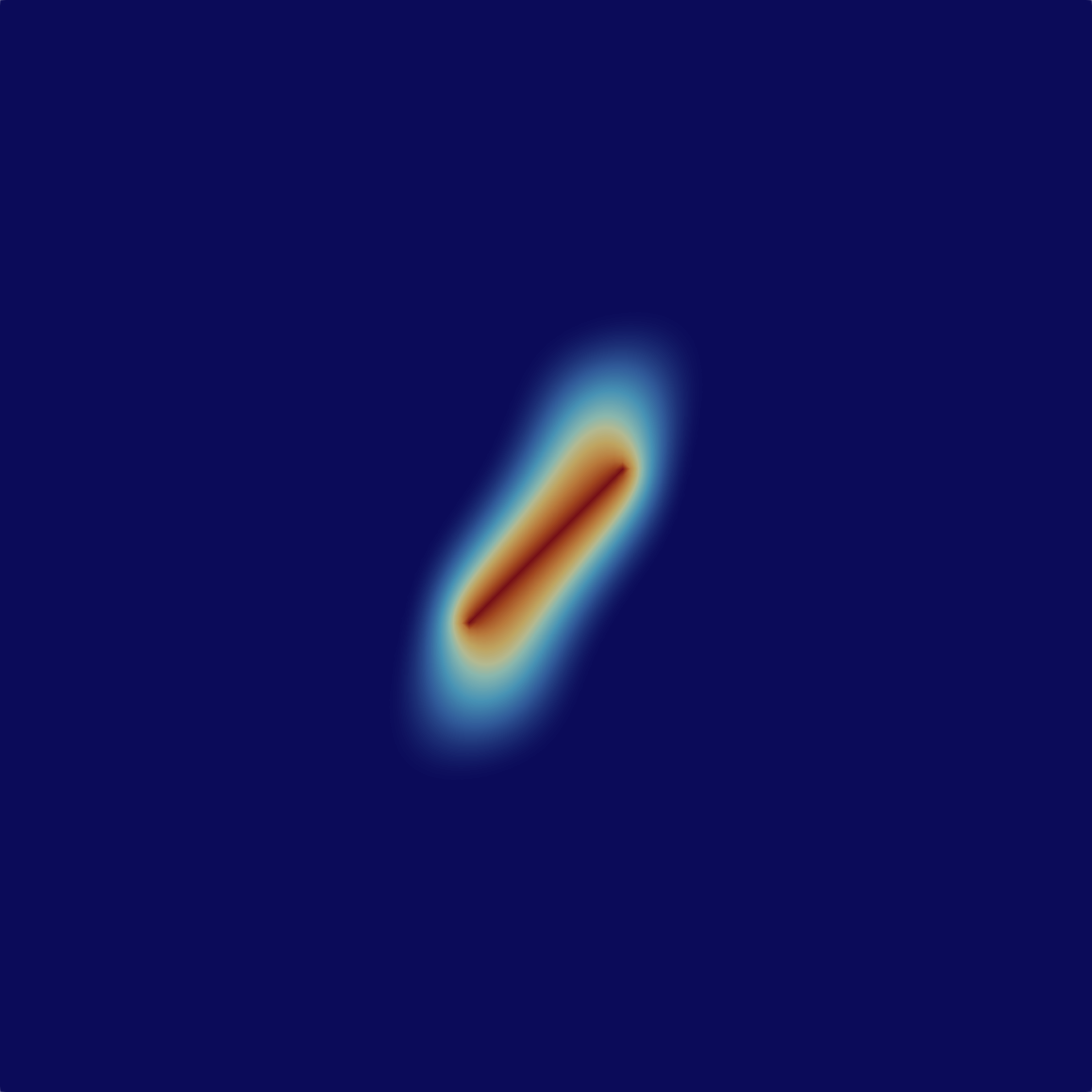}}
        &\raisebox{-.5\height}{\includegraphics[height=2.0cm]{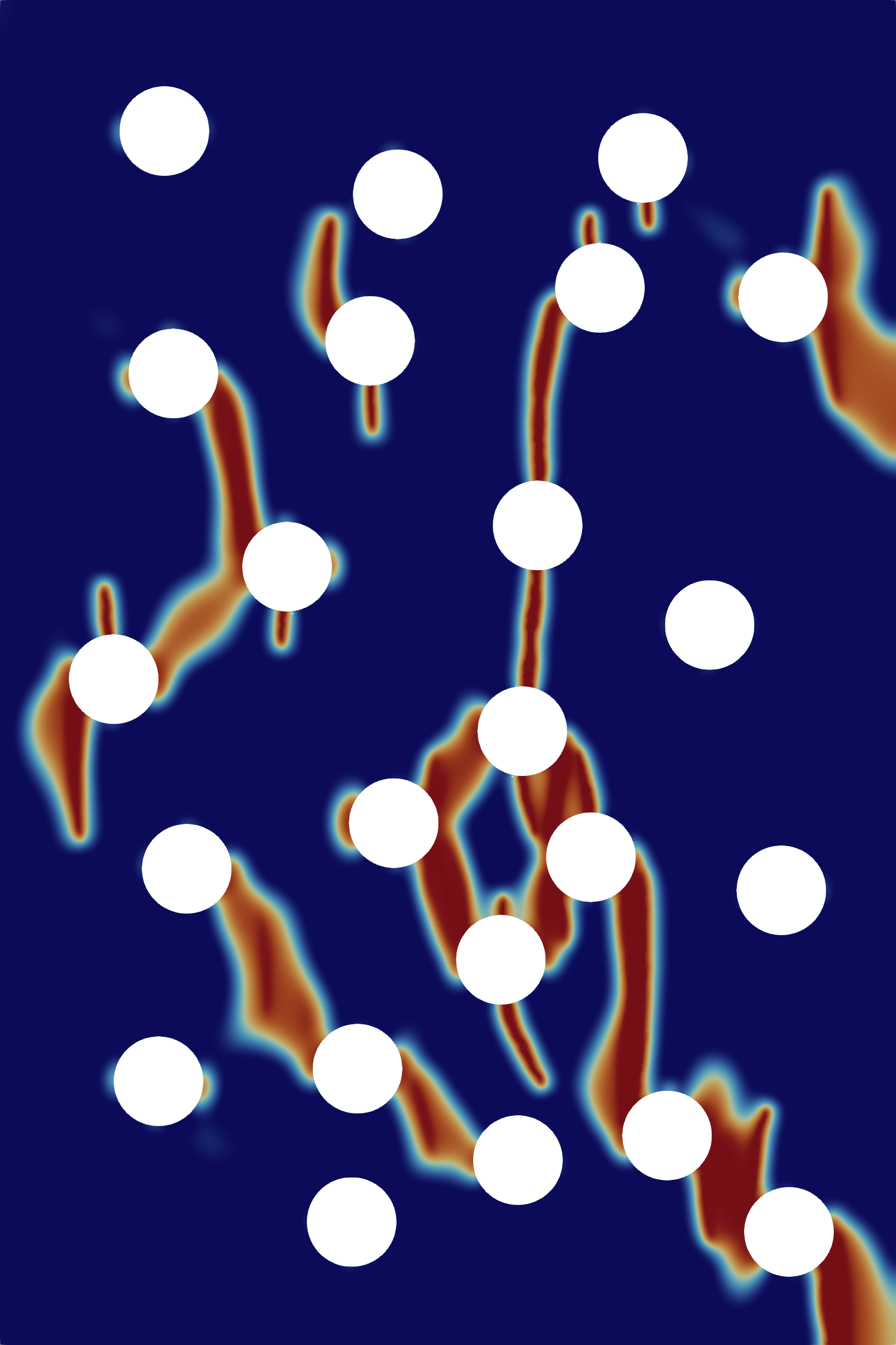}}
        \\[1.25cm]
        no-tension
        &\raisebox{-.5\height}{\includegraphics[height=2.0cm]{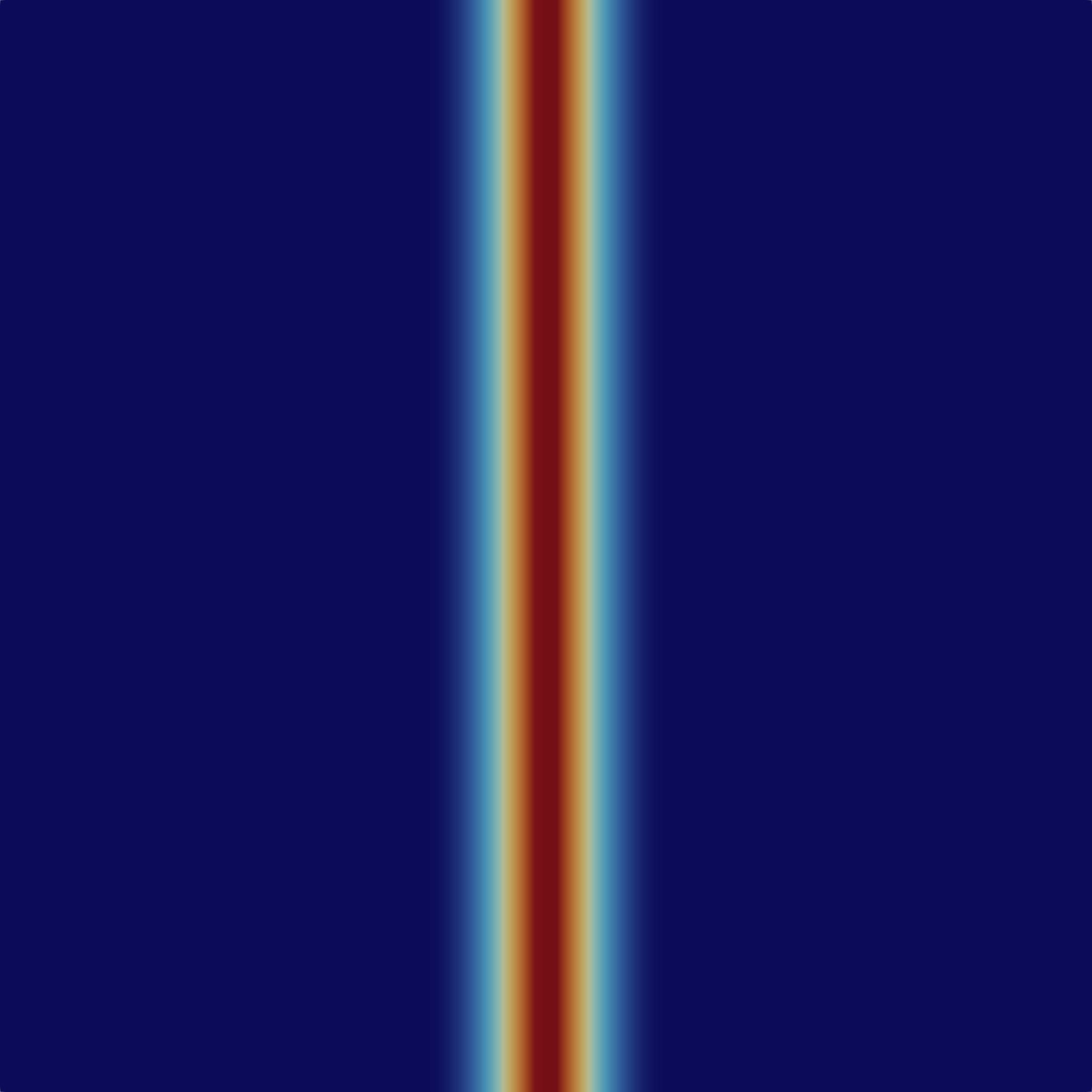}}
        &\raisebox{-.5\height}{\includegraphics[height=2.0cm]{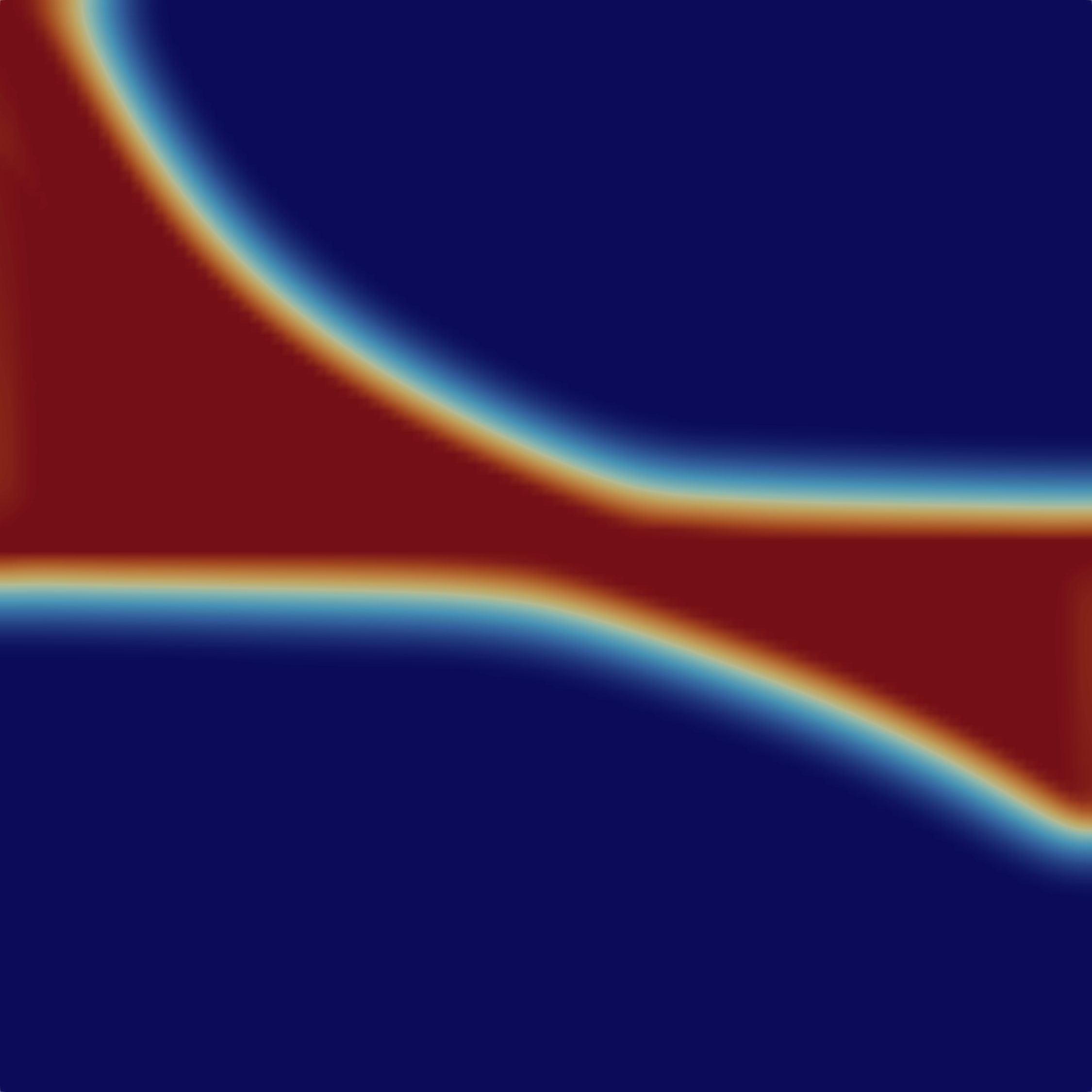}}
        &\raisebox{-.5\height}{\includegraphics[height=2.0cm]{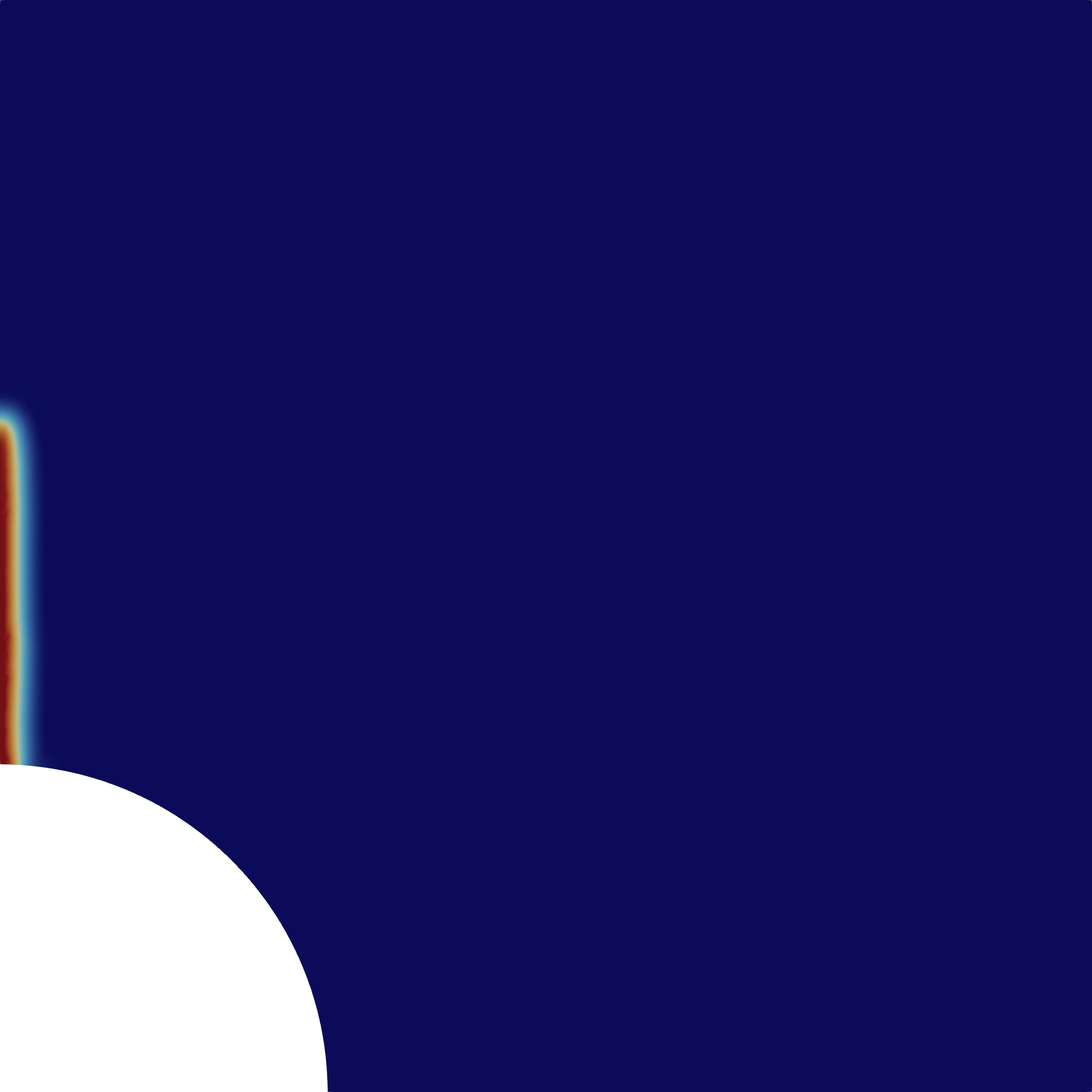}}
        &\raisebox{-.5\height}{\includegraphics[height=2.0cm]{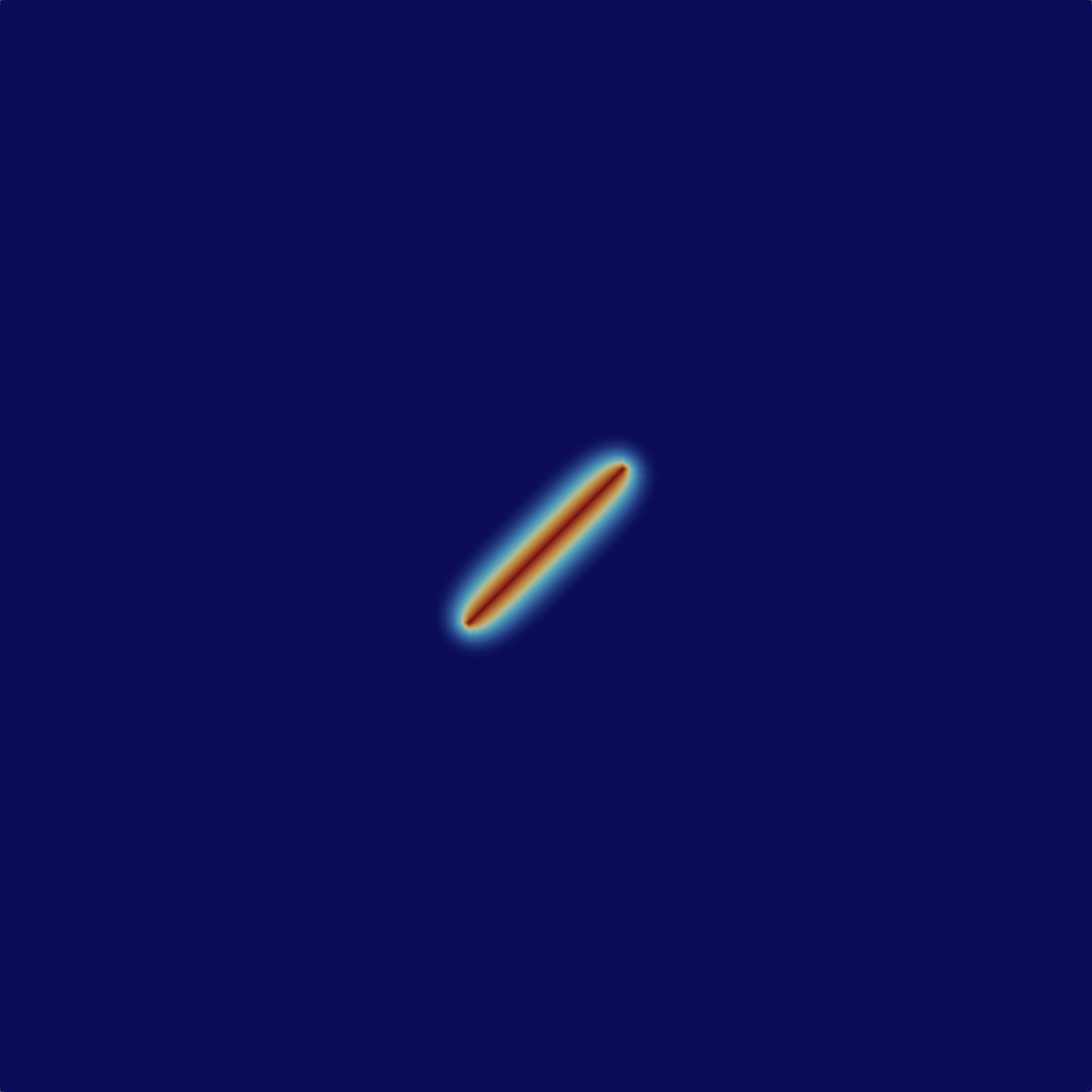}}
        &\raisebox{-.5\height}{\includegraphics[height=2.0cm]{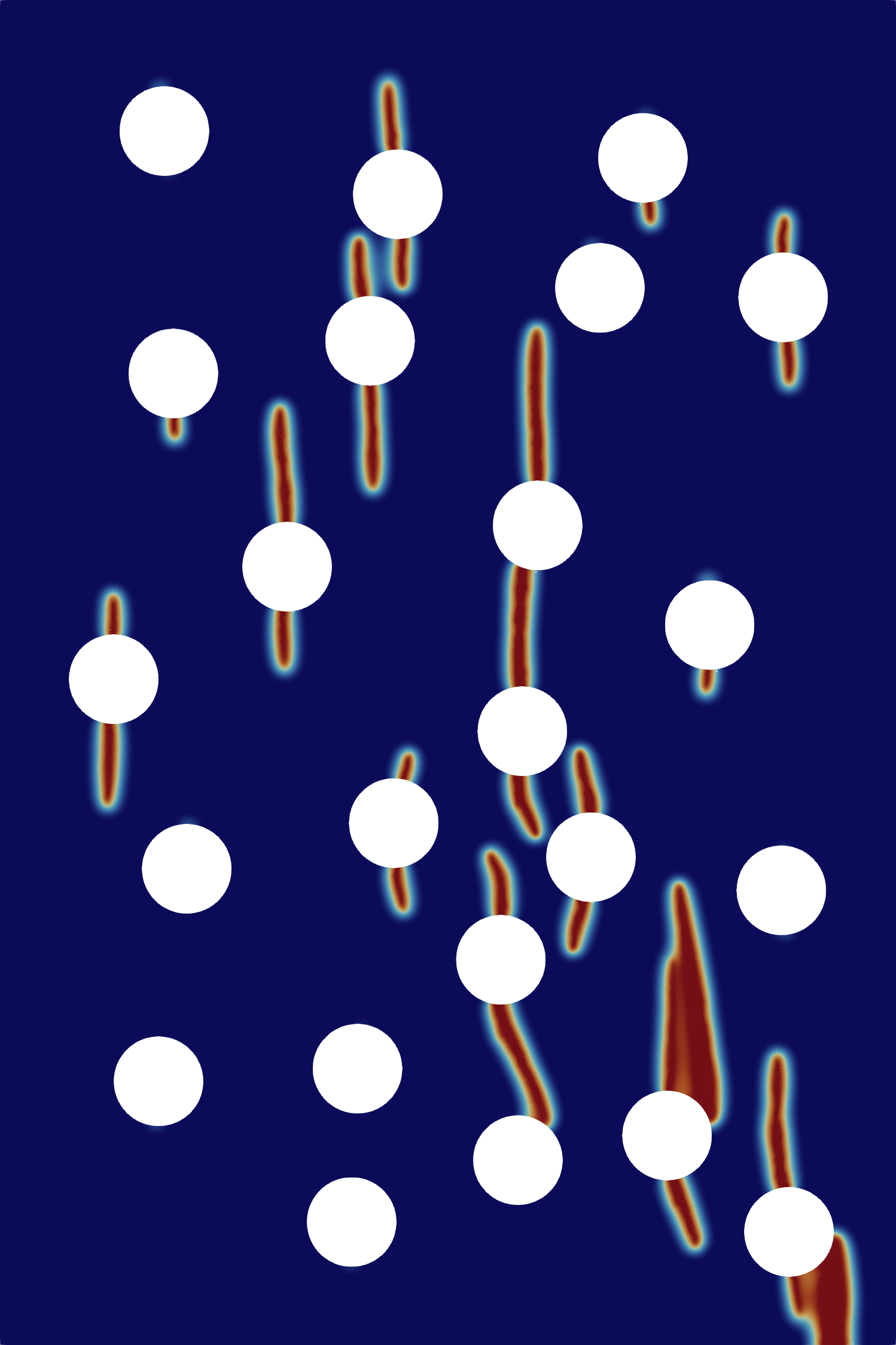}}
        \\[1.25cm]
        DP-like $\gamma\!=\!\sqrt{\!\tfrac{2\mu_0}{\kappa_0}\!}\!\!\!\!\!\!\!\!$
        &\raisebox{-.5\height}{\includegraphics[height=2.0cm]{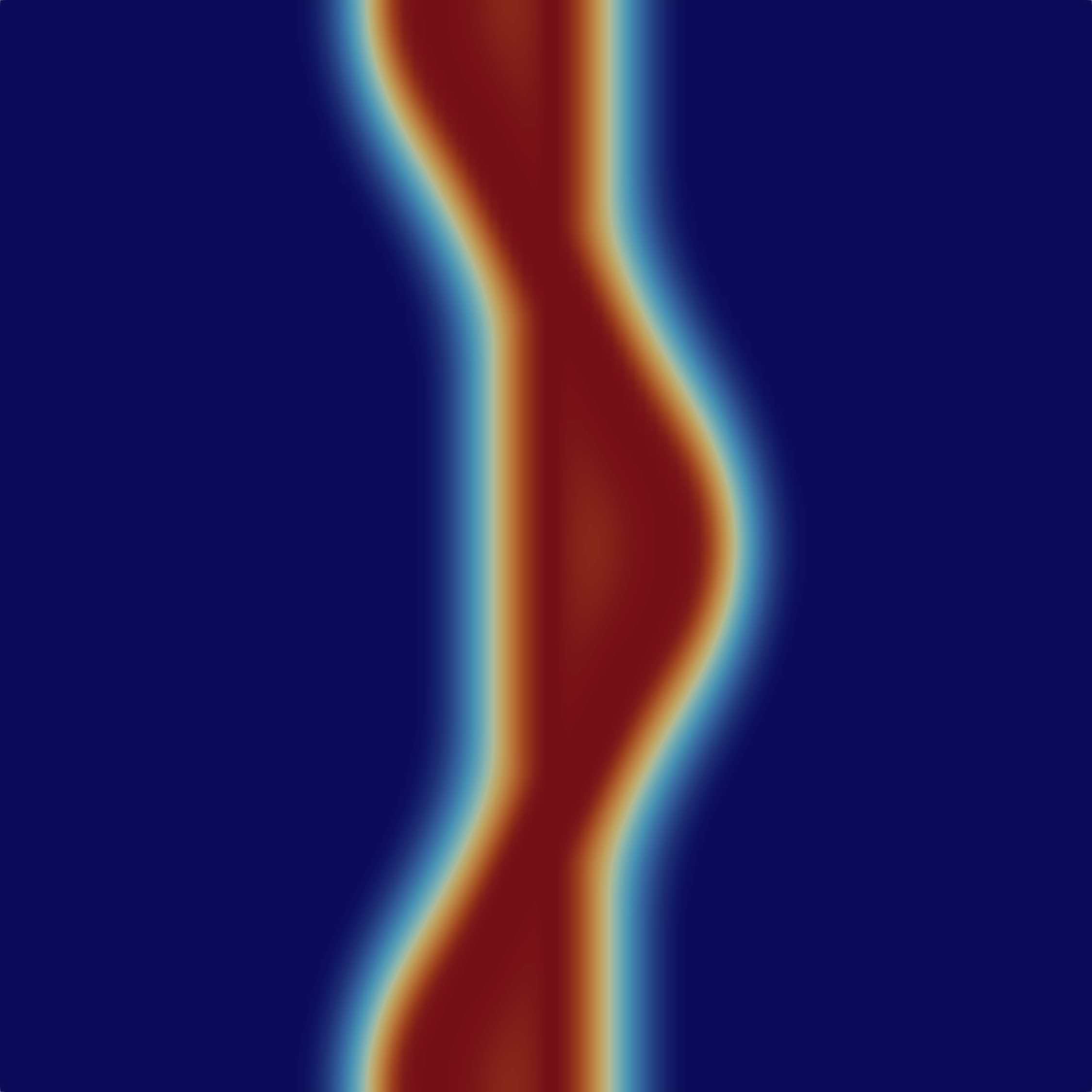}}
        &\raisebox{-.5\height}{\includegraphics[height=2.0cm]{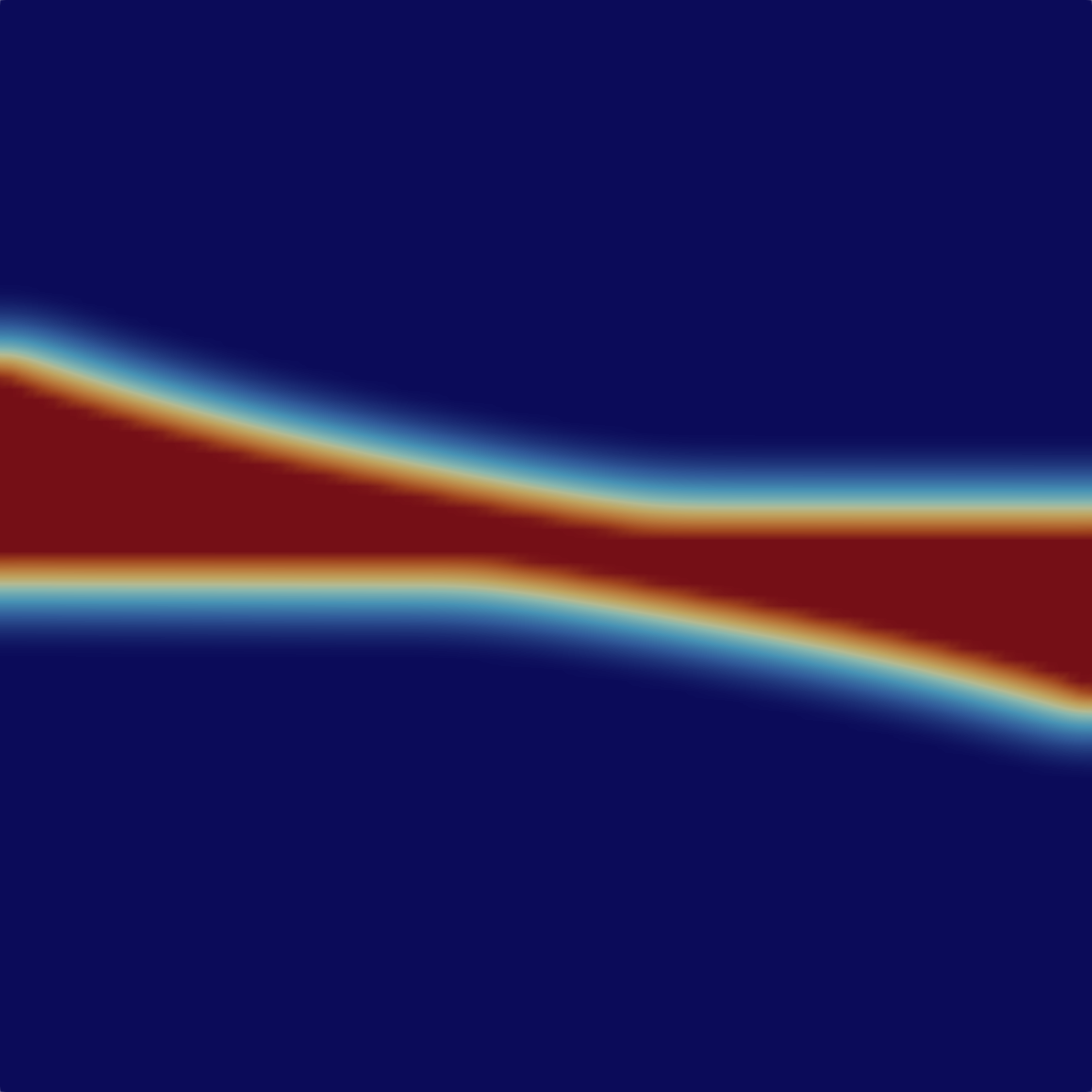}}
        &\raisebox{-.5\height}{\includegraphics[height=2.0cm]{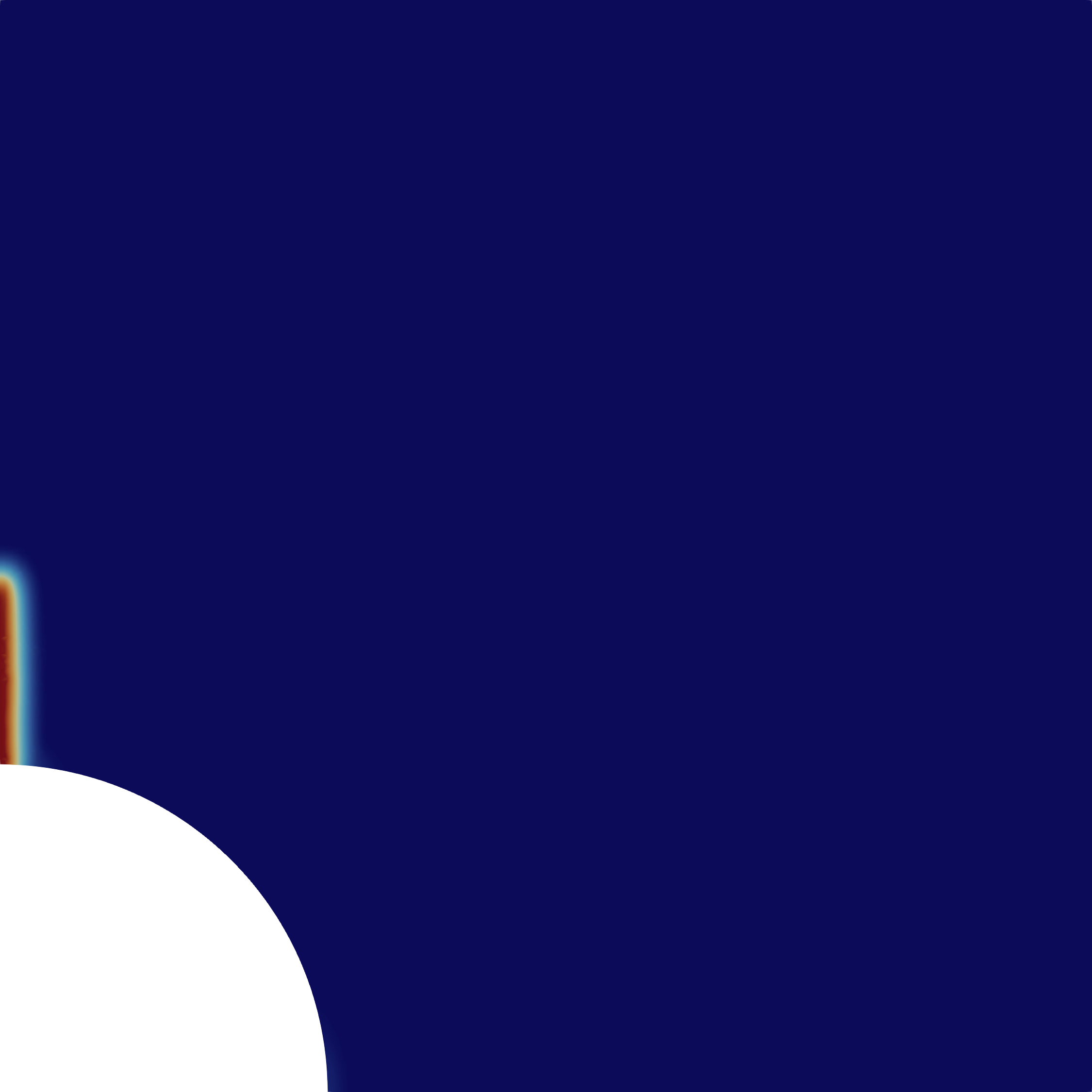}}
        &\raisebox{-.5\height}{\includegraphics[height=2.0cm]{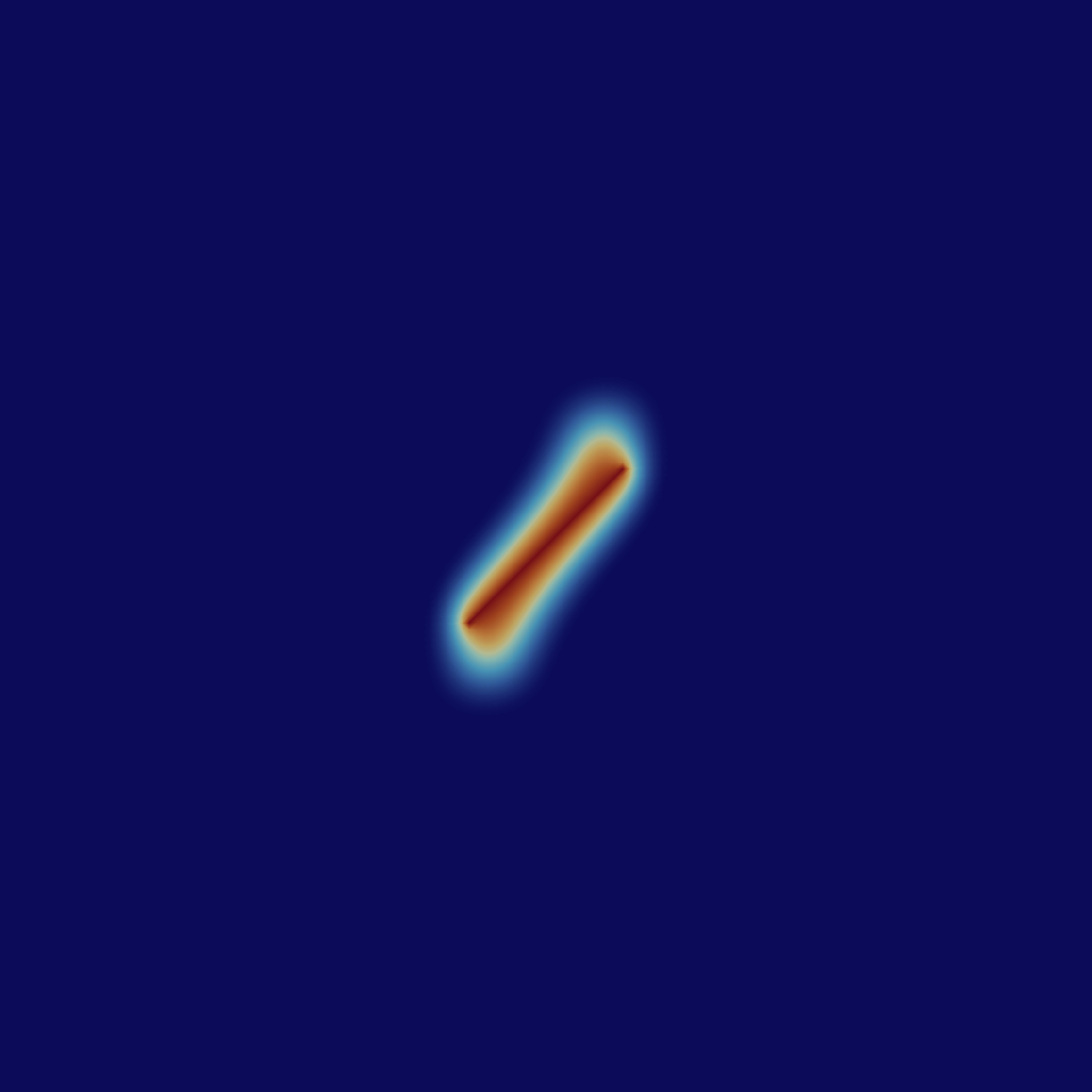}}
        &\raisebox{-.5\height}{\includegraphics[height=2.0cm]{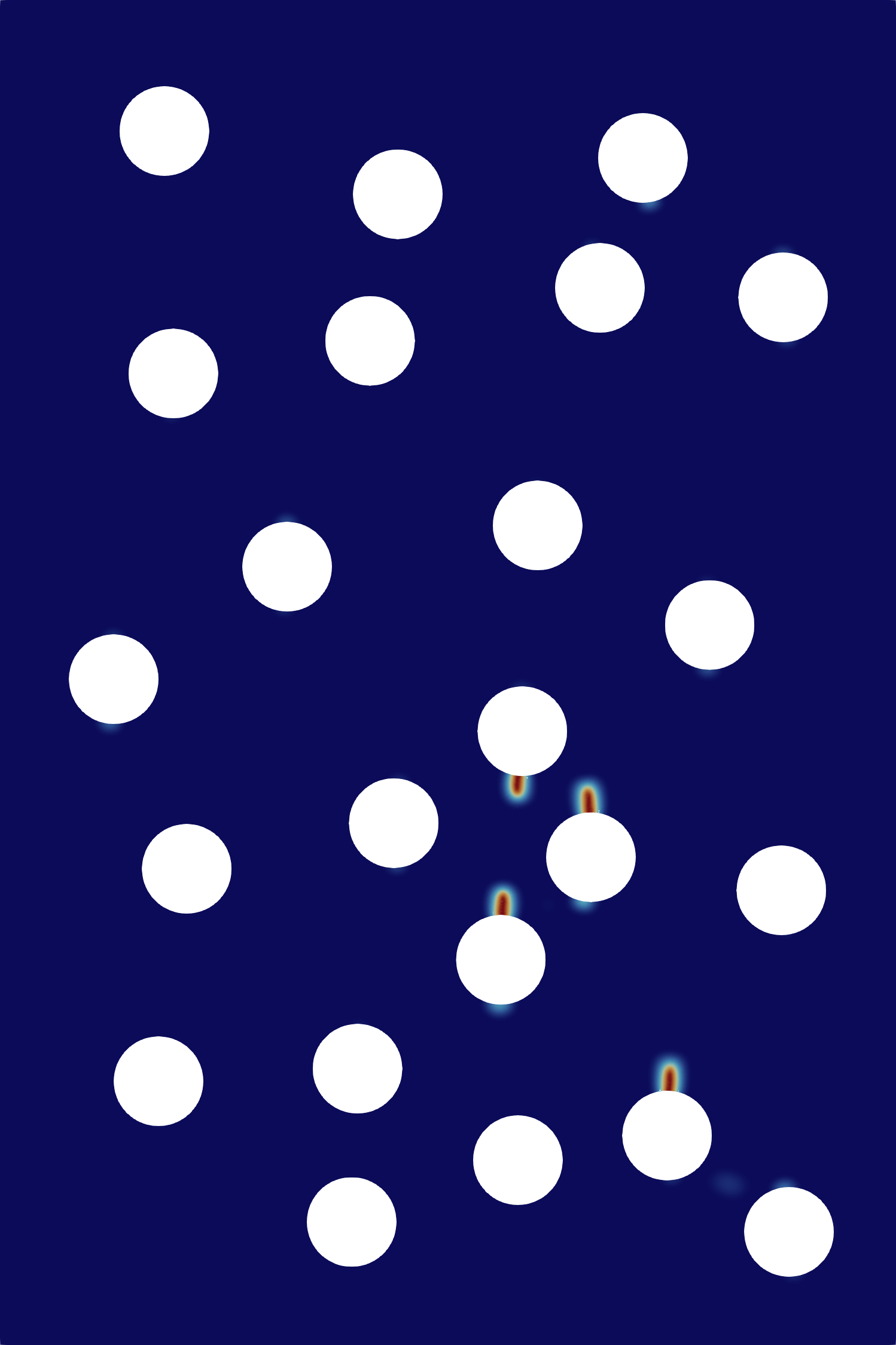}}
        \\
        \bottomrule
    \end{tabularx}
    \begin{center}
        \includegraphics{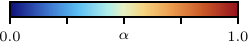}
    \end{center}
\end{table}

\subsection{Implementational details}\label{app:implementational_details}
We rely on \texttt{FEniCSx} v0.9.0 \cite{baratta_2023,scroggs_2022a,scroggs_2022b} and \texttt{PETSc} release 3.23.5 \cite{balay_2024a,balay_2024b,balay_1997} for our implementation of the phase-field fracture model and the algorithms in this study.
We use \texttt{UFL} \cite{alnaes_2014} to define the energy functional and to automatically obtain the residuals and their linearization, which are passed to \texttt{PETSc} through the \texttt{petsc4py} interface \cite{dalcin_2011}.
To solve the two non-linear subproblems of \eqref{eq:1_order}, we rely on the \texttt{SNES} Newton-type solver class of \texttt{PETSc}, namely the standard \texttt{NEWTONLS} for the mechanical problem and the damage problem when using the penalized formulation, or \texttt{SNESVINEWTONRSLS} when using the reduced-space active set strategy for the damage problem (and for the mechanical problem in case of the Brazilian test in Section~\ref{sec:brazilian3D}).
For the linear solve in each Newton iteration, we use the \texttt{MUMPS} package \cite{amestoy_2001,amestoy_2019} with either the Cholesky factorization, or the LU factorization for the damage subproblem in the case of the star-convex model with $\gamma^\star > 0$.
To re-use the phase-field stiffness matrix for the damage problem when adopting the reduced-space active set strategy, we make use of the \texttt{snes\_lag\_jacobian} option of \texttt{PETSc}.

For the sampling of $\|\mathbf{R}_{\mathbf{u}}\|_2 (\lambda)$, $\phi(\lambda)$ and $\phi^\prime(\lambda)$, we employ a custom \texttt{SNESLineSearchPreCheck} routine written in \texttt{C}, which we directly hook to the \texttt{SNESLineSearch} object.

Our implementations are available at \url{https://github.com/jonas-heinzmann/phase_field_exact_linesearch}, where the detailed configurations and versions can also be found.
The bisection line search is available in the \texttt{PETSc} repository at \href{https://gitlab.com/petsc/petsc/-/blob/02fac80991349f649e48b0bbdbfd635de889974f/src/snes/linesearch/impls/bisection/linesearchbisection.c}{\texttt{src/snes/linesearch/impls/bisection/linesearchbisection.c}}.
All computations were run on comparable hardware on the Euler cluster of ETH Zürich.

\subsection{Global convergence of Newton's method with exact line search}\label{app:convergence_proof}
In the following we prove Proposition~\ref{prop:convergence_proof_Newtonextls}, based on and expanding upon \cite{shea2025greedy}.
We define the function
\begin{equation}
    D(\lambda,\mathbf{w}^k) = E(\mathbf{w}^k + \lambda \Delta\mathbf{w}^k) - E(\mathbf{w}^k) \quad\text{.}
\end{equation}
By virtue of the Mean Value Theorem, there exists a scalar $\chi \in [0,1]$ such that, by defining
\begin{equation}
    \boldsymbol{\xi} = \mathbf{w}^k + \chi \, \lambda \, \Delta \mathbf{w}^k \quad\text{,}
\end{equation}
we can write the Taylor expansion as
\begin{equation}\label{eq:taylor}
    D(\lambda,\mathbf{w}^k) = \lambda\, \mathbf{R}_{\mathbf{w}}(\mathbf{w}^k)^\intercal \,\Delta \mathbf{w}^k
   + \frac{\lambda^2}{2}\, {\Delta \mathbf{w}^k}^\intercal \mathbf{K}_{\mathbf{w}\mathbf{w}}(\boldsymbol{\xi}) \,\Delta \mathbf{w}^k \quad\text{.}
\end{equation}
Applying the eigenvalue bounds~\eqref{eq:eigval_bounds} to \eqref{eq:taylor}, we obtain
\begin{equation}
    \begin{aligned}
    D(\lambda,\mathbf{w}^k) 
    &\leq \lambda\, \mathbf{R}_{\mathbf{w}}(\mathbf{w}^k)^\intercal \,\Delta \mathbf{w}^k
    + \frac{\lambda^2}{2}\, \rho_\text{U} \, \Vert \Delta \mathbf{w}^k \Vert^2 \\
    &= -\lambda\, \mathbf{R}_{\mathbf{w}}(\mathbf{w}^k)^\intercal
        \mathbf{K}_{\mathbf{w}\mathbf{w}}^{-1}(\mathbf{w}^k) \,\mathbf{R}_{\mathbf{w}}(\mathbf{w}^k)
    + \frac{\lambda^2}{2}\, \rho_\text{U} \, \Vert \mathbf{K}_{\mathbf{w}\mathbf{w}}^{-1}(\mathbf{w}^k)
        \mathbf{R}_{\mathbf{w}}(\mathbf{w}^k) \Vert^2 \\
    &= -\lambda\, \mathbf{R}_{\mathbf{w}}(\mathbf{w}^k)^\intercal
        \mathbf{K}_{\mathbf{w}\mathbf{w}}^{-1}(\mathbf{w}^k) \,\mathbf{R}_{\mathbf{w}}(\mathbf{w}^k)
    + \frac{\lambda^2}{2}\, \rho_\text{U} \,
        \mathbf{R}_{\mathbf{w}}(\mathbf{w}^k)^\intercal \mathbf{K}_{\mathbf{w}\mathbf{w}}^{-2}(\mathbf{w}^k)
        \mathbf{R}_{\mathbf{w}}(\mathbf{w}^k) \\
    &\leq -\lambda\, \mathbf{R}_{\mathbf{w}}(\mathbf{w}^k)^\intercal
        \mathbf{K}_{\mathbf{w}\mathbf{w}}^{-1}(\mathbf{w}^k) \,\mathbf{R}_{\mathbf{w}}(\mathbf{w}^k)
    + \frac{\lambda^2}{2}\, \frac{\rho_\text{U}}{\rho_\text{L}} \,
        \mathbf{R}_{\mathbf{w}}(\mathbf{w}^k)^\intercal \mathbf{K}_{\mathbf{w}\mathbf{w}}^{-1}(\mathbf{w}^k)
        \mathbf{R}_{\mathbf{w}}(\mathbf{w}^k) \\
    &= - g(\lambda)\, P(\mathbf{w}^k),
    \end{aligned}
\end{equation}
where
\begin{equation}
    g(\lambda) = \lambda \left(1 - \frac{\lambda}{2} \frac{\rho_\text{U}}{\rho_\text{L}}\right)
    \quad\text{, and}\quad
    P(\mathbf{w}^k) = \mathbf{R}_{\mathbf{w}}(\mathbf{w}^k)^\intercal 
    \mathbf{K}_{\mathbf{w}\mathbf{w}}^{-1}(\mathbf{w}^k) \mathbf{R}_{\mathbf{w}}(\mathbf{w}^k) \quad\text{.}
\end{equation}
Hence, it follows that
\begin{equation}\label{eq:D_gP}
    D(\lambda,\mathbf{w}^k) \leq - g(\lambda)\, P(\mathbf{w}^k)\quad\text{,}
\end{equation}
with $P(\mathbf{w}^k) \geq 0$ due to the strict convexity of $E(\mathbf{w})$ with respect to $\mathbf{w}$.

Thanks to the exact line search where we minimize the energy over $\lambda \in [0,1]$, we can write for the left-hand side of~\eqref{eq:D_gP}
\begin{equation}\label{eq:exact_line}
    E(\mathbf{w}^{k+1}) - E(\mathbf{w}^k) = \min_{\omega \in [0,1]} D(\omega, \mathbf{w}^k) 
    \leq D(\lambda, \mathbf{w}^k) \quad \forall \lambda \in [0,1] \quad\text{.}
\end{equation}
Combining \eqref{eq:D_gP} and \eqref{eq:exact_line} gives
\begin{equation}
    E(\mathbf{w}^{k+1}) - E(\mathbf{w}^k) \leq - g(\lambda) \, P(\mathbf{w}^k) \quad \forall \lambda \in [0,1] \quad\text{.}
\end{equation}
In particular, being
\begin{equation}
 \min_{\lambda \in [0,1]} -g(\lambda) = -\frac{1}{2} \, \frac{\rho_\text{L}}{\rho_\text{U}} \quad\text{,}
\end{equation}
it is also
\begin{equation}
    \begin{aligned}
        E(\mathbf{w}^{k+1}) - E(\mathbf{w}^k) \leq -\frac{1}{2} \, \frac{\rho_\text{L}}{\rho_\text{U}} \, P(\mathbf{w}^k)
        &= -\frac{1}{2} \, \frac{\rho_\text{L}}{\rho_\text{U}} \, \mathbf{R}_{\mathbf{w}}(\mathbf{w}^k)^\intercal 
            \mathbf{K}_{\mathbf{w}\mathbf{w}}^{-1}(\mathbf{w}^k) \mathbf{R}_{\mathbf{w}}(\mathbf{w}^k) \\
        &\leq -\frac{1}{2} \, \frac{\rho_\text{L}}{\rho_\text{U}^2} \, \Vert\mathbf{R}_{\mathbf{w}}(\mathbf{w}^k)\Vert^2 \quad\text{.}
    \end{aligned}
\end{equation}
Recursively applying the above inequality and changing the signs, we obtain
\begin{equation}\label{eq:R_dE}
    \frac{1}{2}\,\frac{\rho_\text{L}}{\rho_\text{U}^2}\,\sum_{q=0}^k \Vert\mathbf{R}_{\mathbf{w}}(\mathbf{w}^q)\Vert^2
    \leq E(\mathbf{w}^{0}) - E(\mathbf{w}^{k+1}) \quad\text{.}
\end{equation}
By continuity and coercivity, $E(\mathbf{w})$ is lower bounded.
Due to the strict convexity of $E(\mathbf{w})$ with respect to $\mathbf{w}$, we have $\rho_\text{L} > 0$.
Thus, taking the limit as $k \rightarrow \infty$ in \eqref{eq:R_dE} directly implies
\begin{equation}
    \sum_{k=0}^\infty \Vert\mathbf{R}_{\mathbf{w}}(\mathbf{w}^k)\Vert^2 < \infty \quad\text{,}
\end{equation}
and a positive sequence with finite sum is known to converge to zero, i.e.
\begin{equation}\label{eq:convergence_f}
    \lim_{k \rightarrow \infty} \Vert\mathbf{R}_{\mathbf{w}}(\mathbf{w}^k)\Vert^2 = 0
\end{equation}
which concludes the proof of Proposition~\ref{prop:convergence_proof_Newtonextls}. \qed

\subsection{Gallery of the energy and directional derivative sampled along the search direction}\label{app:linesearch_sampling}
\begin{figure}[t]
    \centering
    \begin{subfigure}[t]{\textwidth}
        \centering
        \includegraphics{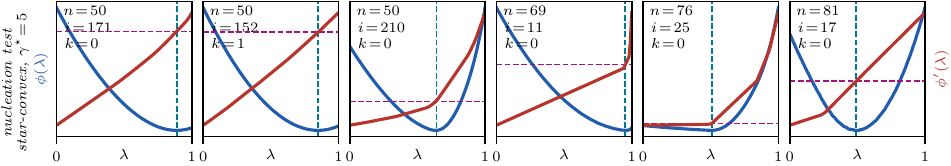}
        \caption{}
        \label{fig:linesearch_sampling_displacement_a}
    \end{subfigure}
    \begin{subfigure}[t]{\textwidth}
        \centering
        \includegraphics{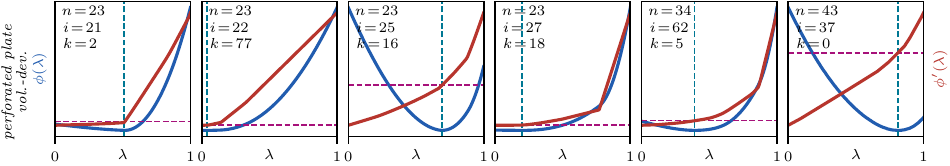}
        \caption{}
        \label{fig:linesearch_sampling_displacement_b}
    \end{subfigure}
    \caption{Sampling of the energy and directional derivative along the Newton direction for the mechanical problem, for selected examples from the nucleation test (a) and the perforated plate (b).
    Here, it is $\phi(\lambda)=E(\mathbf{u}^{i,k}+\lambda \Delta \mathbf{u}^{i,k}, \bm{\alpha}^{i-1})$.
    The value of $\lambda$ which minimizes the energy and the root of the directional derivative are highlighted with a vertical and a horizontal dashed line, respectively.}
    \label{fig:linesearch_sampling_displacement}
\end{figure}

\begin{figure}[t]
    \centering
    \begin{subfigure}[t]{\textwidth}
        \centering
        \includegraphics{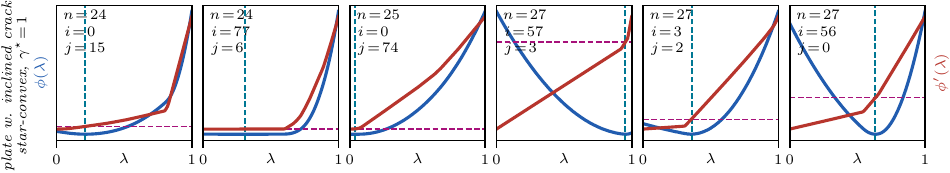}
        \caption{}
        \label{fig:linesearch_sampling_phasefield_b}
    \end{subfigure}
    \begin{subfigure}[t]{\textwidth}
        \centering
        \includegraphics{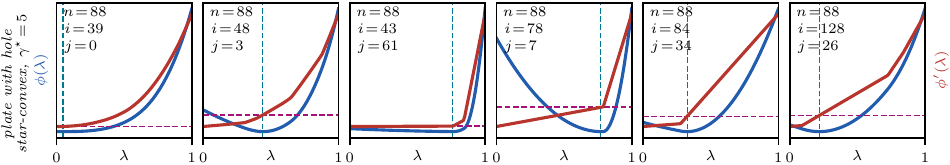}
        \caption{}
        \label{fig:linesearch_sampling_phasefield_d}
    \end{subfigure}
    \caption{Sampling of the energy and directional derivative along the Newton direction for the damage problem with the penalized formulation, for selected examples from the plate with inclined crack test (a) and the plate with hole test (b).
    Here, it is $\phi(\lambda)=\tilde{E}(\mathbf{u}^i,\bm{\alpha}^{i,j}+\lambda \Delta \bm{\alpha}^{i,j})$.
    The value of $\lambda$ which minimizes the energy and the root of the directional derivative are highlighted with a vertical and a horizontal dashed line, respectively.}
    \label{fig:linesearch_sampling_phasefield}
\end{figure}

In Figs. \ref{fig:linesearch_sampling_displacement} and \ref{fig:linesearch_sampling_phasefield}, we report various examples of the energy and its directional derivative along the Newton search direction, either for the mechanical problem or for the damage problem (when using the penalized formulation).
In both cases, the directional derivative shows distinct kinks or regions where the energy is nearly flat.
This behavior can be attributed to the definitions of the strain energy density splits for the mechanical problem and to the penalty term for the damage problem.
These plots further illustrate the necessity of the exact line search strategy proposed in this work.

\subsection{Other line search algorithms}\label{app:other_line_search_algorithms}
In the following, we give short descriptions of the line-search algorithms used for the performance comparisons in Section~\ref{sec:performance_comparison}.
Among the large variety of algorithms proposed in the literature, we select three options available for the \texttt{SNES} solver class of \texttt{PETSc} (release \texttt{3.23.5}) \cite{balay_2024a,balay_2024b,balay_1997}, we refer to the source code for more details.

\subsubsection{Backtracking line search}
\begin{figure}[t]
    \centering
    \begin{subfigure}[t]{0.49\textwidth}
        \centering
        \includegraphics{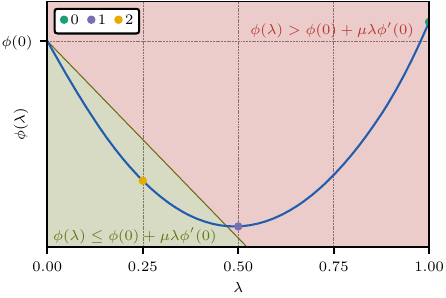}
        \caption{}
        \label{fig:linesearch_backtracking}
    \end{subfigure}
    \begin{subfigure}[t]{0.49\textwidth}
        \centering
        \includegraphics{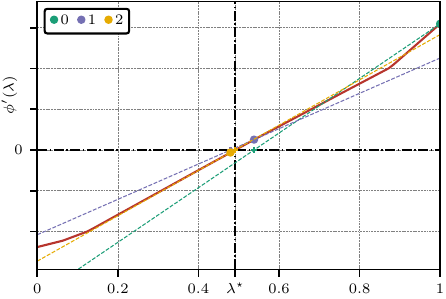}
        \caption{}
        \label{fig:linesearch_secant}
    \end{subfigure}
    \caption{(a) Schematic of the backtracking line search, applied to the example of the energy in Fig.~\ref{fig:toyproblem_linesearch_sampling} at $n=50$, $i=32$, $k=14$.
    The areas fulfilling the sufficient decrease condition are highlighted in green, those not in red. The various iterates $\lambda_l$ are represented by dots, with the first two violating \eqref{eq:armijo}, and only after the second halving step, a sufficient decrease of the energy is achieved. 
    (b) Schematic of a secant-based line search, again for the example of the directional derivative in Fig.~\ref{fig:toyproblem_linesearch_sampling}.
    The iterations and their determined secants are visualized with different colors, while the root in the directional derivative is highlighted with dash-dotted lines.}
    \label{fig:linesearch_backtracking_bisection}
\end{figure}

The first chosen algorithm is a simple \textit{backtracking} line search (\texttt{SNESLineSearchBT} in \texttt{PETSc}).
In the given implementation, its objective is to determine $\lambda$ such that the \textit{Armijo} or \textit{sufficient decrease} condition
\begin{equation}\label{eq:armijo}
    \phi (\lambda) \leq \phi (0) + \mu \lambda \phi^\prime (0)
\end{equation}
is fulfilled.
Here, $\mu > 0$ is a parameter determining the minimal accepted decrement of the objective function.
A straightforward algorithmic strategy to find the desired value of $\lambda$ is to backtrack, i.e. to iteratively half the step size $\lambda_{l+1} = \lambda_l / 2$ starting from the full Newton step $\lambda_0=1$.
This is repeated until either \eqref{eq:armijo} is fulfilled -- as illustrated in Fig.~\ref{fig:linesearch_backtracking} -- or until a maximum number of iterations $l_{\max}$ is reached (we choose $l_{\max}=10$).
As objective function $\phi (\lambda)$ we choose either $\tfrac{1}{2}\|\mathbf{R}_{\mathbf{w}}\|_2^2 (\lambda)$ with the commonly used parameter $\mu=10^{-4}$ \cite{martins_enginnering_2022,asher_first_2011}, or the energy $E(\lambda)$ with $\mu=1$ (since, with the energy, $\mu=10^{-4}$ did not yield satisfactory results).
The computational cost of this algorithm is related to a one-time evaluation of the derivative at the initial point $\phi^\prime (0)$, plus one evaluation of the objective function at $\mathbf{w}^k + \lambda_l \Delta \mathbf{w}^k$ per line search iteration.
The former is $\phi^\prime (0) = \mathbf{R}_{\mathbf{w}}^\intercal \mathbf{K}_{\mathbf{ww}} \Delta \mathbf{w}$ for the residual-based approach, or $\phi^\prime (0) = \mathbf{R}_{\mathbf{w}}^\intercal \Delta \mathbf{w}$ (appropriately adjusted for the reduced-space method) for the energy-based approach.
While the backtracking algorithm guarantees a continuous descent of the energy by virtue of \eqref{eq:armijo}, the algorithm may still stagnate for the case of a complicated objective function in the search direction $\phi(\lambda)$.
Depending on the choice of $\mu$, the line search may even either fail to, or wrongly accept the full step $\Delta \mathbf{w}^k$.
An enhancement of this method is e.g. to employ a quadratic or cubic model to obtain a better iterative update for the step size in an interval of e.g. $\lambda_{l+1} \in [0.1, 0.5] \lambda_l$, thus backtracking more aggressively, as outlined in \cite{dennis_numerical_1996}.
However, preliminary studies in \cite{heinzmann_linesearch_2025} suggest that these variants are not necessarily advantageous for the phase-field problem, which is why they are not used in this work.

\subsubsection{Secant-based line searches}
Secondly, we compare the bisection line search to secant-based methods.
These look for minima in the objective function $\phi (\lambda)$ by employing Newton's method, hence their iterative update reads
\begin{equation}\label{eq:LS_sec}
    \lambda_{l+1} = \lambda_l - \frac{\phi^\prime_l}{\phi^{\prime\prime}_l} \qquad \text{;}
\end{equation}
an illustration of the iterative procedure is given in Fig.~\ref{fig:linesearch_secant}.
In this work, we use three different implementations, depending on the choice of the objective function and on the approximation of the first and second derivative, $\phi^\prime_l$ and $\phi^{\prime\prime}_l$.
\begin{itemize}
    \item In the first implementation, we choose as objective $\|\mathbf{R}_{\mathbf{w}}\|_2^2 (\lambda)$ and employ a finite-difference approximation for the derivatives, namely
    \begin{equation}
        \phi^\prime_l = \frac{3 \phi(\lambda_l) - 4 \phi\left(\frac{\lambda_l + \lambda_{l-1}}{2}\right) + \phi(\lambda_{l-1})}{\lambda_l - \lambda_{l-1}}
    \end{equation}
    for the first derivative, and
    \begin{equation}
        \phi^{\prime\prime}_l = \frac{\phi^\prime_l - \phi^\prime_{l-1}}{\lambda_l - \lambda_{l-1}} \qquad \text{with} \qquad \phi^\prime_{l-1} = \frac{-3 \phi(\lambda_{l-1}) + 4 \phi\left(\frac{\lambda_l + \lambda_{l-1}}{2}\right) - \phi(\lambda_{l})}{\lambda_l - \lambda_{l-1}}
    \end{equation}
    for the second derivative.
    Hence, two additional residual evaluations are required per line search iteration, see \cite{Brune_composing_2015} for more details.
    This is implemented as \texttt{SNESLineSearchL2} in \texttt{PETSc} and does not feature any convergence checks based on an absolute or relative tolerance or the change of $\lambda$ in release \texttt{3.23.5}.
    Instead, it prescribes a maximum number of iterations, $l \leq l_{\max}$.
    As before, we set $l_{\max}=10$.
    The \texttt{PETSc} implementation features an additional check, reversing the step direction in \eqref{eq:LS_sec} based on the sign of $\phi^{\prime\prime}_l$; we refer to the source code for more details.
    \item In the second version, we use the same algorithm as before but with $E(\lambda)$ as objective.
    \item As third option, we use the \textit{critical point} line search \cite{Brune_composing_2015}.
    This still follows the iterative update \eqref{eq:LS_sec}, however $\phi^\prime (\lambda)$ is not approximated, but obtained from the directional derivative \eqref{eq:dir_deriv} (or \eqref{eq:dir_deriv_rs} for the reduced-space solver).
    For the second derivative we use the second-order finite-difference approximation\footnote{In principle, the second derivative could be computed as $\phi^{\prime\prime} (\lambda) = (\Delta \mathbf{w}^k)^\intercal \mathbf{K}_{\mathbf{ww}} (\mathbf{w}^k + \lambda \Delta \mathbf{w}^k) \Delta \mathbf{w}^k$. However, the evaluation of the stiffness matrix at $\mathbf{w}^k + \lambda \Delta \mathbf{w}^k$ would make the line search prohibitively expensive.} 
    \begin{equation}
        \phi^{\prime\prime}_l \approx \frac{3 \phi^\prime(\lambda_l) - 4 \phi^\prime \left( \frac{\lambda_l + \lambda_{l-1}}{2} \right) + \phi^\prime(\lambda_{l-1})}{\lambda_l - \lambda_{l-1}} \qquad\text{.}
    \end{equation}
    This algorithm is implemented in \texttt{PETSc} as \texttt{SNESLineSearchCP} , where convergence is checked based on an absolute tolerance $\phi^\prime (\lambda_l)/ \|\Delta \mathbf{w}\|_2 \leq \mathtt{atol}$, a relative tolerance $\phi^\prime (\lambda_l) / \phi^\prime (0) \leq \mathtt{rtol}$, or $\lambda_l - \lambda_{l-1} \leq \mathtt{ltol}$.
    As with the proposed exact line search based on bisection, we do not use $\mathtt{rtol}$, while we set $\mathtt{atol} = 10^{-12}$, and $\mathtt{ltol} = 10^{-6}$.
    We also set a maximum number of iterations again equal to $l_{\max}=10$.
    The computational effort of this method entails two evaluations of the directional derivative per iteration, i.e. at $\phi^\prime(\lambda_l)$ and at $\phi^\prime \left( (\lambda_l + \lambda_{l-1}) / 2 \right)$.
    Similarly to \texttt{SNESLineSearchL2}, \texttt{SNESLineSearchCP} features additional checks and possibly a change of the iterative update direction in \texttt{PETSc} release \texttt{3.23.5}; we refer to the source code for details.
\end{itemize}
While the backtracking algorithm automatically determines $\lambda^\star \in (0,1]$ with the choice $\lambda_0=1$, we enforce $\lambda^\star\leq 1$ for the secant-based line searches to facilitate the comparison with the proposed bisection line search.
The chosen algorithms may perform better or worse depending on how the parameters are set. 
Our settings aimed at achieving a generally satisfactory behavior for the problems in the obstacle course; a detailed study or optimization of their influence goes beyond the scope of this work.

\end{document}